\documentclass[prd,aps,showpacs,epsf,floats,onecolumn]{revtex4}%
\usepackage{amssymb}
\usepackage{amsfonts}
\usepackage{amsmath}
\usepackage{graphicx}%
\providecommand{\U}[1]{\protect\rule{.1in}{.1in}}
\begin{document}
\title{\textbf{From the Classical Frenet-Serret Apparatus to the Curvature and
Torsion of Quantum-Mechanical Evolutions. Part II. Nonstationary Hamiltonians}}
\author{\textbf{Paul M. Alsing}$^{1}$ and \textbf{Carlo Cafaro}$^{2,3}$}
\affiliation{$^{1}$Air Force Research Laboratory, Information Directorate, Rome, NY 13441, USA}
\affiliation{$^{2}$University at Albany-SUNY, Albany, NY 12222, USA}
\affiliation{$^{3}$SUNY Polytechnic Institute, Utica, NY 13502, USA}

\begin{abstract}
We present a geometric perspective on how to quantify the bending and the
twisting of quantum curves traced by state vectors evolving under
nonstationary Hamiltonians. Specifically, relying on the existing geometric
viewpoint for stationary Hamiltonians, we discuss the generalization of our
theoretical construct to time-dependent quantum-mechanical scenarios where
both time-varying curvature and torsion coefficients play a key role.
Specifically, we present a quantum version of the Frenet-Serret apparatus for
a quantum trajectory in projective Hilbert space traced out by a
parallel-transported pure quantum state evolving unitarily under a
time-dependent Hamiltonian specifying the Schr\"{o}dinger evolution equation.
The time-varying curvature coefficient is specified by the magnitude squared
of the covariant derivative of the tangent vector $\left\vert T\right\rangle $
to the state vector $\left\vert \Psi\right\rangle $ and measures the bending
of the quantum curve. The time-varying torsion coefficient, instead, is given
by the magnitude squared of the projection of the covariant derivative of the
tangent vector $\left\vert T\right\rangle $ to the state vector $\left\vert
\Psi\right\rangle $, orthogonal to $\left\vert T\right\rangle $ and
$\left\vert \Psi\right\rangle $ and, in addition, measures the twisting of the
quantum curve. We find that the time-varying setting exhibits a richer
structure from a statistical standpoint. For instance, unlike the
time-independent configuration, we find that the notion of generalized
variance enters nontrivially in the definition of the torsion of a curve
traced out by a quantum state evolving under a nonstationary Hamiltonian. To
physically illustrate the significance of our construct, we apply it to an
exactly soluble time-dependent two-state Rabi problem specified by a
sinusoidal oscillating time-dependent potential. In this context, we show that
the analytical expressions for the curvature and torsion coefficients are
completely described by only two real three-dimensional vectors, the Bloch
vector that specifies the quantum system and the externally applied
time-varying magnetic field. Although we show that the torsion is identically
zero for an arbitrary time-dependent single-qubit Hamiltonian evolution, we
study the temporal behavior of the curvature coefficient in different
dynamical scenarios, including off-resonance and on-resonance regimes and, in
addition, strong and weak driving configurations. While our formalism applies
to pure quantum states in arbitrary dimensions, the analytic derivation of
associated curvatures and orbit simulations can become quite involved as the
dimension increases. Thus, finally we briefly comment on the possibility of
applying our geometric formalism to higher-dimensional qudit systems that
evolve unitarily under a general nonstationary Hamiltonian.

\end{abstract}

\maketitle

\section{Introduction}

Geometry is a very useful tool in physics since it helps improving our
description and, to a certain extent, our understanding of \ classical and
quantum physical phenomena by providing deep physical insights via geometric
intuition \cite{pettini07}. For example, it is known to geometers that the
classical Frenet-Serret apparatus of a space curve in three-dimensional
Euclidean space characterizes the local geometry of curves and specifies
relevant geometric invariants such as the curvature and the torsion of a curve
\cite{parker77}. In classical Newtonian mechanics, curvature and torsion
coefficients introduced within the Frenet-Serret apparatus can help physicists
describing the geometric properties of classical trajectories of a particle.
In particular, curvature and torsion coefficients of a space curve can be
employed to study the geometry of the cylindrical helix motion of a spin-$1/2$
charged particle (for instance, an electron) in a homogeneous external
magnetic field \cite{consa18}. Since the curvature of a space curve can be
expressed in terms of velocity and acceleration \cite{parker77}, it is
reasonable to expect that the temporal rate of change of the curvature is
somewhat related to the time-derivative of the acceleration, also know as the
jerk. For an interesting discussion on the jerk of space curves corresponding
to electrons moving in a constant magnetic field in three-dimensional
Euclidean space equipped with moving Frenet-Serret frames, we refer to Refs.
\cite{ozen19,ozen20}. The jerk, viewed as a predictor of large accelerations
of short durations, has significant practical applications in engineering,
particularly in robotics and automation \cite{saridis88}. In classical
engineering applications, the goal is to minimize the jerk. This minimization
is justified by the fact that the efficiency of control algorithms is affected
in a negative fashion by the third derivative of the position (i.e., the jerk)
along the desired trajectory \cite{saridis88}. In quantum information
processing, the smart manipulation of quantum states that encode quantum
information about a physical system is an extremely valuable task to
accomplish \cite{nielsen00}. Clever ways of optimally transport quantum source
states to quantum target states requires minimizing the complexity
\cite{brown17,chapman18,brown19,auzzi21} along with maximizing the efficiency
of the quantum processes
\cite{uzdin12,campaioli19,cafaroQR,carloprd22,carlopre22,carlocqg23}. In this
respect, quantum researchers have been recently using minimum jerk trajectory
techniques \cite{wise05} to get optimal paths along which efficiently
transport atoms in optical lattices \cite{liu19,matt23}. Indeed, discovering
efficient ways to manipulate and control atoms with minimal losses is
fundamental for the coherent formation of single ultracold molecules suitable
for investigations in quantum engineering and information processing.

%PMA: make this next part a new paragraph
While studying the problem of parameter estimation in the context of geometric
quantum mechanics, the notion of curvature of a quantum Schr\"{o}dinger
trajectory was proposed in Ref. \cite{brody96} as an extension of the concept
of curvature of a classical exponential family of distributions of interest in
statistical mechanics. In Ref. \cite{brody96}, the curvature of a curve can be
specified by means of a conveniently defined squared acceleration vector of
the curve and represents a measure of the parametric sensitivity
\cite{brody13} that characterizes the parametric estimation problem at study.
In Ref. \cite{laba17}, Laba and Tkachuk presented a definition of curvature
and torsion coefficients of quantum evolutions for pure quantum state evolving
under a stationary Hamiltonian evolution. Limiting the attention to
single-qubit quantum states, the curvature measures the deviation of the
dynamically evolving state vector from the geodesic line on the Bloch sphere
in Ref. \cite{laba17}. \ Instead, the torsion coefficient \ measures the
deviation of the dynamically evolving state vector from a two-dimensional
subspace defined by the instantaneous plane of evolution.

Relying on the concept of Frenet-Serret apparatus and, in part, inspired by
the work by Laba and Tkachuk in Ref. \cite{laba17}, we offered in Ref.
\cite{alsing1} a geometric viewpoint on how to characterize the bending and
the twisting of quantum curves traced by dynamically evolving state vectors.
Specifically, we proposed a quantum version of the Frenet-Serret apparatus for
a quantum curve in projective Hilbert space traced by a parallel-transported
pure quantum state that evolves unitarily under a stationary Hamiltonian. In
our approach, the constant curvature coefficient is expressed as the magnitude
squared of the covariant derivative of the tangent vector $\left\vert
T\right\rangle $ to the state vector $\left\vert \Psi\right\rangle $ and
measures the bending of the quantum curve. Our constant torsion coefficient
measures the twisting of the quantum curve and is given by the magnitude
squared of the projection of the covariant derivative of the tangent vector
$\left\vert T\right\rangle $, perpendicular to both $\left\vert T\right\rangle
$ and $\left\vert \Psi\right\rangle $. For a discussion on the merits and
limitations of our geometric approach to curvature and torsion of quantum
evolutions specified by stationary Hamiltonians, we refer to Ref.
\cite{alsing1}. It suffices to remark here that one of the major limitations
of our work in Ref. \cite{alsing1} is its restricted applicability to
time-independent Hamiltonian evolutions. Therefore, our primary goal in this
paper if to go beyond stationary Hamiltonian evolutions. Unfortunately, it is
known that it is rather complicated to get exact analytical solutions to the
time-dependent Schr\"{o}dinger equations, even for two-level quantum systems.
The first illustrations of analytically solvable two-state time-dependent
problems were reported by Landau-Zener in Refs. \cite{landau32,zener32} and
Rabi in Refs. \cite{rabi37,rabi54}. The effort in discovering analytical
solutions to time-dependent quantum evolutions has been rather intense
throughout the years
\cite{barnes12,barnes13,messina14,grimaudo18,cafaroijqi,grimaudo23}.
Interestingly, this effort has propagated to finding exact analytical
solutions for particular classes of non-stationary Hamiltonians for $d$-level
quantum systems (i.e., qudits) as well, as evident from Ref. \cite{elena20}.

In this paper, our main objectives can be outlined as follows:

\begin{enumerate}
\item[{[i]}] Extend our geometric approach, originally developed for
stationary Hamiltonians in Ref. \cite{alsing1}, to nonstationary Hamiltonians
of arbitrary finite-dimensional quantum systems.

\item[{[ii]}] Study the richer statistical structure underlying the notions of
curvature and torsion of a quantum evolution governed by a nonstationary
Hamiltonian, going beyond the concepts of skewness and kurtosis reported in
Ref. \cite{alsing1}.

\item[{[iii]}] Provide a closed-form expression for the curvature of a
two-level system that evolves under a time-dependent Hamiltonian in terms of
only two real three-dimensional vectors, i.e., the Bloch vector and the
magnetic (field) vector. This is especially useful due to the computability
challenges that emerge, unlike what happens in the stationary setting of Ref.
\cite{alsing1}, in the nonstationary scenario when one attempts to express the
physical time parameter in terms of the arc length parameter. Interestingly,
the fact that the Bloch vector representation of qubits offers a very
intuitive picture of the dynamics of the quantum system was already pointed
out by Feynman and collaborators in Ref. \cite{dick57} while solving maser
(microwave amplification by stimulated emission of radiation) problems by
studying the behavior of an ensemble of noninteracting two-level quantum
systems in the presence of an external perturbation.

\item[{[iv]}] Discuss the physical insights into the dynamics of two-level
quantum systems that can be obtained by applying our geometric approach to a
time-dependent problem of physical relevance. Specifically, we shall apply our
geometric construct to an exactly soluble time-dependent two-state Rabi
problem specified by a sinusoidal oscillating time-dependent potential. This
task is particularly significant since a time-dependent problem has a much
richer structure that a time-independent one Ref. \cite{alsing1}. In many
ways, it is more representative since, for instance, the curvature and the
torsion coefficients can generally exhibit rather complicated temporal behaviors.
\end{enumerate}

The rest of the paper is organized as follows. In Section II, we present the
extension of our geometric characterization of curvature and torsion
coefficients of quantum evolutions specified by arbitrary nonstationary
Hamiltonians. In Section III, we present a statistical interpretation of the
various terms that specify the time-dependent curvature and torsion
coefficients when expressed in terms of expectation values of conveniently
defined operators constructed out of the original nonstationary Hermitian
Hamiltonian operator. After pointing out some computability challenges related
to the definitions of the curvature and torsion coefficients by means of the
projector operator formalism or, alternatively, in terms of the previously
mentioned expectations values, we focus in\ Section IV on nonstationary
Hamiltonians and two-level quantum systems.\ Specifically, we provide a
closed-form expression for the curvature coefficient for a curve traced by a
single-qubit quantum state that evolves under an arbitrary time-dependent
Hamiltonian in terms of only two real three-dimensional vectors with a clear
geometric visualization. These two vectors are the Bloch vector and the
magnetic (field) vector, respectively. In Section V, exploiting part of the
results presented in Section IV, we study the temporal behavior of the
time-dependent curvature coefficient of the quantum curves generate by
evolving a single-qubit quantum state under the Hamiltonian $\mathrm{H}\left(
t\right)  \overset{\text{def}}{=}(\hslash\omega_{0}/2)\sigma_{z}+\hslash
\Omega_{0}\left[  \cos\left(  \omega t\right)  \sigma_{x}+\sin\left(  \omega
t\right)  \sigma_{y}\right]  $ under distinct physical regimes. The physical
parameters that define this Hamiltonian are the resonance frequency of the
atom $\omega_{0}$, the energy difference $\hslash\omega_{0}$ between the
ground and excited states, the Rabi frequency $\Omega_{0}$ and, finally, the
external driving frequency $\omega$. Our final remarks appear in Section VI.
Finally, for the ease of readability of the paper, several technical details
are located in the Appendices.

\section{General formalism}

In this section, we present the extension of our geometric characterization of
curvature and torsion coefficients of quantum evolutions specified by
arbitrary nonstationary Hamiltonians. The state vectors are assumed to belong
to an arbitrary $N$-dimensional complex Hilbert space $\mathcal{H}_{N}$. For
example, $\mathcal{H}_{N}=\mathcal{H}_{d}^{k}$ with $N=d^{k}$ and
$\mathcal{H}_{d}^{k}$ being the Hilbert space of $k$-qudit quantum states
\cite{maiolini}. Furthermore, we assume to consider a time-dependent
Hamiltonian evolution specified by the Schr\"{o}dinger's equation
$i\hslash\partial_{t}\left\vert \psi\left(  t\right)  \right\rangle
=\mathrm{H}\left(  t\right)  \left\vert \psi\left(  t\right)  \right\rangle $.
Finally, to construct the time-dependent analog of set of orthonormal vectors
$\left\{  \left\vert \Psi\left(  s\right)  \right\rangle \text{, }\left\vert
T\left(  s\right)  \right\rangle \text{, }\left\vert N\left(  s\right)
\right\rangle \right\}  $ obtained in the time-independent setting, we follow
closely our time-independent construction in Ref. \cite{alsing1}.

\subsection{Preliminaries}

We begin by recalling that the unit state vector $\left\vert \psi\left(
t\right)  \right\rangle $ is generally such that $\left\langle \psi\left(
t\right)  \left\vert \dot{\psi}\left(  t\right)  \right.  \right\rangle
=(-i/\hslash)\left\langle \psi\left(  t\right)  \left\vert \mathrm{H}\left(
t\right)  \right\vert \psi\left(  t\right)  \right\rangle \neq0$. From
$\left\vert \psi\left(  t\right)  \right\rangle $, we construct the parallel
transported unit state vector $\left\vert \Psi\left(  t\right)  \right\rangle
\overset{\text{def}}{=}e^{i\beta\left(  t\right)  }\left\vert \psi\left(
t\right)  \right\rangle $ with the phase $\beta\left(  t\right)  $ such that
$\left\langle \Psi\left(  t\right)  \left\vert \dot{\Psi}\left(  t\right)
\right.  \right\rangle =0$. Note that $i\hslash\left\vert \dot{\Psi}\left(
t\right)  \right\rangle =\left[  \mathrm{H}\left(  t\right)  -\hslash
\dot{\beta}\left(  t\right)  \right]  \left\vert \Psi\left(  t\right)
\right\rangle $. Then, the condition $\left\langle \Psi\left(  t\right)
\left\vert \dot{\Psi}\left(  t\right)  \right.  \right\rangle =0$ is
equivalent to setting the phase $\beta\left(  t\right)  $ equal to%
\begin{equation}
\beta\left(  t\right)  \overset{\text{def}}{=}\frac{1}{\hslash}\int_{0}%
^{t}\left\langle \psi\left(  t^{\prime}\right)  \left\vert \mathrm{H}\left(
t^{\prime}\right)  \right\vert \psi\left(  t^{\prime}\right)  \right\rangle
dt^{\prime}\text{.}%
\end{equation}
Therefore, the state $\left\vert \Psi\left(  t\right)  \right\rangle $ reduces
to%
\begin{equation}
\left\vert \Psi\left(  t\right)  \right\rangle =e^{(i/\hslash)\int_{0}%
^{t}\left\langle \psi\left(  t^{\prime}\right)  \left\vert \mathrm{H}\left(
t^{\prime}\right)  \right\vert \psi\left(  t^{\prime}\right)  \right\rangle
dt^{\prime}}\left\vert \psi\left(  t\right)  \right\rangle \text{,}%
\end{equation}
and evolves according to the equation $i\hslash\left\vert \dot{\Psi}\left(
t\right)  \right\rangle =\Delta\mathrm{H}\left(  t\right)  \left\vert
\Psi\left(  t\right)  \right\rangle $ with $\Delta\mathrm{H}\left(  t\right)
\overset{\text{def}}{=}\mathrm{H}\left(  t\right)  -\left\langle
\mathrm{H}\left(  t\right)  \right\rangle $. Unlike what happens in the
time-independent setting, the speed of quantum evolution is no longer constant
and satisfies the relation $v^{2}\left(  t\right)  =\left\langle \dot{\Psi
}\left(  t\right)  \left\vert \dot{\Psi}\left(  t\right)  \right.
\right\rangle =\left\langle \left(  \Delta\mathrm{H}\left(  t\right)  \right)
^{2}\right\rangle /\hslash^{2}$. For convenience, we introduce at this stage
the arc length $s=s\left(  t\right)  $ defined in terms of the speed of
evolution $v\left(  t\right)  $ as%
\begin{equation}
s\left(  t\right)  \overset{\text{def}}{=}\int_{0}^{t}v(t^{\prime})dt^{\prime
}\text{,} \label{s-equation}%
\end{equation}
with $ds=v(t)dt$, that is, $\partial_{t}=v(t)\partial_{s}$. Finally,
introducing the unitless operator $\Delta h\left(  t\right)
\overset{\text{def}}{=}\Delta\mathrm{H}\left(  t\right)  /[\hslash
v(t)]=\Delta\mathrm{H}\left(  t\right)  /\sqrt{\left\langle \left(
\Delta\mathrm{H}\left(  t\right)  \right)  ^{2}\right\rangle }$, the unit
tangent vector $\left\vert T\left(  s\right)  \right\rangle
\overset{\text{def}}{=}\partial_{s}\left\vert \Psi\left(  s\right)
\right\rangle =\left\vert \Psi^{\prime}\left(  s\right)  \right\rangle $
reduces to $\left\vert T\left(  s\right)  \right\rangle =-i\Delta h\left(
s\right)  \left\vert \Psi\left(  s\right)  \right\rangle $. For completeness,
we remark that it is straightforward to verify that $\left\langle T\left(
s\right)  \left\vert T\left(  s\right)  \right.  \right\rangle =1$ by
construction. For later use, we also point out that $\partial_{s}\left\langle
\Delta h(s)\right\rangle =\left\langle \Delta h^{\prime}(s)\right\rangle $
and, indeed, this relation can be generalized to an arbitrary power of
differentiation \cite{footnote1}.
%%===========================================
%% PMA footnote
%\footnote{\bf This result follows from the fact that
%$\partial_t \langle  H(t) \rangle  \equiv \partial_t \langle \psi(t)| H(t) |\psi(t)\rangle=
%\langle \partial_t\psi(t)| H(t) |\psi(t)\rangle +  \langle \psi(t)| H(t) |\partial_t\psi(t)\rangle +
%\langle \psi(t)|  \partial_tH(t) |\psi(t)\rangle =   \langle \dot{H}(t)\rangle$
%since the first two terms cancel via Schr\"odinger's equation $i \partial_t|\psi(t)\rangle|  \equiv H(t) |\psi(t)\rangle$. Note, for an arbitrary time-dependent operator $\mathcal{O}(t)$ one has in general
%$\partial_t \langle \mathcal{O}(t) \rangle = i\, \langle [H(t),\mathcal{O}(t)] \rangle
%+ \langle  \partial_t \mathcal{O}(t)\rangle$}.
%%===========================================
For instance, to the second power, we have $\partial_{s}^{2}\left\langle
\Delta h(s)\right\rangle =\left\langle \Delta h^{\prime\prime}(s)\right\rangle
$. From the tangent vector $\left\vert T\left(  s\right)  \right\rangle
=-i\Delta h\left(  s\right)  \left\vert \Psi\left(  s\right)  \right\rangle $,
we can construct $\left\vert T^{\prime}\left(  s\right)  \right\rangle
\overset{\text{def}}{=}\partial_{s}\left\vert T\left(  s\right)  \right\rangle
$. After some algebra, we get $\left\vert T^{\prime}\left(  s\right)
\right\rangle =-i\Delta h(s)\left\vert \Psi^{\prime}\left(  s\right)
\right\rangle -i\Delta h^{\prime}(s)\left\vert \Psi\left(  s\right)
\right\rangle $ with
\begin{equation}
\left\langle T^{\prime}\left(  s\right)  \left\vert T^{\prime}\left(
s\right)  \right.  \right\rangle =\left\langle \left(  \Delta h^{\prime
}(s)\right)  ^{2}\right\rangle +\left\langle \left(  \Delta h(s)\right)
^{4}\right\rangle -2i\operatorname{Re}\left[  \left\langle \Delta h^{\prime
}(s)\left(  \Delta h(s)\right)  ^{2}\right\rangle \right]  \neq1\text{,}%
\end{equation}
in general. We are now ready to introduce the curvature and torsion
coefficients for quantum evolutions generated by time-dependent Hamiltonians.
Indeed, having at our disposal the vectors $\left\vert \Psi\left(  s\right)
\right\rangle $, $\left\vert T\left(  s\right)  \right\rangle $, and
$\left\vert T^{\prime}\left(  s\right)  \right\rangle $, we can introduce the
curvature coefficient $\kappa_{\mathrm{AC}}^{2}\left(  s\right)
\overset{\text{def}}{=}\left\langle \tilde{N}_{\ast}\left(  s\right)
\left\vert \tilde{N}_{\ast}\left(  s\right)  \right.  \right\rangle $ with
$\left\vert \tilde{N}_{\ast}\left(  s\right)  \right\rangle
\overset{\text{def}}{=}\mathrm{P}^{\left(  \Psi\right)  }\left\vert T^{\prime
}\left(  s\right)  \right\rangle $, $\mathrm{P}^{\left(  \Psi\right)
}\overset{\text{def}}{=}\mathrm{I}-\left\vert \Psi\left(  s\right)
\right\rangle \left\langle \Psi\left(  s\right)  \right\vert $, and
$\mathrm{I}$ being the identity operator in $\mathcal{H}_{N}$. The subscript
\textquotedblleft\textrm{AC}\textquotedblright\textrm{\ }means Alsing and
Cafaro here. Note that the curvature coefficient $\kappa_{\mathrm{AC}}%
^{2}\left(  s\right)  $ can be rewritten as%
\begin{equation}
\kappa_{\mathrm{AC}}^{2}\left(  s\right)  \overset{\text{def}}{=}\left\Vert
\mathrm{D}\left\vert T(s)\right\rangle \right\Vert ^{2}=\left\Vert
\mathrm{D}^{2}\left\vert \Psi\left(  s\right)  \right\rangle \right\Vert
^{2}\text{,} \label{peggio}%
\end{equation}
where $\mathrm{D}\overset{\text{def}}{=}\mathrm{P}^{\left(  \Psi\right)
}d/ds=\left(  \mathrm{I}-\left\vert \Psi\right\rangle \left\langle
\Psi\right\vert \right)  d/ds$ with $\mathrm{D}\left\vert T(s)\right\rangle
\overset{\text{def}}{=}\mathrm{P}^{\left(  \Psi\right)  }\left\vert T^{\prime
}(s)\right\rangle $ is the covariant derivative
\cite{carlocqg23,samuel88,paulPRA23}. Observe that $\kappa_{\mathrm{AC}}%
^{2}\left(  s\right)  $ in Eq. (\ref{peggio}) is the magnitude squared of the
second covariant derivative of the state vector $\left\vert \Psi\left(
s\right)  \right\rangle $ that traces out the quantum Schr\"{o}dinger
trajectory in the projective Hilbert space. Moreover, the torsion coefficient
$\tau_{\mathrm{AC}}^{2}\left(  s\right)  \overset{\text{def}}{=}\left\langle
\tilde{N}\left(  s\right)  \left\vert \tilde{N}\left(  s\right)  \right.
\right\rangle $ with $\left\vert \tilde{N}\left(  s\right)  \right\rangle
\overset{\text{def}}{=}\mathrm{P}^{\left(  T\right)  }\mathrm{P}^{\left(
\Psi\right)  }\left\vert T^{\prime}\left(  s\right)  \right\rangle $ and
$\mathrm{P}^{\left(  T\right)  }\overset{\text{def}}{=}\mathrm{I}-\left\vert
T\left(  s\right)  \right\rangle \left\langle T\left(  s\right)  \right\vert
$. For clarity, we remark that $\left\vert \tilde{N}_{\ast}\left(  s\right)
\right\rangle $ denotes here a vector that is neither normalized to one nor
orthogonal to the vector $\left\vert T\left(  s\right)  \right\rangle $. The
vector $\left\vert \tilde{N}\left(  s\right)  \right\rangle $, although not
properly normalized, is orthogonal to $\left\vert T\left(  s\right)
\right\rangle $. Finally, $\left\vert N\left(  s\right)  \right\rangle $ is a
unit vector that is orthogonal to $\left\vert T\left(  s\right)  \right\rangle
$. In analogy to the curvature coefficient, the torsion coefficient
$\tau_{\mathrm{AC}}^{2}\left(  s\right)  $ can be recast in an alternative
fashion once one introduces the covariant derivative $\mathrm{D}%
\overset{\text{def}}{=}\mathrm{P}^{\left(  \Psi\right)  }d/ds=\left(
\mathrm{I}-\left\vert \Psi\right\rangle \left\langle \Psi\right\vert \right)
d/ds$ with \textrm{D}$\left\vert T(s)\right\rangle \overset{\text{def}%
}{=}\mathrm{P}^{\left(  \Psi\right)  }\left\vert T^{\prime}(s)\right\rangle $.
Moreover, the torsion coefficient $\tau_{\mathrm{AC}}^{2}\left(  s\right)
\overset{\text{def}}{=}\left\langle \tilde{N}\left(  s\right)  \left\vert
\tilde{N}\left(  s\right)  \right.  \right\rangle $ with $\left\vert \tilde
{N}\left(  s\right)  \right\rangle \overset{\text{def}}{=}\mathrm{P}^{\left(
T\right)  }\mathrm{P}^{\left(  \Psi\right)  }\left\vert T^{\prime}\left(
s\right)  \right\rangle $ and $\mathrm{P}^{\left(  T\right)  }%
\overset{\text{def}}{=}\mathrm{I}-\left\vert T\left(  s\right)  \right\rangle
\left\langle T\left(  s\right)  \right\vert $ can be rewritten as%
\begin{equation}
\tau_{\mathrm{AC}}^{2}\left(  s\right)  \overset{\text{def}}{=}\left\Vert
\mathrm{P}^{(T)}\mathrm{D}\left\vert T(s)\right\rangle \right\Vert
^{2}=\left\Vert \mathrm{P}^{(T)}\mathrm{D}^{2}\left\vert \Psi\left(  s\right)
\right\rangle \right\Vert ^{2}\text{,} \label{torcina}%
\end{equation}
where \textrm{D}$\left\vert T(s)\right\rangle =$ $\mathrm{D}^{2}\left\vert
\Psi\left(  s\right)  \right\rangle =\mathrm{P}^{\left(  \Psi\right)
}\left\vert T^{\prime}\left(  s\right)  \right\rangle $. From Eq.
(\ref{torcina}), we see that $\tau_{\mathrm{AC}}^{2}\left(  s\right)  $ can be
viewed as the magnitude squared of the projection of the covariant derivative
of the tangent vector $\left\vert T\right\rangle $, orthogonal to both
$\left\vert T\right\rangle $ and $\left\vert \Psi\right\rangle $. Finally, the
set of three orthonormal vectors needed to quantify the curvature and the
torsion of a quantum evolution becomes $\left\{  \left\vert \Psi\left(
s\right)  \right\rangle \text{, }\left\vert T\left(  s\right)  \right\rangle
\text{, }\left\vert N\left(  s\right)  \right\rangle \right\}  $ with
$\left\vert N\left(  s\right)  \right\rangle \overset{\text{def}}{=}$
$\left\vert \tilde{N}\left(  s\right)  \right\rangle /\sqrt{\left\langle
\tilde{N}\left(  s\right)  \left\vert \tilde{N}\left(  s\right)  \right.
\right\rangle }$. Note that although $\dim_{%
%TCIMACRO{\U{2102} }%
%BeginExpansion
\mathbb{C}
%EndExpansion
}\mathcal{H}_{N}$
%PMA
can have arbitrary dimension, we restrict our attention to the
three-dimensional complex subspace spanned by $\left\{  \left\vert \Psi\left(
s\right)  \right\rangle \text{, }\left\vert T\left(  s\right)  \right\rangle
\text{, }\left\vert N\left(  s\right)  \right\rangle \right\}  $. This is
consistent with the classical geometric viewpoint according to which the
curvature and torsion coefficients can be interpreted as the lowest and
second-lowest, respectively, members of a family of generalized curvatures
functions \cite{alvarez19}. This geometric perspective, valid for curves in
higher-dimensional spaces, requires a set of $m$ orthonormal vectors in order
to generate $\left(  m-1\right)  $-generalized curvature functions
\cite{alvarez19}.

\subsection{Curvature and Torsion}

We point out that the reparametrization of a quantum curve by its arc length
$s$, while always possible in principle, may be very challenging in actuality.
In analogy to the classical case of space curves in $%
%TCIMACRO{\U{211d} }%
%BeginExpansion
\mathbb{R}
%EndExpansion
^{3}$ \cite{parker77}, there are essentially two obstacles. First, $s\left(
t\right)  $ in Eq. (\ref{s-equation}) may not be computable in closed-form.
Second, even if we have $s=s\left(  t\right)  $, it may not be possible to
invert this relation and, thus, find $t=t\left(  s\right)  $ to be used in
$\left\vert \Psi\left(  s\right)  \right\rangle \overset{\text{def}%
}{=}\left\vert \Psi\left(  t(s)\right)  \right\rangle $. Therefore, the direct
calculation of the time-dependent curvature and torsion coefficients in Eqs.
(\ref{peggio}) and (\ref{torcina}) based on the projection operators formalism
appears to be generally troublesome. For this reason, in what follows we
present alternative expressions for $\kappa_{\mathrm{AC}}^{2}\left(  s\right)
$ and $\tau_{\mathrm{AC}}^{2}\left(  s\right)  $ specified in terms of
expectation values (taken with respect to the state $\left\vert \Psi\left(
t\right)  \right\rangle $, or $\left\vert \psi\left(  t\right)  \right\rangle
$) that can be evaluated without the need of finding the relation $t=t\left(
s\right)  $.

\subsubsection{Curvature}

In what follows, for ease of notation, we shall avoid making explicit the
$s$-dependence of the various operators and/or expectation values. So, for
instance, $\Delta h\left(  s\right)  $ will be simply denoted as $\Delta h$.
Let us begin by constructing the curvature coefficient $\kappa_{\mathrm{AC}%
}^{2}\left(  s\right)  \overset{\text{def}}{=}\left\langle \tilde{N}_{\ast
}\left(  s\right)  \left\vert \tilde{N}_{\ast}\left(  s\right)  \right.
\right\rangle $ with $\left\vert \tilde{N}_{\ast}\left(  s\right)
\right\rangle \overset{\text{def}}{=}\mathrm{P}^{\left(  \Psi\right)
}\left\vert T^{\prime}\left(  s\right)  \right\rangle $ and $\mathrm{P}%
^{\left(  \Psi\right)  }\overset{\text{def}}{=}\mathrm{I}-\left\vert
\Psi\left(  s\right)  \right\rangle \left\langle \Psi\left(  s\right)
\right\vert $. After some algebra, we get%
\begin{equation}
\left\vert \tilde{N}_{\ast}\right\rangle =-\left\{  \left[  \left(  \Delta
h\right)  ^{2}-\left\langle \left(  \Delta h\right)  ^{2}\right\rangle
\right]  +i\left[  \Delta h^{\prime}-\left\langle \Delta h^{\prime
}\right\rangle \right]  \right\}  \left\vert \Psi\right\rangle \text{.}
\label{nas}%
\end{equation}
To calculate $\kappa_{\mathrm{AC}}^{2}\left(  s\right)  \overset{\text{def}%
}{=}\left\langle \tilde{N}_{\ast}\left(  s\right)  \left\vert \tilde{N}_{\ast
}\left(  s\right)  \right.  \right\rangle $, it is convenient to introduce the
Hermitian operator $\hat{\alpha}_{1}\overset{\text{def}}{=}\left(  \Delta
h\right)  ^{2}-\left\langle \left(  \Delta h\right)  ^{2}\right\rangle $ and
the anti-Hermitian operator $\hat{\beta}_{1}\overset{\text{def}}{=}i\left[
\Delta h^{\prime}-\left\langle \Delta h^{\prime}\right\rangle \right]  $ with
$\hat{\beta}_{1}^{\dagger}=-\hat{\beta}_{1}$. Then, $\left\vert \tilde
{N}_{\ast}\right\rangle =-\left(  \hat{\alpha}_{1}+\hat{\beta}_{1}\right)
\left\vert \Psi\right\rangle $ and $\left\langle \tilde{N}_{\ast}\left(
s\right)  \left\vert \tilde{N}_{\ast}\left(  s\right)  \right.  \right\rangle
$ equals $\left\langle \hat{\alpha}_{1}^{2}\right\rangle -\left\langle
\hat{\beta}_{1}^{2}\right\rangle +\left\langle \left[  \hat{\alpha}_{1}\text{,
}\hat{\beta}_{1}\right]  \right\rangle $ with $\left[  \hat{\alpha}_{1}\text{,
}\hat{\beta}_{1}\right]  \overset{\text{def}}{=}\hat{\alpha}_{1}\hat{\beta
}_{1}-\hat{\beta}_{1}\hat{\alpha}_{1}$ being the quantum commutator between
$\hat{\alpha}_{1}$ and $\hat{\beta}_{1}$. Note that since $\hat{\alpha}_{1}$
and $\hat{\beta}_{1}$ are Hermitian and anti-Hermitian operators,
respectively, their commutator $\left[  \hat{\alpha}_{1}\text{, }\hat{\beta
}_{1}\right]  $ is a Hermitian operator. Thus, its expectation value
$\left\langle \left[  \hat{\alpha}_{1}\text{, }\hat{\beta}_{1}\right]
\right\rangle $ is a real number. Using the definitions of $\hat{\alpha}_{1}$
and $\hat{\beta}_{1}$, we obtain $\left\langle \hat{\alpha}_{1}^{2}%
\right\rangle =\left\langle (\Delta h)^{4}\right\rangle -\left\langle (\Delta
h)^{2}\right\rangle ^{2}$, $\left\langle \hat{\beta}_{1}^{2}\right\rangle
=-\left[  \left\langle (\Delta h^{\prime})^{2}\right\rangle -\left\langle
\Delta h^{\prime}\right\rangle ^{2}\right]  $, and $\left\langle \left[
\hat{\alpha}_{1}\text{, }\hat{\beta}_{1}\right]  \right\rangle =i\left\langle
\left[  (\Delta h)^{2}\text{, }\Delta h^{\prime}\right]  \right\rangle $.
Observe that $\left\langle \left[  (\Delta h)^{2}\text{, }\Delta h^{\prime
}\right]  \right\rangle $ is purely imaginary since $\left[  (\Delta
h)^{2}\text{, }\Delta h^{\prime}\right]  $ is a anti-Hermitian operator. For
completeness, we emphasize that $\left[  (\Delta h)^{2}\text{, }\Delta
h^{\prime}\right]  $ is not a null operator, in general. Indeed, $\left[
(\Delta h)^{2}\text{, }\Delta h^{\prime}\right]  =\Delta h\left[  \Delta
h\text{, }\Delta h^{\prime}\right]  +\left[  \Delta h\text{, }\Delta
h^{\prime}\right]  \Delta h$ with $\left[  \Delta h\text{, }\Delta h^{\prime
}\right]  =\left[  \mathrm{H}\text{, }\mathrm{H}^{\prime}\right]  $. Then,
focusing for simplicity on time-dependent qubit Hamiltonians of the form
\textrm{H}$\left(  s\right)  =\vec{h}\left(  s\right)  \cdot\vec{\sigma}$, the
commutator $\left[  \mathrm{H}\text{, }\mathrm{H}^{\prime}\right]  =2i(\vec
{h}\times\vec{h}^{\prime})\cdot\vec{\sigma}$ does not vanish since $\vec{h}$
and $\vec{h}^{\prime}$ are not generally collinear. In conclusion, the
curvature coefficient $\kappa_{\mathrm{AC}}^{2}\left(  s\right)  $ in Eq.
(\ref{peggio}) becomes in the time-dependent setting%
\begin{equation}
\kappa_{\mathrm{AC}}^{2}\left(  s\right)  =\left\langle (\Delta h)^{4}%
\right\rangle -\left\langle (\Delta h)^{2}\right\rangle ^{2}+\left[
\left\langle (\Delta h^{\prime})^{2}\right\rangle -\left\langle \Delta
h^{\prime}\right\rangle ^{2}\right]  +i\left\langle \left[  (\Delta
h)^{2}\text{, }\Delta h^{\prime}\right]  \right\rangle \text{.}
\label{curvatime}%
\end{equation}
From Eq. (\ref{curvatime}), we note that $\kappa_{\mathrm{AC}}^{2}\left(
s\right)  $ reduces to its time-independent expression $\left\langle (\Delta
h)^{4}\right\rangle -\left\langle (\Delta h)^{2}\right\rangle ^{2}$ when the
Hamiltonian \textrm{H} is constant and, as a consequence, $\Delta h^{\prime}$
becomes the null operator.

\subsubsection{Torsion}

We are now ready to calculate the expression of the torsion coefficient
$\tau_{\mathrm{AC}}^{2}\left(  s\right)  \overset{\text{def}}{=}\left\langle
\tilde{N}\left(  s\right)  \left\vert \tilde{N}\left(  s\right)  \right.
\right\rangle $ with $\left\vert \tilde{N}\left(  s\right)  \right\rangle
\overset{\text{def}}{=}\mathrm{P}^{\left(  T\right)  }\mathrm{P}^{\left(
\Psi\right)  }\left\vert T^{\prime}\left(  s\right)  \right\rangle $ and
$\mathrm{P}^{\left(  T\right)  }\overset{\text{def}}{=}\mathrm{I}-\left\vert
T\left(  s\right)  \right\rangle \left\langle T\left(  s\right)  \right\vert
$. We observe that $\left\vert T\left(  s\right)  \right\rangle \left\langle
T\left(  s\right)  \right\vert =\Delta h\left\vert \Psi\right\rangle
\left\langle \Psi\right\vert \Delta h$ and $\left\vert \tilde{N}\left(
s\right)  \right\rangle \overset{\text{def}}{=}\mathrm{P}^{\left(  T\right)
}\mathrm{P}^{\left(  \Psi\right)  }\left\vert T^{\prime}\left(  s\right)
\right\rangle =\mathrm{P}^{\left(  T\right)  }\left\vert \tilde{N}_{\ast
}\left(  s\right)  \right\rangle $ with $\left\vert \tilde{N}_{\ast}\left(
s\right)  \right\rangle $ in\ Eq. (\ref{nas}). Then, after some algebra, we
arrive at%
\begin{equation}
\left\vert \tilde{N}\right\rangle =-\left\{  \left[  \left(  \Delta h\right)
^{2}-\left\langle \left(  \Delta h\right)  ^{2}\right\rangle -\Delta
h\left\langle \left(  \Delta h\right)  ^{3}\right\rangle \right]  +i\left[
\Delta h^{\prime}-\left\langle \Delta h^{\prime}\right\rangle -\Delta
h\left\langle \left(  \Delta h\right)  \left(  \Delta h^{\prime}\right)
\right\rangle \right]  \right\}  \left\vert \Psi\right\rangle \text{.}
\label{nas2}%
\end{equation}
To calculate $\tau_{\mathrm{AC}}^{2}\left(  s\right)  $ in Eq. (\ref{torcina})
, it is useful to introduce the Hermitian operator $\hat{\alpha}%
_{2}\overset{\text{def}}{=}\left(  \Delta h\right)  ^{2}-\left\langle \left(
\Delta h\right)  ^{2}\right\rangle -\Delta h\left\langle \left(  \Delta
h\right)  ^{3}\right\rangle $ and the anti-Hermitian operator $\hat{\beta}%
_{2}\overset{\text{def}}{=}i\left[  \Delta h^{\prime}-\left\langle \Delta
h^{\prime}\right\rangle -\Delta h\left\langle \left(  \Delta h\right)  \left(
\Delta h^{\prime}\right)  \right\rangle \right]  $ with $\hat{\beta}%
_{2}^{\dagger}=-\hat{\beta}_{2}$. Then, $\left\vert \tilde{N}\right\rangle
=-\left(  \hat{\alpha}_{2}+\hat{\beta}_{2}\right)  \left\vert \Psi
\right\rangle $ and $\left\langle \tilde{N}\left(  s\right)  \left\vert
\tilde{N}\left(  s\right)  \right.  \right\rangle $ is equal to $\left\langle
\hat{\alpha}_{2}^{2}\right\rangle -\left\langle \hat{\beta}_{2}^{2}%
\right\rangle +\left\langle \left[  \hat{\alpha}_{2}\text{, }\hat{\beta}%
_{2}\right]  \right\rangle $ with $\left[  \hat{\alpha}_{2}\text{, }\hat
{\beta}_{2}\right]  \overset{\text{def}}{=}\hat{\alpha}_{2}\hat{\beta}%
_{2}-\hat{\beta}_{2}\hat{\alpha}_{2}$ denoting the quantum commutator between
$\hat{\alpha}_{2}$ and $\hat{\beta}_{2}$. In analogy to our previous
discussion on the calculation of the curvature coefficient in
Eq.\ (\ref{curvatime}), we note that since $\hat{\alpha}_{2}$ and $\hat{\beta
}_{2}$ are Hermitian and anti-Hermitian operators, respectively, their
commutator $\left[  \hat{\alpha}_{2}\text{, }\hat{\beta}_{2}\right]  $ is a
Hermitian operator. Thus, its expectation value $\left\langle \left[
\hat{\alpha}_{2}\text{, }\hat{\beta}_{2}\right]  \right\rangle $ is a real
number. Exploiting \ the definitions of $\hat{\alpha}_{2}$ and $\hat{\beta
}_{2}$, we get $\left\langle \hat{\alpha}_{2}^{2}\right\rangle =\left\langle
(\Delta h)^{4}\right\rangle -\left\langle (\Delta h)^{2}\right\rangle
^{2}-\left\langle (\Delta h)^{3}\right\rangle ^{2}$, $\left\langle \hat{\beta
}_{2}^{2}\right\rangle =-\left[  \left\langle (\Delta h^{\prime}%
)^{2}\right\rangle -\left\langle \Delta h^{\prime}\right\rangle ^{2}%
-\left\langle \left(  \Delta h^{\prime}\right)  \left(  \Delta h\right)
\right\rangle \left\langle \left(  \Delta h\right)  \left(  \Delta h^{\prime
}\right)  \right\rangle \right]  $, and $\left\langle \left[  \hat{\alpha}%
_{2}\text{, }\hat{\beta}_{2}\right]  \right\rangle =i\left\{  \left\langle
\left[  (\Delta h)^{2}\text{, }\Delta h^{\prime}\right]  \right\rangle
-\left\langle (\Delta h)^{3}\right\rangle \left\langle \left[  \Delta h\text{,
}\Delta h^{\prime}\right]  \right\rangle \right\}  $. Observe that
$\left\langle \left[  (\Delta h)^{2}\text{, }\Delta h^{\prime}\right]
\right\rangle $ and $\left\langle \left[  \Delta h\text{, }\Delta h^{\prime
}\right]  \right\rangle $ are purely imaginary numbers since $\left[  (\Delta
h)^{2}\text{, }\Delta h^{\prime}\right]  $ and $\left[  \Delta h\text{,
}\Delta h^{\prime}\right]  $ are anti-Hermitian operators. In conclusion, the
torsion coefficient $\tau_{\mathrm{AC}}^{2}\left(  s\right)  $ in Eq.
(\ref{torcina}) becomes in the time-dependent setting%
\begin{align}
\tau_{\mathrm{AC}}^{2}\left(  s\right)   &  =\left\langle (\Delta
h)^{4}\right\rangle -\left\langle (\Delta h)^{2}\right\rangle ^{2}%
-\left\langle (\Delta h)^{3}\right\rangle ^{2}+\left[  \left\langle (\Delta
h^{\prime})^{2}\right\rangle -\left\langle \Delta h^{\prime}\right\rangle
^{2}-\left\langle \left(  \Delta h^{\prime}\right)  \left(  \Delta h\right)
\right\rangle \left\langle \left(  \Delta h\right)  \left(  \Delta h^{\prime
}\right)  \right\rangle \right]  +\nonumber\\
&  +i\left\{  \left\langle \left[  (\Delta h)^{2}\text{, }\Delta h^{\prime
}\right]  \right\rangle -\left\langle (\Delta h)^{3}\right\rangle \left\langle
\left[  \Delta h\text{, }\Delta h^{\prime}\right]  \right\rangle \right\}
\text{.} \label{torsiontime}%
\end{align}
From Eq. (\ref{torsiontime}), we notice that $\tau_{\mathrm{AC}}^{2}\left(
s\right)  $ reduces to its time-independent form $\left\langle (\Delta
h)^{4}\right\rangle -\left\langle (\Delta h)^{2}\right\rangle ^{2}%
-\left\langle (\Delta h)^{3}\right\rangle ^{2}$ when the Hamiltonian
\textrm{H} is time-independent and, consequently, the operator $\Delta
h^{\prime}$ vanishes.

\section{Statistical interpretation}

In Ref. \cite{alsing1}, we interpreted the time-independent curvature
$\kappa_{\mathrm{AC}}^{2}$ and torsion $\tau_{\mathrm{AC}}^{2}$ coefficients
in terms of concepts of use in statistical mathematics, namely the kurtosis
$\alpha_{4}$ and the skewness $\alpha_{3}$ of a probability distribution
function \cite{ernest44,sharma15,pearson16}. Specifically, we showed that
$\kappa_{\mathrm{AC}}^{2}=\alpha_{4}-1$ and $\tau_{\mathrm{AC}}^{2}=\alpha
_{4}-1-\alpha_{3}^{2}$, where $\alpha_{4}\overset{\text{def}}{=}m_{4}%
/m_{2}^{2}$, $\alpha_{3}\overset{\text{def}}{=}m_{3}/m_{2}^{3/2}$,
$m_{r}\overset{\text{def}}{=}(1/n)\sum_{i=1}^{n}\left(  x_{i}-\bar{x}\right)
^{r}$ is the $r$-th central moment, and $\bar{x}$ is the arithmetic mean of
$n$ real numbers $x_{i}$ with $1\leq i\leq n$. In this section, we aim to
extend our statistical interpretation to the time-dependent coefficients
$\kappa_{\mathrm{AC}}^{2}$ and torsion $\tau_{\mathrm{AC}}^{2}$. Specifically,
we present a statistical interpretation of the various terms that specify the
time-dependent curvature and torsion coefficients when expressed in terms of
expectation values of conveniently defined operators constructed out of the
original nonstationary Hermitian Hamiltonian operator \textrm{H}$\left(
t\right)  $. We shall find that $\kappa_{\mathrm{AC}}^{2}\left(  t\right)  $
and $\tau_{\mathrm{AC}}^{2}\left(  t\right)  $ are given by%
\begin{equation}
\kappa_{\mathrm{AC}}^{2}\left(  t\right)  =\left(  \mathrm{\alpha}%
_{4}-1\right)  +\frac{\left(  \sigma_{\mathrm{\dot{H}}}\right)  ^{2}-\left(
\dot{\sigma}_{\mathrm{H}}\right)  ^{2}}{v^{4}}+i\frac{\mathrm{cov}\left(
\mathrm{H}\text{, }\left[  \mathrm{H}\text{, \textrm{\.{H}}}\right]  \right)
+\mathrm{cov}\left(  \left[  \mathrm{H}\text{, \textrm{\.{H}}}\right]  \text{,
}\mathrm{H}\right)  }{v^{4}}\text{,} \label{form1}%
\end{equation}
and%
\begin{equation}
\tau_{\mathrm{AC}}^{2}\left(  t\right)  =\left(  \mathrm{\alpha}%
_{4}-1-\mathrm{\alpha}_{3}^{2}\right)  +\left[  \frac{\left(  \sigma
_{\mathrm{\dot{H}}}\right)  ^{2}-\left(  \dot{\sigma}_{\mathrm{H}}\right)
^{2}}{v^{4}}+\frac{1}{4}\frac{\left\langle \left[  \mathrm{H}\text{,
}\mathrm{\dot{H}}\right]  \right\rangle ^{2}}{v^{6}}\right]  +i\left[
\frac{\mathrm{cov}\left(  \mathrm{H}\text{, }\left[  \mathrm{H}\text{,
}\mathrm{\dot{H}}\right]  \right)  +\mathrm{cov}\left(  \left[  \mathrm{H}%
\text{, }\mathrm{\dot{H}}\right]  \text{, }\mathrm{H}\right)  }{v^{4}%
}-\mathrm{\alpha}_{3}\frac{\left\langle \left[  \mathrm{H}\text{,
}\mathrm{\dot{H}}\right]  \right\rangle }{v^{3}}\right]  \text{,}
\label{form2}%
\end{equation}
respectively. Here we have defined the covariance between two operators $A$
and $B$ as $\text{cov}(A,B) = \langle A B \rangle- \langle A\rangle\langle B
\rangle$ \cite{footnote2}.
%% PMA
%\textbf{Here we have defined the covariance between two operators $A$ and $B$ as
%$cov(A,B) = \langle A B \rangle -  \langle A\rangle \langle B \rangle$}
%\footnote{The expressions on the righthand side of Eq.(\ref{form1}) and  Eq.(\ref{form2}) could be further simplified if one instead used the symmetric covariance between two operators $A$ and $B$ defined as
%$cov^{(sym)}(A,B) = \frac{1}{2}\langle A B + B A \rangle -  \langle A\rangle \langle B \rangle$.}.
%%
From Eqs. (\ref{form1}) and (\ref{form2}), we notice that nonstationary case
exhibits a richer statistical structure that underlies the notions of
curvature and torsion. Indeed, in addition to the kurtosis $\alpha_{4}$ and
the skewness $\alpha_{3}$, there is the emergence of terms such as the speed
of quantum evolution ($v=\sigma_{\mathrm{H}}$), the variance of the first
derivative of the Hamiltonian ($\sigma_{\mathrm{\dot{H}}}^{2}$), the
acceleration squared of the quantum evolution ($\dot{\sigma}_{\mathrm{H}}^{2}%
$), \ the covariance $\mathrm{cov}\left(  \mathrm{H}\text{, }\left[
\mathrm{H}\text{, }\mathrm{\dot{H}}\right]  \right)  $ between the Hamiltonian
($\mathrm{H}$) and the commutator of the Hamiltonian with the first derivative
($\left[  \mathrm{H}\text{, }\mathrm{\dot{H}}\right]  $) and, finally, the
expectation value $\left\langle \left[  \mathrm{H}\text{, }\mathrm{\dot{H}%
}\right]  \right\rangle $ of the commutator of the Hamiltonian ($\mathrm{H}$)
and the first derivative of the Hamiltonian ($\mathrm{\dot{H}}$). In what
follows, we derive \ and discuss the meaning of Eqs. (\ref{form1}) and
(\ref{form2}).

\subsection{Curvature}

In terms of the arc length parameter $s$ and the unitless operator $\Delta h$,
the general expression of the curvature coefficient $\kappa_{\mathrm{AC}}%
^{2}\left(  s\right)  $ is given by,%
\begin{equation}
\kappa_{\mathrm{AC}}^{2}\left(  s\right)  =\left\langle \left(  \Delta
h\right)  ^{4}\right\rangle -\left\langle \left(  \Delta h\right)
^{2}\right\rangle ^{2}+\left[  \left\langle \left(  \Delta h^{\prime}\right)
^{2}\right\rangle -\left\langle \Delta h^{\prime}\right\rangle ^{2}\right]
+i\left\langle \left[  \left(  \Delta h\right)  ^{2}\text{, }\Delta h^{\prime
}\right]  \right\rangle \text{.} \label{KS1}%
\end{equation}
Then, setting $\hslash=1$ and recalling that $s=s\left(  t\right)  $, $\Delta
h\overset{\text{def}}{=}\Delta\mathrm{H}/v$ with $v^{2}\overset{\text{def}%
}{=}\left\langle \left(  \Delta\mathrm{H}\right)  ^{2}\right\rangle $,
$\kappa_{\mathrm{AC}}^{2}\left(  s\right)  $ in Eq. (\ref{KS1}) can be recast
as
\begin{equation}
\kappa_{\mathrm{AC}}^{2}\left(  t\right)  =\frac{\left\langle \left(
\Delta\mathrm{H}\right)  ^{4}\right\rangle -\left\langle \left(
\Delta\mathrm{H}\right)  ^{2}\right\rangle ^{2}}{\left\langle \left(
\Delta\mathrm{H}\right)  ^{2}\right\rangle ^{2}}+\frac{\left\langle \left(
\Delta\mathrm{\dot{H}}\right)  ^{2}\right\rangle -\left(  \partial_{t}%
\sqrt{\left\langle \left(  \Delta\mathrm{H}\right)  ^{2}\right\rangle
}\right)  ^{2}}{\left\langle \left(  \Delta\mathrm{H}\right)  ^{2}%
\right\rangle ^{2}}+i\frac{\left\langle \left[  \left(  \Delta\mathrm{H}%
\right)  ^{2}\text{, }\Delta\mathrm{\dot{H}}\right]  \right\rangle
}{\left\langle \left(  \Delta\mathrm{H}\right)  ^{2}\right\rangle ^{2}%
}\text{.} \label{KS2}%
\end{equation}
In our time-independent analysis in Ref. \cite{alsing1}, we showed that the
first term in Eq. (\ref{KS2}) reduces to%
\begin{equation}
\frac{\left\langle \left(  \Delta\mathrm{H}\right)  ^{4}\right\rangle
-\left\langle \left(  \Delta\mathrm{H}\right)  ^{2}\right\rangle ^{2}%
}{\left\langle \left(  \Delta\mathrm{H}\right)  ^{2}\right\rangle ^{2}%
}=\mathrm{\alpha}_{4}-1\text{.} \label{KS3}%
\end{equation}
Furthermore, focusing on the second term in Eq. (\ref{KS2}), we note that
denoting $\left\langle \left(  \Delta\mathrm{\dot{H}}\right)  ^{2}%
\right\rangle $ is the variance $\left(  \sigma_{\mathrm{\dot{H}}}\right)
^{2}$ of the time derivative $\mathrm{\dot{H}}$ of the Hamiltonian \textrm{H},
$\left\langle \left(  \Delta\mathrm{H}\right)  ^{2}\right\rangle ^{2}$ is the
variance squared of the Hamiltonian \textrm{H }and equals $v^{4}$ and,
finally,
%PMA add $(\dot{\sigma}_H)^2 = to the next inline equation
$(\dot{\sigma}_{H})^{2} \equiv\left(  \partial_{t}\sqrt{\left\langle \left(
\Delta\mathrm{H}\right)  ^{2}\right\rangle }\right)  ^{2}$ is the square of
the the time derivative of the speed (i.e., the acceleration) of quantum
evolution in projective Hilbert space.
%PMA removed the following end of sentence
%and can be written as $\left(  \dot{\sigma
%}_{\mathrm{H}}\right)  ^{2}$.
Therefore, the second term in Eq. (\ref{KS2}) can rewritten as%
\begin{equation}
\frac{\left\langle \left(  \Delta\mathrm{\dot{H}}\right)  ^{2}\right\rangle
-\left(  \partial_{t}\sqrt{\left\langle \left(  \Delta\mathrm{H}\right)
^{2}\right\rangle }\right)  ^{2}}{\left\langle \left(  \Delta\mathrm{H}%
\right)  ^{2}\right\rangle ^{2}}=\frac{\left(  \sigma_{\mathrm{\dot{H}}%
}\right)  ^{2}-\left(  \dot{\sigma}_{\mathrm{H}}\right)  ^{2}}{v^{4}}\text{.}
\label{KS4}%
\end{equation}
Interestingly, we point out that the positivity of $\left\langle \left(
\Delta h^{\prime}\right)  ^{2}\right\rangle $ leads to $\left(  \dot{\sigma
}_{\mathrm{H}}\right)  ^{2}\leq\left(  \sigma_{\mathrm{\dot{H}}}\right)  ^{2}%
$. This inequality, in turn, implies that the acceleration $\dot{\sigma
}_{\mathrm{H}}$ of the quantum evolution in projective Hilbert space is upper
bounded by the standard deviation $\sigma_{\mathrm{\dot{H}}}$ of the
time-derivative $\mathrm{\dot{H}}$ of the Hamiltonian operator $\mathrm{H}$.
For a formal proof of this inequality along with a discussion on its physical
significance, we refer to Ref. \cite{carloPLA}. We focus now on the last term
in Eq. (\ref{KS1}), namely $i\left\langle \left[  \left(  \Delta h\right)
^{2}\text{, }\Delta h^{\prime}\right]  \right\rangle =i\left\langle \left[
\left(  \Delta\mathrm{H}\right)  ^{2}\text{, }\Delta\mathrm{\dot{H}}\right]
\right\rangle /\left\langle \left(  \Delta\mathrm{H}\right)  ^{2}\right\rangle
^{2}$. First, note that this term is real since the expectation value of a
commutator between two Hermitian operators (i.e., a anti-Hermitian operator)
is a purely imaginary number. So, the presence of the imaginary unit $i$ makes
the expression $i\left\langle \left[  \left(  \Delta h\right)  ^{2}\text{,
}\Delta h^{\prime}\right]  \right\rangle $ real. Then, note that the
commutator $\left[  \left(  \Delta\mathrm{H}\right)  ^{2}\text{, }%
\Delta\mathrm{\dot{H}}\right]  $ can be rewritten as%
\begin{align}
\left[  \left(  \Delta\mathrm{H}\right)  ^{2}\text{, }\Delta\mathrm{\dot{H}%
}\right]   &  =\left(  \Delta\mathrm{H}\right)  ^{2}\left(  \Delta
\mathrm{\dot{H}}\right)  -\left(  \Delta\mathrm{\dot{H}}\right)  \left(
\Delta\mathrm{H}\right)  ^{2}\nonumber\\
&  =\left(  \mathrm{H}-\left\langle \mathrm{H}\right\rangle \right)
^{2}\left(  \mathrm{\dot{H}}-\left\langle \mathrm{\dot{H}}\right\rangle
\right)  -\left(  \mathrm{\dot{H}}-\left\langle \mathrm{\dot{H}}\right\rangle
\right)  \left(  \mathrm{H}-\left\langle \mathrm{H}\right\rangle \right)
^{2}\nonumber\\
&  =\left[  \mathrm{H}^{2}+\left\langle \mathrm{H}\right\rangle ^{2}%
-2\mathrm{H}\left\langle \mathrm{H}\right\rangle \right]  \left(
\mathrm{\dot{H}}-\left\langle \mathrm{\dot{H}}\right\rangle \right)  -\left(
\mathrm{\dot{H}}-\left\langle \mathrm{\dot{H}}\right\rangle \right)  \left[
\mathrm{H}^{2}+\left\langle \mathrm{H}\right\rangle ^{2}-2\mathrm{H}%
\left\langle \mathrm{H}\right\rangle \right] \nonumber\\
&  =\mathrm{H}^{2}\mathrm{\dot{H}-}\left\langle \mathrm{\dot{H}}\right\rangle
\mathrm{H}^{2}+\left\langle \mathrm{H}\right\rangle ^{2}\mathrm{\dot{H}%
}-\left\langle \mathrm{H}\right\rangle ^{2}\left\langle \mathrm{\dot{H}%
}\right\rangle -2\left\langle \mathrm{H}\right\rangle \mathrm{H\dot{H}%
}+2\left\langle \mathrm{H}\right\rangle \left\langle \mathrm{\dot{H}%
}\right\rangle \mathrm{H}+\nonumber\\
&  -\mathrm{\dot{H}H}^{2}-\mathrm{\dot{H}}\left\langle \mathrm{H}\right\rangle
^{2}+2\left\langle \mathrm{H}\right\rangle \mathrm{\dot{H}H}+\left\langle
\mathrm{\dot{H}}\right\rangle \mathrm{H}^{2}+\left\langle \mathrm{\dot{H}%
}\right\rangle \left\langle \mathrm{H}\right\rangle ^{2}-2\mathrm{H}%
\left\langle \mathrm{\dot{H}}\right\rangle \left\langle \mathrm{H}%
\right\rangle \text{,} \label{b1}%
\end{align}
that is, the expectation value $\left\langle \left[  \left(  \Delta
\mathrm{H}\right)  ^{2}\text{, }\Delta\mathrm{\dot{H}}\right]  \right\rangle $
reduces to
\begin{align}
\left\langle \left[  \left(  \Delta\mathrm{H}\right)  ^{2}\text{, }%
\Delta\mathrm{\dot{H}}\right]  \right\rangle  &  =\left\langle \mathrm{H}%
^{2}\mathrm{\dot{H}}\right\rangle -\left\langle \mathrm{\dot{H}}\right\rangle
\left\langle \mathrm{H}^{2}\right\rangle +\left\langle \mathrm{H}\right\rangle
^{2}\left\langle \mathrm{\dot{H}}\right\rangle -\left\langle \mathrm{H}%
\right\rangle ^{2}\left\langle \mathrm{\dot{H}}\right\rangle -2\left\langle
\mathrm{H}\right\rangle \left\langle \mathrm{H\dot{H}}\right\rangle
+2\left\langle \mathrm{H}\right\rangle ^{2}\left\langle \mathrm{\dot{H}%
}\right\rangle +\nonumber\\
&  -\left\langle \mathrm{\dot{H}H}^{2}\right\rangle -\left\langle
\mathrm{\dot{H}}\right\rangle \left\langle \mathrm{H}\right\rangle
^{2}+2\left\langle \mathrm{H}\right\rangle \left\langle \mathrm{\dot{H}%
H}\right\rangle +\left\langle \mathrm{\dot{H}}\right\rangle \left\langle
\mathrm{H}^{2}\right\rangle +\left\langle \mathrm{\dot{H}}\right\rangle
\left\langle \mathrm{H}\right\rangle ^{2}-2\mathrm{H}\left\langle
\mathrm{\dot{H}}\right\rangle \left\langle \mathrm{H}\right\rangle \nonumber\\
&  =\left\langle \mathrm{H}^{2}\mathrm{\dot{H}}\right\rangle -\left\langle
\mathrm{\dot{H}H}^{2}\right\rangle -2\left\langle \mathrm{H}\right\rangle
\left\langle \mathrm{H\dot{H}}\right\rangle +2\left\langle \mathrm{H}%
\right\rangle \left\langle \mathrm{\dot{H}H}\right\rangle \nonumber\\
&  =\left\langle \left[  \mathrm{H}^{2}\text{, }\mathrm{\dot{H}}\right]
\right\rangle -2\left\langle \mathrm{H}\right\rangle \left\langle \left[
\mathrm{H}\text{, }\mathrm{\dot{H}}\right]  \right\rangle \text{.} \label{b2}%
\end{align}
Further manipulation of Eq. (\ref{b2}) \ yields%
\begin{align}
\left\langle \left[  \left(  \Delta\mathrm{H}\right)  ^{2}\text{, }%
\Delta\mathrm{\dot{H}}\right]  \right\rangle  &  =\left\langle \left[
\mathrm{H}^{2}\text{, }\mathrm{\dot{H}}\right]  \right\rangle -2\left\langle
\mathrm{H}\right\rangle \left\langle \left[  \mathrm{H}\text{, }%
\mathrm{\dot{H}}\right]  \right\rangle \nonumber\\
&  =\left\langle \mathrm{H}\left[  \mathrm{H}\text{, }\mathrm{\dot{H}}\right]
+\left[  \mathrm{H}\text{, }\mathrm{\dot{H}}\right]  \mathrm{H}\right\rangle
-\left\langle \mathrm{H}\right\rangle \left\langle \left[  \mathrm{H}\text{,
}\mathrm{\dot{H}}\right]  \right\rangle -\left\langle \mathrm{H}\right\rangle
\left\langle \left[  \mathrm{H}\text{, }\mathrm{\dot{H}}\right]  \right\rangle
\nonumber\\
&  =\left[  \left\langle \mathrm{H}\left[  \mathrm{H}\text{, }\mathrm{\dot{H}%
}\right]  \right\rangle -\left\langle \mathrm{H}\right\rangle \left\langle
\left[  \mathrm{H}\text{, }\mathrm{\dot{H}}\right]  \right\rangle \right]
+\left[  \left\langle \left[  \mathrm{H}\text{, }\mathrm{\dot{H}}\right]
\mathrm{H}\right\rangle -\left\langle \left[  \mathrm{H}\text{, }%
\mathrm{\dot{H}}\right]  \right\rangle \left\langle \mathrm{H}\right\rangle
\right] \nonumber\\
&  =\mathrm{cov}\left(  \mathrm{H}\text{, }\left[  \mathrm{H}\text{,
}\mathrm{\dot{H}}\right]  \right)  +\mathrm{cov}\left(  \left[  \mathrm{H}%
\text{, }\mathrm{\dot{H}}\right]  \text{, }\mathrm{H}\right)  \text{,}%
\end{align}
that is, $\left\langle \left[  \left(  \Delta\mathrm{H}\right)  ^{2}\text{,
}\Delta\mathrm{\dot{H}}\right]  \right\rangle =\mathrm{cov}\left(
\mathrm{H}\text{, }\left[  \mathrm{H}\text{, }\mathrm{\dot{H}}\right]
\right)  +\mathrm{cov}\left(  \left[  \mathrm{H}\text{, }\mathrm{\dot{H}%
}\right]  \text{, }\mathrm{H}\right)  $. Recall, for clarity, that
\textrm{cov}$\left(  A\text{, }B\right)  \overset{\text{def}}{=}\left\langle
\left(  \Delta A\right)  \left(  \Delta B\right)  \right\rangle =\left\langle
AB\right\rangle -\left\langle A\right\rangle \left\langle B\right\rangle $. In
conclusion, we found that%
\begin{equation}
i\left\langle \left[  \left(  \Delta h\right)  ^{2}\text{, }\Delta h^{\prime
}\right]  \right\rangle =i\frac{\left\langle \left[  \left(  \Delta
\mathrm{H}\right)  ^{2}\text{, }\Delta\mathrm{\dot{H}}\right]  \right\rangle
}{v^{4}}=i\frac{\mathrm{cov}\left(  \mathrm{H}\text{, }\left[  \mathrm{H}%
\text{, }\mathrm{\dot{H}}\right]  \right)  +\mathrm{cov}\left(  \left[
\mathrm{H}\text{, }\mathrm{\dot{H}}\right]  \text{, }\mathrm{H}\right)
}{v^{4}}\text{.} \label{love}%
\end{equation}
To justify the presence of the imaginary unit $i$ in the last term in Eq.
(\ref{love}), we observe that%
\begin{equation}
\mathrm{cov}\left(  \mathrm{H}\text{, }\left[  \mathrm{H}\text{, }%
\mathrm{\dot{H}}\right]  \right)  +\mathrm{cov}\left(  \left[  \mathrm{H}%
\text{, }\mathrm{\dot{H}}\right]  \text{, }\mathrm{H}\right)  =\left\langle
\left\{  \mathrm{H}\text{, }\left[  \mathrm{H}\text{, }\mathrm{\dot{H}%
}\right]  \right\}  \right\rangle -2\left\langle \mathrm{H}\right\rangle
\left\langle \left[  \mathrm{H}\text{, }\mathrm{\dot{H}}\right]  \right\rangle
\text{.} \label{loveyou}%
\end{equation}
Therefore, the presence of the complex imaginary unit $i$ in Eq. (\ref{love})
is consistent with the fact that $\left\langle \left[  \mathrm{H}\text{,
}\mathrm{\dot{H}}\right]  \right\rangle $ and $\left\langle \left\{
\mathrm{H}\text{, }\left[  \mathrm{H}\text{, }\mathrm{\dot{H}}\right]
\right\}  \right\rangle $ in Eq. (\ref{loveyou}) are purely imaginary numbers.
As a matter of fact, the anti-commutator between a Hermitian and a
anti-Hermitian operator is a anti-Hermitian operator. From Eq. (\ref{love}),
we see that the term $i\left\langle \left[  \left(  \Delta h\right)
^{2}\text{, }\Delta h^{\prime}\right]  \right\rangle $ can be expressed in
terms of a superposition of covariance terms between the Hamiltonian and the
commutator of the Hamiltonian with its first derivative. It encodes
information about the correlational structure that specifies the quantum
evolution under observation. In conclusion, substituting Eqs. (\ref{KS3}),
(\ref{KS4}), and (\ref{love}) into Eq. (\ref{KS2}), we recover Eq.
(\ref{form1}).

\subsection{Torsion}

Using the arc length parameter $s$ and the unitless operator $\Delta h$, the
general expression of the torsion coefficient $\tau_{\mathrm{AC}}^{2}\left(
s\right)  $ is given by%
\begin{align}
\tau_{\mathrm{AC}}^{2}\left(  s\right)   &  =\left\langle \left(  \Delta
h\right)  ^{4}\right\rangle -\left\langle \left(  \Delta h\right)
^{2}\right\rangle ^{2}-\left\langle \left(  \Delta h\right)  ^{3}\right\rangle
^{2}+\left[  \left\langle \left(  \Delta h^{\prime}\right)  ^{2}\right\rangle
-\left\langle \Delta h^{\prime}\right\rangle ^{2}-\left\langle \left(  \Delta
h^{\prime}\right)  \left(  \Delta h\right)  \right\rangle \left\langle \left(
\Delta h\right)  \left(  \Delta h^{\prime}\right)  \right\rangle \right]
+\nonumber\\
& \nonumber\\
&  +i\left[  \left\langle \left[  \left(  \Delta h\right)  ^{2}\text{, }\Delta
h^{\prime}\right]  \right\rangle -\left\langle \left(  \Delta h\right)
^{3}\right\rangle \left\langle \left[  \Delta h\text{, }\Delta h^{\prime
}\right]  \right\rangle \right]  \text{.} \label{apple}%
\end{align}
From our discussion in Ref. \cite{alsing1}, the term $\left\langle \left(
\Delta h\right)  ^{4}\right\rangle -\left\langle \left(  \Delta h\right)
^{2}\right\rangle ^{2}-\left\langle \left(  \Delta h\right)  ^{3}\right\rangle
^{2}$ in the time-dependent expression of $\tau_{\mathrm{AC}}^{2}\left(
s\right)  $ in Eq. (\ref{apple}) can be recast as $\mathrm{\alpha}%
_{4}-1-\mathrm{\alpha}_{3}^{2}$. Moreover, from our analysis presented in the
previous subsection in Eqs. (\ref{KS4}) and (\ref{love}), we also have
$\left\langle \left(  \Delta h^{\prime}\right)  ^{2}\right\rangle
-\left\langle \Delta h^{\prime}\right\rangle ^{2}=\left\langle \left(  \Delta
h^{\prime}\right)  ^{2}\right\rangle =\left[  \left(  \sigma_{\mathrm{\dot{H}%
}}\right)  ^{2}-\left(  \dot{\sigma}_{\mathrm{H}}\right)  ^{2}\right]  /v^{4}$
and $i\left\langle \left[  \left(  \Delta h\right)  ^{2}\text{, }\Delta
h^{\prime}\right]  \right\rangle =i\left[  \mathrm{cov}\left(  \mathrm{H}%
\text{, }\left[  \mathrm{H}\text{, }\mathrm{\dot{H}}\right]  \right)
+\mathrm{cov}\left(  \left[  \mathrm{H}\text{, }\mathrm{\dot{H}}\right]
\text{, }\mathrm{H}\right)  \right]  /v^{4}$, respectively. Therefore, we only
need to focus on the terms $\left\langle \left(  \Delta h^{\prime}\right)
\left(  \Delta h\right)  \right\rangle \left\langle \left(  \Delta h\right)
\left(  \Delta h^{\prime}\right)  \right\rangle $ and $i\left\langle \left(
\Delta h\right)  ^{3}\right\rangle \left\langle \left[  \Delta h\text{,
}\Delta h^{\prime}\right]  \right\rangle $ in Eq. (\ref{apple}). Recalling
that $\Delta h\overset{\text{def}}{=}\Delta\mathrm{H}/v$ and $v^{2}%
\overset{\text{def}}{=}\left\langle \left(  \Delta\mathrm{H}\right)
^{2}\right\rangle $, $\left\langle \left(  \Delta h^{\prime}\right)  \left(
\Delta h\right)  \right\rangle \left\langle \left(  \Delta h\right)  \left(
\Delta h^{\prime}\right)  \right\rangle $ can be written as%
\begin{align}
\left\langle \left(  \Delta h^{\prime}\right)  \left(  \Delta h\right)
\right\rangle \left\langle \left(  \Delta h\right)  \left(  \Delta h^{\prime
}\right)  \right\rangle  &  =\left(  \frac{\left\langle \left(  \Delta
\mathrm{\dot{H}}\right)  \left(  \Delta\mathrm{H}\right)  \right\rangle
}{v^{3}}-\frac{\dot{v}}{v^{2}}\right)  \left(  \frac{\left\langle \left(
\Delta\mathrm{H}\right)  \left(  \Delta\mathrm{\dot{H}}\right)  \right\rangle
}{v^{3}}-\frac{\dot{v}}{v^{2}}\right) \nonumber\\
&  =\frac{1}{v^{4}}\left(  \frac{\left\langle \left(  \Delta\mathrm{\dot{H}%
}\right)  \left(  \Delta\mathrm{H}\right)  \right\rangle }{v}-\dot{v}\right)
\left(  \frac{\left\langle \left(  \Delta\mathrm{H}\right)  \left(
\Delta\mathrm{\dot{H}}\right)  \right\rangle }{v}-\dot{v}\right) \nonumber\\
&  =\frac{1}{v^{6}}\left(  \left\langle \left(  \Delta\mathrm{\dot{H}}\right)
\left(  \Delta\mathrm{H}\right)  \right\rangle -v\dot{v}\right)  \left(
\left\langle \left(  \Delta\mathrm{H}\right)  \left(  \Delta\mathrm{\dot{H}%
}\right)  \right\rangle -v\dot{v}\right)  \text{,}%
\end{align}
that is,%
\begin{equation}
\left\langle \left(  \Delta h^{\prime}\right)  \left(  \Delta h\right)
\right\rangle \left\langle \left(  \Delta h\right)  \left(  \Delta h^{\prime
}\right)  \right\rangle =\frac{1}{v^{6}}\left(  \left\langle \left(
\Delta\mathrm{\dot{H}}\right)  \left(  \Delta\mathrm{H}\right)  \right\rangle
-v\dot{v}\right)  \left(  \left\langle \left(  \Delta\mathrm{H}\right)
\left(  \Delta\mathrm{\dot{H}}\right)  \right\rangle -v\dot{v}\right)
\text{.} \label{he1}%
\end{equation}
Then, noting that $v$ $\dot{v}=\left\langle \left(  \Delta\mathrm{\dot{H}%
}\right)  \left(  \Delta\mathrm{H}\right)  \right\rangle /2+\left\langle
\left(  \Delta\mathrm{H}\right)  \left(  \Delta\mathrm{\dot{H}}\right)
\right\rangle /2$ , Eq. (\ref{he1}) yields%
\begin{align}
\left\langle \left(  \Delta h^{\prime}\right)  \left(  \Delta h\right)
\right\rangle \left\langle \left(  \Delta h\right)  \left(  \Delta h^{\prime
}\right)  \right\rangle  &  =\frac{1}{v^{6}}\left(  \left\langle \left(
\Delta\mathrm{\dot{H}}\right)  \left(  \Delta\mathrm{H}\right)  \right\rangle
-\frac{\left\langle \left(  \Delta\mathrm{\dot{H}}\right)  \left(
\Delta\mathrm{H}\right)  \right\rangle }{2}-\frac{\left\langle \left(
\Delta\mathrm{H}\right)  \left(  \Delta\mathrm{\dot{H}}\right)  \right\rangle
}{2}\right)  \cdot\nonumber\\
&  \cdot\left(  \left\langle \left(  \Delta\mathrm{H}\right)  \left(
\Delta\mathrm{\dot{H}}\right)  \right\rangle -\frac{\left\langle \left(
\Delta\mathrm{\dot{H}}\right)  \left(  \Delta\mathrm{H}\right)  \right\rangle
}{2}-\frac{\left\langle \left(  \Delta\mathrm{H}\right)  \left(
\Delta\mathrm{\dot{H}}\right)  \right\rangle }{2}\right) \nonumber\\
&  =\frac{1}{v^{6}}\left(  \frac{\left\langle \left(  \Delta\mathrm{\dot{H}%
}\right)  \left(  \Delta\mathrm{H}\right)  \right\rangle }{2}-\frac
{\left\langle \left(  \Delta\mathrm{H}\right)  \left(  \Delta\mathrm{\dot{H}%
}\right)  \right\rangle }{2}\right)  \left(  \frac{\left\langle \left(
\Delta\mathrm{H}\right)  \left(  \Delta\mathrm{\dot{H}}\right)  \right\rangle
}{2}-\frac{\left\langle \left(  \Delta\mathrm{\dot{H}}\right)  \left(
\Delta\mathrm{H}\right)  \right\rangle }{2}\right) \nonumber\\
&  =-\frac{1}{v^{6}}\left(  \frac{\left\langle \left(  \Delta\mathrm{H}%
\right)  \left(  \Delta\mathrm{\dot{H}}\right)  \right\rangle }{2}%
-\frac{\left\langle \left(  \Delta\mathrm{\dot{H}}\right)  \left(
\Delta\mathrm{H}\right)  \right\rangle }{2}\right)  ^{2}\nonumber\\
&  =-\frac{1}{4v^{6}}\left\langle \left[  \Delta\mathrm{H}\text{, }%
\Delta\mathrm{\dot{H}}\right]  \right\rangle ^{2}\nonumber\\
&  =-\frac{1}{4v^{6}}\left\langle \left[  \mathrm{H}\text{, }\mathrm{\dot{H}%
}\right]  \right\rangle ^{2}\text{,}%
\end{align}
that is,%
\begin{equation}
\left\langle \left(  \Delta h^{\prime}\right)  \left(  \Delta h\right)
\right\rangle \left\langle \left(  \Delta h\right)  \left(  \Delta h^{\prime
}\right)  \right\rangle =-\frac{1}{4v^{6}}\left\langle \left[  \mathrm{H}%
\text{, }\mathrm{\dot{H}}\right]  \right\rangle ^{2}\text{.} \label{showme}%
\end{equation}
Interestingly, observe that $\left\langle \left(  \Delta h^{\prime}\right)
\left(  \Delta h\right)  \right\rangle \left\langle \left(  \Delta h\right)
\left(  \Delta h^{\prime}\right)  \right\rangle \geq0$ since $\left\langle
\left[  \mathrm{H}\text{, }\mathrm{\dot{H}}\right]  \right\rangle $ in\ Eq.
(\ref{showme}) is purely imaginary. We now focus on the last term in Eq.
(\ref{apple}) that needs to be addressed, namely $i\left\langle \left(  \Delta
h\right)  ^{3}\right\rangle \left\langle \left[  \Delta h\text{, }\Delta
h^{\prime}\right]  \right\rangle $. We have,
\begin{align}
i\left\langle \left(  \Delta h\right)  ^{3}\right\rangle \left\langle \left[
\Delta h\text{, }\Delta h^{\prime}\right]  \right\rangle  &  =i\left\langle
\left(  \Delta\mathrm{H}\right)  ^{3}\right\rangle \frac{\left\langle \left[
\Delta\mathrm{H}\text{, }\Delta\mathrm{\dot{H}}\right]  \right\rangle
}{\left\langle \left(  \Delta\mathrm{H}\right)  ^{2}\right\rangle ^{3}%
}\nonumber\\
&  =i\left\langle \left(  \Delta\mathrm{H}\right)  ^{3}\right\rangle
\frac{\left\langle \left[  \mathrm{H}\text{, }\mathrm{\dot{H}}\right]
\right\rangle }{\left\langle \left(  \Delta\mathrm{H}\right)  ^{2}%
\right\rangle ^{3}}\nonumber\\
&  =i\frac{\left\langle \left(  \Delta\mathrm{H}\right)  ^{3}\right\rangle
}{\left\langle \left(  \Delta\mathrm{H}\right)  ^{2}\right\rangle ^{3/2}}%
\frac{\left\langle \left[  \mathrm{H}\text{, }\mathrm{\dot{H}}\right]
\right\rangle }{\left\langle \left(  \Delta\mathrm{H}\right)  ^{2}%
\right\rangle ^{3/2}}\nonumber\\
&  =i\mathrm{\alpha}_{3}\frac{\left\langle \left[  \mathrm{H}\text{,
}\mathrm{\dot{H}}\right]  \right\rangle }{v^{3}}\text{,}%
\end{align}
that is,%
\begin{equation}
i\left\langle \left(  \Delta h\right)  ^{3}\right\rangle \left\langle \left[
\Delta h\text{, }\Delta h^{\prime}\right]  \right\rangle =i\mathrm{\alpha}%
_{3}\frac{\left\langle \left[  \mathrm{H}\text{, }\mathrm{\dot{H}}\right]
\right\rangle }{v^{3}}\text{.} \label{toni}%
\end{equation}
From Eq. (\ref{toni}), we note for completeness that $i\left\langle \left(
\Delta h\right)  ^{3}\right\rangle \left\langle \left[  \Delta h\text{,
}\Delta h^{\prime}\right]  \right\rangle $ is a real quantity since
$\left\langle \left[  \mathrm{H}\text{, }\mathrm{\dot{H}}\right]
\right\rangle $ is purely imaginary. In conclusion, exploiting our results
from the previous subsection and substituting Eqs. (\ref{showme}) and
(\ref{toni}) into Eq. (\ref{apple}), we recover Eq. (\ref{form2}). For a
detailed discussion on a statistical perspective on the vanishing of the
time-dependent torsion coefficient for two-level systems, we refer to Appendix A.

\section{Nonstationary Hamiltonians and two-level systems}

Although the expressions for the time-dependent curvature and torsion
coefficients proposed in Section III are formally applicable to arbitrary
finite-dimensional quantum systems, it can be rather challenging to find
closed-form expressions for them. This challenge is not only caused by the
difficulty with the analytical integration of Schr\"{o}dinger's evolution
equation for arbitrary time-varying potentials, it is also due to the delicate
parametrization relations between the arc length parameter $s$ and the
physical time $t$. As a matter of fact, even assuming that one is able to
arrive at $s=s\left(  t\right)  $, it may be impossible inverting the relation
and finding $t=t\left(  s\right)  $. However, this latter relation is
necessary in order to calculate curvature and torsion coefficients via the
projector operator formalism. The other approach, that is to say the one based
upon the calculation of expectation values, requires at least the knowledge of
\ the state vector $\left\vert \psi\left(  t\right)  \right\rangle $
satisfying the Schr\"{o}dinger's evolution equation and, despite exhibiting an
insightful statistical interpretation as discussed in the previous section, it
lacks a neat and transparent geometric appeal.

Motivated by these drawbacks, we focus on this section on the simplest (yet
nontrivial) time-dependent type of problem, i.e., the two-dimensional
one.\ Specifically, we focus on nonstationary Hamiltonians and two-level
quantum systems. In particular, we provide a closed-form expression for the
curvature coefficient for a curve traced by a single-qubit quantum state that
evolves under an arbitrary time-dependent Hamiltonian in terms of only two
real three-dimensional vectors with a clear geometric visualization. These two
vectors are the Bloch vector $\mathbf{a}\left(  t\right)  $ and the magnetic
field vector $\mathbf{m}\left(  t\right)  $, respectively. While the former
vector enters via $\rho\left(  t\right)  =$ $\left\vert \psi\left(  t\right)
\right\rangle \left\langle \psi\left(  t\right)  \right\vert
\overset{\text{def}}{=}\left[  \mathrm{I}+\mathbf{a}\left(  t\right)
\cdot\vec{\sigma}\right]  /2$, the latter enters the nonstationary Hamiltonian
\textrm{H}$\left(  t\right)  \overset{\text{def}}{=}\mathbf{m}\left(
t\right)  \cdot\vec{\sigma}$.

\subsection{Curvature}

In what follows, we would like to express the curvature coefficient
$\kappa_{\mathrm{AC}}^{2}$ given by%
\begin{align}
\kappa_{\mathrm{AC}}^{2}  &  =\left\langle \left(  \Delta h\right)
^{4}\right\rangle -\left\langle \left(  \Delta h\right)  ^{2}\right\rangle
^{2}+\left[  \left\langle \left(  \Delta h^{\prime}\right)  ^{2}\right\rangle
-\left\langle \Delta h^{\prime}\right\rangle ^{2}\right]  +i\left\langle
\left[  \left(  \Delta h\right)  ^{2}\text{, }\Delta h^{\prime}\right]
\right\rangle \nonumber\\
&  =\frac{\left\langle \left(  \Delta\mathrm{H}\right)  ^{4}\right\rangle
-\left\langle \left(  \Delta\mathrm{H}\right)  ^{2}\right\rangle ^{2}%
}{\left\langle \left(  \Delta\mathrm{H}\right)  ^{2}\right\rangle ^{2}}%
+\frac{\left\langle \left(  \Delta\mathrm{\dot{H}}\right)  ^{2}\right\rangle
-\left(  \partial_{t}\sqrt{\left\langle \left(  \Delta\mathrm{H}\right)
^{2}\right\rangle }\right)  ^{2}}{\left\langle \left(  \Delta\mathrm{H}%
\right)  ^{2}\right\rangle ^{2}}+i\frac{\left\langle \left[  \left(
\Delta\mathrm{H}\right)  ^{2}\text{, }\Delta\mathrm{\dot{H}}\right]
\right\rangle }{\left\langle \left(  \Delta\mathrm{H}\right)  ^{2}%
\right\rangle ^{2}}\text{,} \label{raccia1}%
\end{align}
in terms of the vectors $\mathbf{a}$ and $\mathbf{m}$. Note that an arbitrary
qubit observable $Q=q_{0}\mathrm{I}+\mathbf{q\cdot}\vec{\sigma}$ with
$q_{0}\in%
%TCIMACRO{\U{211d} }%
%BeginExpansion
\mathbb{R}
%EndExpansion
$ and $\mathbf{q\in%
%TCIMACRO{\U{211d} }%
%BeginExpansion
\mathbb{R}
%EndExpansion
}^{3}$ has a corresponding expectation value given by $\left\langle
Q\right\rangle _{\rho}=q_{0}+\mathbf{a\cdot q}$, where $\rho
\overset{\text{def}}{=}\left[  \mathrm{I}+\mathbf{a}\cdot\vec{\sigma}\right]
/2$. From the time-independent analysis in Ref. \cite{alsing1}, we recall that
the term $\left\langle \left(  \Delta h\right)  ^{4}\right\rangle
-\left\langle \left(  \Delta h\right)  ^{2}\right\rangle ^{2}$ in Eq.
(\ref{raccia1}) becomes
\begin{equation}
\left\langle \left(  \Delta h\right)  ^{4}\right\rangle -\left\langle \left(
\Delta h\right)  ^{2}\right\rangle ^{2}=\frac{\left\langle \left(
\Delta\mathrm{H}\right)  ^{4}\right\rangle -\left\langle \left(
\Delta\mathrm{H}\right)  ^{2}\right\rangle ^{2}}{\left\langle \left(
\Delta\mathrm{H}\right)  ^{2}\right\rangle ^{2}}=4\frac{\left\langle
\mathrm{H}\right\rangle ^{2}}{\left\langle \left(  \Delta\mathrm{H}\right)
^{2}\right\rangle }=4\frac{\left(  \mathbf{a\cdot m}\right)  ^{2}}%
{\mathbf{m}^{2}-\left(  \mathbf{a\cdot m}\right)  ^{2}}\text{.} \label{azz1}%
\end{equation}
Let us focus our attention on the term $\left\langle \left(  \Delta h^{\prime
}\right)  ^{2}\right\rangle -\left\langle \Delta h^{\prime}\right\rangle ^{2}%
$. We begin by noticing that%
\begin{equation}
\Delta h^{\prime}=\partial_{s}\left(  \Delta h\right)  =\frac{1}{v}%
\partial_{t}\left(  \frac{\Delta\mathrm{H}}{v}\right)  =\frac{1}{v}\left[
\frac{\Delta\mathrm{\dot{H}}}{v}-\frac{\Delta\mathrm{H}}{v^{2}}\dot{v}\right]
=\frac{\Delta\mathrm{\dot{H}}}{v^{2}}-\frac{\Delta\mathrm{H}}{v^{3}}\dot
{v}\text{,} \label{iman}%
\end{equation}
where $\left\langle \Delta h^{\prime}\right\rangle =0$ since $\left\langle
\Delta\mathrm{H}\right\rangle =0$ and $\left\langle \Delta\mathrm{\dot{H}%
}\right\rangle =0$. Recall that $\Delta h\overset{\text{def}}{=}%
\Delta\mathrm{H}/v$ and $v\overset{\text{def}}{=}\sqrt{\left\langle \left(
\Delta\mathrm{H}\right)  ^{2}\right\rangle }$ (where we set $\hslash=1$) is
the speed of quantum evolution in projective Hilbert space. We need to
calculate$\left\langle \left(  \Delta h^{\prime}\right)  ^{2}\right\rangle $.
Therefore, using Eq. (\ref{iman}), we observe that%
\begin{align}
\left(  \Delta h^{\prime}\right)  ^{2}  &  =\left(  \Delta h^{\prime}\right)
\left(  \Delta h^{\prime}\right) \nonumber\\
&  =\left(  \frac{\Delta\mathrm{\dot{H}}}{v^{2}}-\frac{\Delta\mathrm{H}}%
{v^{3}}\dot{v}\right)  \left(  \frac{\Delta\mathrm{\dot{H}}}{v^{2}}%
-\frac{\Delta\mathrm{H}}{v^{3}}\dot{v}\right) \nonumber\\
&  =\frac{\left(  \Delta\mathrm{\dot{H}}\right)  ^{2}}{v^{4}}-\frac{\left(
\Delta\mathrm{\dot{H}}\right)  \left(  \Delta\mathrm{H}\right)  }{v^{5}}%
\dot{v}-\frac{\left(  \Delta\mathrm{H}\right)  \left(  \Delta\mathrm{\dot{H}%
}\right)  }{v^{5}}\dot{v}+\frac{\dot{v}^{2}}{v^{6}}\left(  \Delta
\mathrm{H}\right)  ^{2}\nonumber\\
&  =\frac{\left(  \Delta\mathrm{\dot{H}}\right)  ^{2}}{v^{4}}-\frac{\dot{v}%
}{v^{5}}\left[  \left(  \Delta\mathrm{\dot{H}}\right)  \left(  \Delta
\mathrm{H}\right)  +\left(  \Delta\mathrm{H}\right)  \left(  \Delta
\mathrm{\dot{H}}\right)  \right]  +\frac{\dot{v}^{2}}{v^{6}}\left(
\Delta\mathrm{H}\right)  ^{2}\text{,} \label{ma3}%
\end{align}
that is,%
\begin{equation}
\left(  \Delta h^{\prime}\right)  ^{2}=\frac{\left(  \Delta\mathrm{\dot{H}%
}\right)  ^{2}}{v^{4}}-\frac{\dot{v}}{v^{5}}\left[  \left(  \Delta
\mathrm{\dot{H}}\right)  \left(  \Delta\mathrm{H}\right)  +\left(
\Delta\mathrm{H}\right)  \left(  \Delta\mathrm{\dot{H}}\right)  \right]
+\frac{\dot{v}^{2}}{v^{6}}\left(  \Delta\mathrm{H}\right)  ^{2}\text{.}
\label{ma3b}%
\end{equation}
We seek now an expression for $\dot{v}$ in Eq. (\ref{ma3}). Given that
$v\overset{\text{def}}{=}\sqrt{\left\langle \left(  \Delta\mathrm{H}\right)
^{2}\right\rangle }$, we get%
\begin{align}
\dot{v}  &  =\partial_{t}\left(  \sqrt{\left\langle \left(  \Delta
\mathrm{H}\right)  ^{2}\right\rangle }\right) \nonumber\\
&  =\frac{\partial_{t}\left(  \left\langle \left(  \Delta\mathrm{H}\right)
^{2}\right\rangle \right)  }{2\sqrt{\left\langle \left(  \Delta\mathrm{H}%
\right)  ^{2}\right\rangle }}\nonumber\\
&  =\frac{\left\langle \left(  \Delta\mathrm{\dot{H}}\right)  \left(
\Delta\mathrm{H}\right)  +\left(  \Delta\mathrm{H}\right)  \left(
\Delta\mathrm{\dot{H}}\right)  \right\rangle }{2\sqrt{\left\langle \left(
\Delta\mathrm{H}\right)  ^{2}\right\rangle }}\nonumber\\
&  =\frac{\left\langle \left(  \Delta\mathrm{\dot{H}}\right)  \left(
\Delta\mathrm{H}\right)  +\left(  \Delta\mathrm{H}\right)  \left(
\Delta\mathrm{\dot{H}}\right)  \right\rangle }{2v}\text{,}%
\end{align}
that is,%
\begin{equation}
2v\dot{v}=\left\langle \left(  \Delta\mathrm{\dot{H}}\right)  \left(
\Delta\mathrm{H}\right)  +\left(  \Delta\mathrm{H}\right)  \left(
\Delta\mathrm{\dot{H}}\right)  \right\rangle \text{.} \label{ma2}%
\end{equation}
Substituting Eq. (\ref{ma2}) into $\left\langle \left(  \Delta h^{\prime
}\right)  ^{2}\right\rangle $ with $\left(  \Delta h^{\prime}\right)  ^{2}$ in
Eq. (\ref{ma3b}), we get%
\begin{align}
\left\langle \left(  \Delta h^{\prime}\right)  ^{2}\right\rangle  &
=\frac{\left\langle \left(  \Delta\mathrm{\dot{H}}\right)  ^{2}\right\rangle
}{v^{4}}-\frac{\dot{v}}{v^{5}}\left\langle \left(  \Delta\mathrm{\dot{H}%
}\right)  \left(  \Delta\mathrm{H}\right)  +\left(  \Delta\mathrm{H}\right)
\left(  \Delta\mathrm{\dot{H}}\right)  \right\rangle +\frac{\dot{v}^{2}}%
{v^{6}}\left\langle \left(  \Delta\mathrm{H}\right)  ^{2}\right\rangle
\nonumber\\
&  =\frac{\left\langle \left(  \Delta\mathrm{\dot{H}}\right)  ^{2}%
\right\rangle }{v^{4}}-\frac{\dot{v}}{v^{5}}\left(  2v\dot{v}\right)
+\frac{\dot{v}^{2}}{v^{6}}v^{2}\nonumber\\
&  =\frac{\left\langle \left(  \Delta\mathrm{\dot{H}}\right)  ^{2}%
\right\rangle }{v^{4}}-\frac{\dot{v}^{2}}{v^{4}}\text{,}%
\end{align}
that is,%
\begin{equation}
\left\langle \left(  \Delta h^{\prime}\right)  ^{2}\right\rangle
=\frac{\left\langle \left(  \Delta\mathrm{\dot{H}}\right)  ^{2}\right\rangle
-\dot{v}^{2}}{v^{4}}\text{.} \label{m3}%
\end{equation}
Observing that $\left\langle \left(  \Delta\mathrm{\dot{H}}\right)
^{2}\right\rangle $, $v^{2}$, and $\dot{v}$ are given by
\begin{equation}
\left\langle \left(  \Delta\mathrm{\dot{H}}\right)  ^{2}\right\rangle
=\mathbf{\dot{m}}^{2}-\left[  \partial_{t}\left(  \mathbf{a\cdot m}\right)
\right]  ^{2}\text{, }v^{2}=\mathbf{m}^{2}-\left(  \mathbf{a\cdot m}\right)
^{2}\text{, and }\dot{v}=\frac{\mathbf{m\cdot\dot{m}-}\left(  \mathbf{a\cdot
m}\right)  \left[  \partial_{t}\left(  \mathbf{a\cdot m}\right)  \right]
}{\sqrt{\mathbf{m}^{2}-\left(  \mathbf{a\cdot m}\right)  ^{2}}}\text{,}
\label{vvdot}%
\end{equation}
respectively, the term $\left\langle \left(  \Delta h^{\prime}\right)
^{2}\right\rangle $ in Eq. (\ref{m3}) can be written as%
\begin{align}
\left\langle \left(  \Delta h^{\prime}\right)  ^{2}\right\rangle  &
=\frac{\left\langle \left(  \Delta\mathrm{\dot{H}}\right)  ^{2}\right\rangle
-\dot{v}^{2}}{v^{4}}\nonumber\\
&  =\frac{\left(  \mathbf{\dot{m}}^{2}-\left[  \partial_{t}\left(
\mathbf{a\cdot m}\right)  \right]  ^{2}\right)  -\left(  \frac{\mathbf{m\cdot
\dot{m}-}\left(  \mathbf{a\cdot m}\right)  \left[  \partial_{t}\left(
\mathbf{a\cdot m}\right)  \right]  }{\sqrt{\mathbf{m}^{2}-\left(
\mathbf{a\cdot m}\right)  ^{2}}}\right)  ^{2}}{\left[  \mathbf{m}^{2}-\left(
\mathbf{a\cdot m}\right)  ^{2}\right]  ^{2}}\nonumber\\
&  =\frac{\left(  \mathbf{\dot{m}}^{2}-\left[  \partial_{t}\left(
\mathbf{a\cdot m}\right)  \right]  ^{2}\right)  \left(  \mathbf{m}^{2}-\left(
\mathbf{a\cdot m}\right)  ^{2}\right)  -\left(  \mathbf{m\cdot\dot{m}-}\left(
\mathbf{a\cdot m}\right)  \left[  \partial_{t}\left(  \mathbf{a\cdot
m}\right)  \right]  \right)  ^{2}}{\left[  \mathbf{m}^{2}-\left(
\mathbf{a\cdot m}\right)  ^{2}\right]  ^{3}}\nonumber\\
&  =\frac{\mathbf{\dot{m}}^{2}\mathbf{m}^{2}-\mathbf{\dot{m}}^{2}\left(
\mathbf{a\cdot m}\right)  ^{2}-\mathbf{m}^{2}\left[  \partial_{t}\left(
\mathbf{a\cdot m}\right)  \right]  ^{2}-\left(  \mathbf{m\cdot\dot{m}}\right)
^{2}+2\left(  \mathbf{m\cdot\dot{m}}\right)  \left(  \mathbf{a\cdot m}\right)
\left[  \partial_{t}\left(  \mathbf{a\cdot m}\right)  \right]  }{\left[
\mathbf{m}^{2}-\left(  \mathbf{a\cdot m}\right)  ^{2}\right]  ^{3}}\nonumber\\
&  =\frac{\left[  \mathbf{\dot{m}}^{2}\mathbf{m}^{2}-\left(  \mathbf{m\cdot
\dot{m}}\right)  ^{2}\right]  -\left[  \partial_{t}\left(  \mathbf{a\cdot
m}\right)  \mathbf{m-}\left(  \mathbf{a\cdot m}\right)  \mathbf{\dot{m}%
}\right]  ^{2}}{\left[  \mathbf{m}^{2}-\left(  \mathbf{a\cdot m}\right)
^{2}\right]  ^{3}}\text{,}%
\end{align}
that is,%
\begin{equation}
\left\langle \left(  \Delta h^{\prime}\right)  ^{2}\right\rangle
=\frac{\left\langle \left(  \Delta\mathrm{\dot{H}}\right)  ^{2}\right\rangle
-\dot{v}^{2}}{v^{4}}=\frac{\left[  \mathbf{\dot{m}}^{2}\mathbf{m}^{2}-\left(
\mathbf{m\cdot\dot{m}}\right)  ^{2}\right]  -\left[  \partial_{t}\left(
\mathbf{a\cdot m}\right)  \mathbf{m-}\left(  \mathbf{a\cdot m}\right)
\mathbf{\dot{m}}\right]  ^{2}}{\left[  \mathbf{m}^{2}-\left(  \mathbf{a\cdot
m}\right)  ^{2}\right]  ^{3}}\text{.} \label{m4}%
\end{equation}
Finally, since $\mathbf{\dot{a}=}2\mathbf{m\times a}$ (for details, see
Appendix B), we have $\partial_{t}\left(  \mathbf{a\cdot m}\right)
=\mathbf{\dot{a}\cdot m+a\cdot\dot{m}=a\cdot\dot{m}}$ since $\mathbf{\dot
{a}\cdot m=}0$. Therefore, $\left\langle \left(  \Delta h^{\prime}\right)
^{2}\right\rangle $ in Eq. (\ref{m4}) reduces to%
\begin{equation}
\left\langle \left(  \Delta h^{\prime}\right)  ^{2}\right\rangle
=\frac{\left[  \mathbf{m}^{2}\mathbf{\dot{m}}^{2}-\left(  \mathbf{m\cdot
\dot{m}}\right)  ^{2}\right]  -\left[  \left(  \mathbf{a\cdot\dot{m}}\right)
\mathbf{m-}\left(  \mathbf{a\cdot m}\right)  \mathbf{\dot{m}}\right]  ^{2}%
}{\left[  \mathbf{m}^{2}-\left(  \mathbf{a\cdot m}\right)  ^{2}\right]  ^{3}%
}\text{.} \label{azz2}%
\end{equation}
We need to focus now on the last term in Eq. (\ref{raccia1}), $i\left\langle
\left[  \left(  \Delta h\right)  ^{2}\text{, }\Delta h^{\prime}\right]
\right\rangle =i\left\langle \left[  \left(  \Delta\mathrm{H}\right)
^{2}\text{, }\Delta\mathrm{\dot{H}}\right]  \right\rangle /\left\langle
\left(  \Delta\mathrm{H}\right)  ^{2}\right\rangle ^{2}$. Let us begin by
simplifying the expression for $\left[  \left(  \Delta\mathrm{H}\right)
^{2}\text{, }\Delta\mathrm{\dot{H}}\right]  $. We have,%
\begin{align}
\left(  \Delta\mathrm{H}\right)  ^{2}  &  =\left[  \left(  \mathbf{m\cdot
\sigma}\right)  -\left(  \mathbf{a\cdot m}\right)  \mathrm{I}\right]
^{2}\nonumber\\
&  =\left[  \left(  \mathbf{m\cdot\sigma}\right)  -\left(  \mathbf{a\cdot
m}\right)  \mathrm{I}\right]  \left[  \left(  \mathbf{m\cdot\sigma}\right)
-\left(  \mathbf{a\cdot m}\right)  \mathrm{I}\right] \nonumber\\
&  =\left(  \mathbf{m\cdot\sigma}\right)  ^{2}+\left(  \mathbf{a\cdot
m}\right)  ^{2}\mathrm{I}-2\left(  \mathbf{a\cdot m}\right)  \left(
\mathbf{m\cdot\sigma}\right) \nonumber\\
&  =\left[  \mathbf{m}^{2}+\left(  \mathbf{a\cdot m}\right)  ^{2}\right]
\mathrm{I}-2\left(  \mathbf{a\cdot m}\right)  \left(  \mathbf{m\cdot\sigma
}\right)  \label{work1}%
\end{align}
and,%
\begin{equation}
\Delta\mathrm{\dot{H}}=\mathrm{\dot{H}}-\left\langle \mathrm{\dot{H}%
}\right\rangle =\left(  \mathbf{\dot{m}\cdot\sigma}\right)  -\partial
_{t}\left(  \mathbf{a\cdot m}\right)  \mathrm{I}\text{.} \label{work2}%
\end{equation}
Using Eqs. (\ref{work1}) and (\ref{work2}), the commutator $\left[  \left(
\Delta\mathrm{H}\right)  ^{2}\text{, }\Delta\mathrm{\dot{H}}\right]  $ becomes%
\begin{align}
\left[  \left(  \Delta\mathrm{H}\right)  ^{2}\text{, }\Delta\mathrm{\dot{H}%
}\right]   &  =\left[  \left(  \mathbf{m}^{2}+\left(  \mathbf{a\cdot
m}\right)  ^{2}\right)  \mathrm{I}-2\left(  \mathbf{a\cdot m}\right)  \left(
\mathbf{m\cdot\sigma}\right)  \text{, }\left(  \mathbf{\dot{m}\cdot\sigma
}\right)  -\partial_{t}\left(  \mathbf{a\cdot m}\right)  \mathrm{I}\right]
\nonumber\\
&  =\left[  -2\left(  \mathbf{a\cdot m}\right)  \left(  \mathbf{m\cdot\sigma
}\right)  \text{, }\mathbf{\dot{m}\cdot\sigma}\right] \nonumber\\
&  =-2\left(  \mathbf{a\cdot m}\right)  \left[  \left(  \mathbf{m\cdot\sigma
}\right)  \text{, }\mathbf{\dot{m}\cdot\sigma}\right] \nonumber\\
&  =-2\left(  \mathbf{a\cdot m}\right)  2i\left(  \mathbf{m\times\dot{m}%
}\right)  \mathbf{\cdot\sigma}\nonumber\\
&  =-4i\left(  \mathbf{a\cdot m}\right)  \left(  \mathbf{m\times\dot{m}%
}\right)  \mathbf{\cdot\sigma}\text{,} \label{work3}%
\end{align}
that is,%
\begin{equation}
\left[  \left(  \Delta\mathrm{H}\right)  ^{2}\text{, }\Delta\mathrm{\dot{H}%
}\right]  =-4i\left(  \mathbf{a\cdot m}\right)  \left(  \mathbf{m\times\dot
{m}}\right)  \mathbf{\cdot\sigma} \label{work4}%
\end{equation}
From Eq. (\ref{work4}), we get that the expectation value $\left\langle
\left[  \left(  \Delta\mathrm{H}\right)  ^{2}\text{, }\Delta\mathrm{\dot{H}%
}\right]  \right\rangle $ becomes
\begin{align}
\left\langle \left[  \left(  \Delta\mathrm{H}\right)  ^{2}\text{, }%
\Delta\mathrm{\dot{H}}\right]  \right\rangle  &  =-4i\left(  \mathbf{a\cdot
m}\right)  \left\langle \left(  \mathbf{m\times\dot{m}}\right)  \mathbf{\cdot
\sigma}\right\rangle \nonumber\\
&  =-4i\left(  \mathbf{a\cdot m}\right)  \mathrm{tr}\left[  \frac
{\mathrm{I}+\mathbf{a\cdot\sigma}}{2}\left(  \mathbf{m\times\dot{m}}\right)
\mathbf{\cdot\sigma}\right] \nonumber\\
&  =-4i\left(  \mathbf{a\cdot m}\right)  \mathrm{tr}\left[  \frac{\left(
\mathbf{m\times\dot{m}}\right)  \mathbf{\cdot\sigma}+\left(  \mathbf{a\cdot
\sigma}\right)  \left[  \left(  \mathbf{m\times\dot{m}}\right)  \mathbf{\cdot
\sigma}\right]  }{2}\right] \nonumber\\
&  =-4i\left(  \mathbf{a\cdot m}\right)  \left[  \mathbf{a\cdot}\left(
\mathbf{m\times\dot{m}}\right)  \right]  \text{,}%
\end{align}
that is,%
\begin{equation}
\frac{i\left\langle \left[  \left(  \Delta\mathrm{H}\right)  ^{2}\text{,
}\Delta\mathrm{\dot{H}}\right]  \right\rangle }{\left\langle \left(
\Delta\mathrm{H}\right)  ^{2}\right\rangle ^{2}}=4\frac{\left(  \mathbf{a\cdot
m}\right)  \left[  \mathbf{a\cdot}\left(  \mathbf{m\times\dot{m}}\right)
\right]  }{\left[  \mathbf{m}^{2}-\left(  \mathbf{a\cdot m}\right)
^{2}\right]  ^{2}}\text{.} \label{azz3}%
\end{equation}
Finally, combining Eqs. (\ref{azz1}), (\ref{azz2}), and (\ref{azz3}),
$\kappa_{\mathrm{AC}}^{2}$ in Eq. (\ref{raccia1}) reduces to%
\begin{equation}
\kappa_{\mathrm{AC}}^{2}\left(  \mathbf{a}\text{, }\mathbf{m}\right)
=4\frac{\left(  \mathbf{a\cdot m}\right)  ^{2}}{\mathbf{m}^{2}-\left(
\mathbf{a\cdot m}\right)  ^{2}}+\frac{\left[  \mathbf{m}^{2}\mathbf{\dot{m}%
}^{2}-\left(  \mathbf{m\cdot\dot{m}}\right)  ^{2}\right]  -\left[  \left(
\mathbf{a\cdot\dot{m}}\right)  \mathbf{m-}\left(  \mathbf{a\cdot m}\right)
\mathbf{\dot{m}}\right]  ^{2}}{\left[  \mathbf{m}^{2}-\left(  \mathbf{a\cdot
m}\right)  ^{2}\right]  ^{3}}+4\frac{\left(  \mathbf{a\cdot m}\right)  \left[
\mathbf{a\cdot}\left(  \mathbf{m\times\dot{m}}\right)  \right]  }{\left[
\mathbf{m}^{2}-\left(  \mathbf{a\cdot m}\right)  ^{2}\right]  ^{2}}\text{.}
\label{XXX}%
\end{equation}
Eq. (\ref{XXX}) ends our derivation by expressing the curvature coefficient
$\kappa_{\mathrm{AC}}^{2}$ as a function of two three-dimensional vectors,
namely the Bloch vector $\mathbf{a}$ and the magnetic field vector
$\mathbf{m}$.

\subsection{Torsion}

In what follows, we would like to express the torsion coefficient
$\tau_{\mathrm{AC}}^{2}$ with%
\begin{align}
\tau_{\mathrm{AC}}^{2}\left(  s\right)   &  =\left\langle \left(  \Delta
h\right)  ^{4}\right\rangle -\left\langle \left(  \Delta h\right)
^{2}\right\rangle ^{2}-\left\langle \left(  \Delta h\right)  ^{3}\right\rangle
^{2}+\left[  \left\langle \left(  \Delta h^{\prime}\right)  ^{2}\right\rangle
-\left\langle \Delta h^{\prime}\right\rangle ^{2}-\left\langle \left(  \Delta
h^{\prime}\right)  \left(  \Delta h\right)  \right\rangle \left\langle \left(
\Delta h\right)  \left(  \Delta h^{\prime}\right)  \right\rangle \right]
+\nonumber\\
& \nonumber\\
&  +i\left[  \left\langle \left[  \left(  \Delta h\right)  ^{2}\text{, }\Delta
h^{\prime}\right]  \right\rangle -\left\langle \left(  \Delta h\right)
^{3}\right\rangle \left\langle \left[  \Delta h\text{, }\Delta h^{\prime
}\right]  \right\rangle \right]  \text{,} \label{torsione1}%
\end{align}
in terms of the vectors $\mathbf{a}$ and $\mathbf{m}$. From the
time-independent analysis in Ref. \cite{alsing1}, we have that the term
$\left\langle \left(  \Delta h\right)  ^{4}\right\rangle -\left\langle \left(
\Delta h\right)  ^{2}\right\rangle ^{2}-\left\langle \left(  \Delta h\right)
^{3}\right\rangle ^{2}$ vanishes
%PMA
identically, independent of the time dependence of $\mathbf{a}$ and
$\mathbf{m}$. Therefore $\tau_{\mathrm{AC}}^{2}\left(  s\right)  $ in Eq.
(\ref{torsione1}) reduces to%
\begin{equation}
\tau_{\mathrm{AC}}^{2}\left(  s\right)  =\left\langle \left(  \Delta
h^{\prime}\right)  ^{2}\right\rangle -\left\langle \left(  \Delta h^{\prime
}\right)  \left(  \Delta h\right)  \right\rangle \left\langle \left(  \Delta
h\right)  \left(  \Delta h^{\prime}\right)  \right\rangle +i\left[
\left\langle \left[  \left(  \Delta h\right)  ^{2}\text{, }\Delta h^{\prime
}\right]  \right\rangle -\left\langle \left(  \Delta h\right)  ^{3}%
\right\rangle \left\langle \left[  \Delta h\text{, }\Delta h^{\prime}\right]
\right\rangle \right]  \text{.} \label{aq}%
\end{equation}
Recalling that (setting $\hslash=1$) $\Delta h\overset{\text{def}}{=}%
\Delta\mathrm{H}/v$, $\partial_{s}\overset{\text{def}}{=}v^{-1}\partial_{t}$,
and $v\overset{\text{def}}{=}\sqrt{\left\langle \left(  \Delta\mathrm{H}%
\right)  ^{2}\right\rangle }$, $\tau_{\mathrm{AC}}^{2}\left(  s\right)  $ in
Eq. (\ref{aq}) can be recast as
\begin{align}
\tau_{\mathrm{AC}}^{2}\left(  t\right)   &  =\left[  \frac{\left\langle
\left(  \Delta\mathrm{\dot{H}}\right)  ^{2}\right\rangle -\left(  \partial
_{t}\sqrt{\left\langle \left(  \Delta\mathrm{H}\right)  ^{2}\right\rangle
}\right)  ^{2}}{\left\langle \left(  \Delta\mathrm{H}\right)  ^{2}%
\right\rangle ^{2}}-\left(  \frac{\left\langle \left(  \Delta\mathrm{\dot{H}%
}\right)  \left(  \Delta\mathrm{H}\right)  \right\rangle }{\left\langle
\left(  \Delta\mathrm{H}\right)  ^{2}\right\rangle ^{3/2}}-\frac{\partial
_{t}\sqrt{\left\langle \left(  \Delta\mathrm{H}\right)  ^{2}\right\rangle }%
}{\left\langle \left(  \Delta\mathrm{H}\right)  ^{2}\right\rangle }\right)
\left(  \frac{\left\langle \left(  \Delta\mathrm{H}\right)  \left(
\Delta\mathrm{\dot{H}}\right)  \right\rangle }{\left\langle \left(
\Delta\mathrm{H}\right)  ^{2}\right\rangle ^{3/2}}-\frac{\partial_{t}%
\sqrt{\left\langle \left(  \Delta\mathrm{H}\right)  ^{2}\right\rangle }%
}{\left\langle \left(  \Delta\mathrm{H}\right)  ^{2}\right\rangle }\right)
\right]  +\nonumber\\
& \nonumber\\
&  +i\left[  \frac{\left\langle \left[  \left(  \Delta\mathrm{H}\right)
^{2}\text{, }\Delta\mathrm{\dot{H}}\right]  \right\rangle }{\left\langle
\left(  \Delta\mathrm{H}\right)  ^{2}\right\rangle ^{2}}-\left\langle \left(
\Delta\mathrm{H}\right)  ^{3}\right\rangle \frac{\left\langle \left[
\Delta\mathrm{H}\text{, }\Delta\mathrm{\dot{H}}\right]  \right\rangle
}{\left\langle \left(  \Delta\mathrm{H}\right)  ^{2}\right\rangle ^{3}%
}\right]  \text{.} \label{aq2}%
\end{align}
Therefore, in order to express $\tau_{\mathrm{AC}}^{2}\left(  t\right)  $ in
Eq. (\ref{aq2}) by means of the the vectors $\mathbf{a}$ and $\mathbf{m}$, we
need to focus on%
\begin{equation}
\tau_{\mathrm{AC}}^{2}\left(  t\right)  =T_{1}\left(  t\right)  +T_{2}\left(
t\right)  +T_{3}\left(  t\right)  +T_{4}\left(  t\right)  \text{,}%
\end{equation}
where $T_{1}\left(  t\right)  $, $T_{2}\left(  t\right)  $, $T_{3}\left(
t\right)  $, and $T_{4}\left(  t\right)  $ are given by
\begin{align}
&  T_{1}\left(  t\right)  \overset{\text{def}}{=}\frac{\left\langle \left(
\Delta\mathrm{\dot{H}}\right)  ^{2}\right\rangle -\left(  \partial_{t}%
\sqrt{\left\langle \left(  \Delta\mathrm{H}\right)  ^{2}\right\rangle
}\right)  ^{2}}{\left\langle \left(  \Delta\mathrm{H}\right)  ^{2}%
\right\rangle ^{2}}\text{,}\nonumber\\
&  T_{2}\left(  t\right)  \overset{\text{def}}{=}-\left(  \frac{\left\langle
\left(  \Delta\mathrm{\dot{H}}\right)  \left(  \Delta\mathrm{H}\right)
\right\rangle }{\left\langle \left(  \Delta\mathrm{H}\right)  ^{2}%
\right\rangle ^{3/2}}-\frac{\partial_{t}\sqrt{\left\langle \left(
\Delta\mathrm{H}\right)  ^{2}\right\rangle }}{\left\langle \left(
\Delta\mathrm{H}\right)  ^{2}\right\rangle }\right)  \left(  \frac
{\left\langle \left(  \Delta\mathrm{H}\right)  \left(  \Delta\mathrm{\dot{H}%
}\right)  \right\rangle }{\left\langle \left(  \Delta\mathrm{H}\right)
^{2}\right\rangle ^{3/2}}-\frac{\partial_{t}\sqrt{\left\langle \left(
\Delta\mathrm{H}\right)  ^{2}\right\rangle }}{\left\langle \left(
\Delta\mathrm{H}\right)  ^{2}\right\rangle }\right)  \text{,}\nonumber\\
&  T_{3}\left(  t\right)  \overset{\text{def}}{=}i\frac{\left\langle \left[
\left(  \Delta\mathrm{H}\right)  ^{2}\text{, }\Delta\mathrm{\dot{H}}\right]
\right\rangle }{\left\langle \left(  \Delta\mathrm{H}\right)  ^{2}%
\right\rangle ^{2}}\text{,}\nonumber\\
&  T_{4}\left(  t\right)  \overset{\text{def}}{=}-i\left\langle \left(
\Delta\mathrm{H}\right)  ^{3}\right\rangle \frac{\left\langle \left[
\Delta\mathrm{H}\text{, }\Delta\mathrm{\dot{H}}\right]  \right\rangle
}{\left\langle \left(  \Delta\mathrm{H}\right)  ^{2}\right\rangle ^{3}%
}\text{.} \label{co1}%
\end{align}
We start by noting that both $T_{1}\left(  t\right)  $ and $T_{3}\left(
t\right)  $ in Eq. (\ref{co1}) appear in the calculation of the time-dependent
curvature coefficient $\kappa_{\mathrm{AC}}^{2}\left(  t\right)  $. Therefore,
from our previous calculation, $T_{1}\left(  t\right)  $ and $T_{3}\left(
t\right)  $ are given by%
\begin{equation}
T_{1}\left(  t\right)  =\frac{\left\langle \left(  \Delta\mathrm{\dot{H}%
}\right)  ^{2}\right\rangle -\dot{v}^{2}}{v^{4}}=\frac{\left[  \mathbf{\dot
{m}}^{2}\mathbf{m}^{2}-\left(  \mathbf{m\cdot\dot{m}}\right)  ^{2}\right]
-\left[  \partial_{t}\left(  \mathbf{a\cdot m}\right)  \mathbf{m-}\left(
\mathbf{a\cdot m}\right)  \mathbf{\dot{m}}\right]  ^{2}}{\left[
\mathbf{m}^{2}-\left(  \mathbf{a\cdot m}\right)  ^{2}\right]  ^{3}} \label{mm}%
\end{equation}
and,%
\begin{equation}
T_{3}\left(  t\right)  =\frac{i\left\langle \left[  \left(  \Delta
\mathrm{H}\right)  ^{2}\text{, }\Delta\mathrm{\dot{H}}\right]  \right\rangle
}{\left\langle \left(  \Delta\mathrm{H}\right)  ^{2}\right\rangle ^{2}}%
=4\frac{\left(  \mathbf{a\cdot m}\right)  \left[  \mathbf{a\cdot}\left(
\mathbf{m\times\dot{m}}\right)  \right]  }{\left[  \mathbf{m}^{2}-\left(
\mathbf{a\cdot m}\right)  ^{2}\right]  ^{2}}\text{,} \label{work5}%
\end{equation}
respectively. Moreover, from the expression of $T_{2}\left(  t\right)  $ in
Eq. (\ref{co1}), we note that%
\begin{equation}
T_{2}\left(  t\right)  =\left[  \frac{\left\langle \left(  \Delta
\mathrm{\dot{H}}\right)  \left(  \Delta\mathrm{H}\right)  \right\rangle
}{v^{3}}-\frac{\dot{v}}{v^{2}}\right]  \left[  \frac{\left\langle \left(
\Delta\mathrm{H}\right)  \left(  \Delta\mathrm{\dot{H}}\right)  \right\rangle
}{v^{3}}-\frac{\dot{v}}{v^{2}}\right]  \text{,} \label{love1}%
\end{equation}
where $\left\langle \left(  \Delta\mathrm{\dot{H}}\right)  \left(
\Delta\mathrm{H}\right)  \right\rangle $ and $\left\langle \left(
\Delta\mathrm{H}\right)  \left(  \Delta\mathrm{\dot{H}}\right)  \right\rangle
$ are given by
\begin{equation}
\left\langle \left(  \Delta\mathrm{\dot{H}}\right)  \left(  \Delta
\mathrm{H}\right)  \right\rangle =\left\langle \left(  \mathrm{\dot{H}%
}\right)  \left(  \mathrm{H}\right)  \right\rangle -\left\langle
\mathrm{\dot{H}}\right\rangle \left\langle \mathrm{H}\right\rangle
=\mathbf{m\cdot\dot{m}-}i\mathbf{a\cdot}\left(  \mathbf{m\times\dot{m}%
}\right)  -\left(  \mathbf{a\cdot m}\right)  \left[  \partial_{t}\left(
\mathbf{a\cdot m}\right)  \right]  \text{,} \label{love2}%
\end{equation}
and,%
\begin{equation}
\left\langle \left(  \Delta\mathrm{H}\right)  \left(  \Delta\mathrm{\dot{H}%
}\right)  \right\rangle =\left\langle \left(  \mathrm{H}\right)  \left(
\mathrm{\dot{H}}\right)  \right\rangle -\left\langle \mathrm{H}\right\rangle
\left\langle \mathrm{\dot{H}}\right\rangle =\mathbf{m\cdot\dot{m}%
+}i\mathbf{a\cdot}\left(  \mathbf{m\times\dot{m}}\right)  -\left(
\mathbf{a\cdot m}\right)  \left[  \partial_{t}\left(  \mathbf{a\cdot
m}\right)  \right]  \text{,} \label{love3}%
\end{equation}
respectively. From Eqs. (\ref{love2}) and (\ref{love3}), we get%
\begin{equation}
\left\langle \left(  \Delta\mathrm{\dot{H}}\right)  \left(  \Delta
\mathrm{H}\right)  \right\rangle \left\langle \left(  \Delta\mathrm{H}\right)
\left(  \Delta\mathrm{\dot{H}}\right)  \right\rangle =\left[  \mathbf{m\cdot
\dot{m}}-\left(  \mathbf{a\cdot m}\right)  \partial_{t}\left(  \mathbf{a\cdot
m}\right)  \right]  ^{2}+\left[  \mathbf{a\cdot}\left(  \mathbf{m\times\dot
{m}}\right)  \right]  ^{2}\text{,} \label{love4}%
\end{equation}
and,%
\begin{equation}
\left\langle \left(  \Delta\mathrm{\dot{H}}\right)  \left(  \Delta
\mathrm{H}\right)  \right\rangle +\left\langle \left(  \Delta\mathrm{H}%
\right)  \left(  \Delta\mathrm{\dot{H}}\right)  \right\rangle =2\left[
\mathbf{m\cdot\dot{m}}-\left(  \mathbf{a\cdot m}\right)  \partial_{t}\left(
\mathbf{a\cdot m}\right)  \right]  \text{.} \label{love5}%
\end{equation}
Inserting Eqs. (\ref{love4}) and (\ref{love5}) into Eq. (\ref{love1}) along
with recalling the expressions of $v$ and $\dot{v}$, the quantity
$\left\langle \left(  \Delta h^{\prime}\right)  \left(  \Delta h\right)
\right\rangle \left\langle \left(  \Delta h\right)  \left(  \Delta h^{\prime
}\right)  \right\rangle $ with $T_{2}(s)\overset{\text{def}}{=}-\left\langle
\left(  \Delta h^{\prime}\right)  \left(  \Delta h\right)  \right\rangle
\left\langle \left(  \Delta h\right)  \left(  \Delta h^{\prime}\right)
\right\rangle $ becomes%
\begin{align}
\left\langle \left(  \Delta h^{\prime}\right)  \left(  \Delta h\right)
\right\rangle \left\langle \left(  \Delta h\right)  \left(  \Delta h^{\prime
}\right)  \right\rangle  &  =\frac{\left\langle \left(  \Delta\mathrm{\dot{H}%
}\right)  \left(  \Delta\mathrm{H}\right)  \right\rangle \left\langle \left(
\Delta\mathrm{H}\right)  \left(  \Delta\mathrm{\dot{H}}\right)  \right\rangle
}{v^{6}}-\frac{\dot{v}}{v^{5}}\left[  \left\langle \left(  \Delta
\mathrm{\dot{H}}\right)  \left(  \Delta\mathrm{H}\right)  \right\rangle
+\left\langle \left(  \Delta\mathrm{H}\right)  \left(  \Delta\mathrm{\dot{H}%
}\right)  \right\rangle \right]  +\frac{\dot{v}^{2}}{v^{4}}\nonumber\\
& \nonumber\\
&  =\frac{\left[  \mathbf{m\cdot\dot{m}}-\left(  \mathbf{a\cdot m}\right)
\partial_{t}\left(  \mathbf{a\cdot m}\right)  \right]  ^{2}+\left[
\mathbf{a\cdot}\left(  \mathbf{m\times\dot{m}}\right)  \right]  ^{2}}{\left[
\mathbf{m}^{2}-\left(  \mathbf{a\cdot m}\right)  ^{2}\right]  ^{3}%
}+\nonumber\\
&  +\frac{\left[  \mathbf{m\cdot\dot{m}}-\left(  \mathbf{a\cdot m}\right)
\partial_{t}\left(  \mathbf{a\cdot m}\right)  \right]  ^{2}}{\left[
\mathbf{m}^{2}-\left(  \mathbf{a\cdot m}\right)  ^{2}\right]  ^{3}%
}+\nonumber\\
&  -\frac{\left[  \mathbf{m\cdot\dot{m}}-\left(  \mathbf{a\cdot m}\right)
\partial_{t}\left(  \mathbf{a\cdot m}\right)  \right]  }{\left[
\mathbf{m}^{2}-\left(  \mathbf{a\cdot m}\right)  ^{2}\right]  ^{3}}2\left[
\mathbf{m\cdot\dot{m}}-\left(  \mathbf{a\cdot m}\right)  \partial_{t}\left(
\mathbf{a\cdot m}\right)  \right]  \text{,}%
\end{align}
that is, after some algebra,%
\begin{equation}
T_{2}\left(  t\right)  =-\frac{\left[  \mathbf{a\cdot}\left(  \mathbf{m\times
\dot{m}}\right)  \right]  ^{2}}{\left[  \mathbf{m}^{2}-\left(  \mathbf{a\cdot
m}\right)  ^{2}\right]  ^{3}}\text{.} \label{term2}%
\end{equation}
The last term that needs to be calculated is $T_{4}\left(  t\right)  $ in Eq.
(\ref{co1}). Recall that $\Delta\mathrm{H}=\mathrm{H}-\left\langle
\mathrm{H}\right\rangle =\mathbf{m\cdot\sigma-}\left(  \mathbf{a\cdot
m}\right)  \mathrm{I}$ and $\Delta\mathrm{\dot{H}}=\mathrm{\dot{H}%
}-\left\langle \mathrm{\dot{H}}\right\rangle =\mathbf{\dot{m}\cdot
\sigma-\partial}_{t}\left(  \mathbf{a\cdot m}\right)  \mathrm{I}$. Therefore,
we have
\begin{equation}
\left[  \Delta\mathrm{H}\text{, }\Delta\mathrm{\dot{H}}\right]  =\left[
\mathbf{m\cdot\sigma-}\left(  \mathbf{a\cdot m}\right)  \mathrm{I}\text{,
}\mathbf{\dot{m}\cdot\sigma-\partial}_{t}\left(  \mathbf{a\cdot m}\right)
\mathrm{I}\right]  =2i\left(  \mathbf{m\times\dot{m}}\right)  \cdot
\mathbf{\sigma}\text{,}%
\end{equation}
that is,%
\begin{equation}
\left\langle \left[  \Delta\mathrm{H}\text{, }\Delta\mathrm{\dot{H}}\right]
\right\rangle =2i\left\langle \left(  \mathbf{m\times\dot{m}}\right)
\cdot\mathbf{\sigma}\right\rangle =2i\left[  \mathbf{a\cdot}\left(
\mathbf{m\times\dot{m}}\right)  \right]  \text{.} \label{t5}%
\end{equation}
Using Eq. (\ref{t5}) along with recalling that $\left\langle \left(
\Delta\mathrm{H}\right)  ^{3}\right\rangle =2\left(  \mathbf{a\cdot m}\right)
\left[  \mathbf{m}^{2}-\left(  \mathbf{a\cdot m}\right)  ^{2}\right]  $ and
$\left\langle \left(  \Delta\mathrm{H}\right)  ^{2}\right\rangle
=\mathbf{m}^{2}-\left(  \mathbf{a\cdot m}\right)  ^{2}$, $T_{4}\left(
t\right)  $ in Eq. (\ref{co1}) reduces to%
\begin{align}
T_{4}\left(  t\right)   &  =-i\left\langle \left(  \Delta\mathrm{H}\right)
^{3}\right\rangle \frac{\left\langle \left[  \Delta\mathrm{H}\text{, }%
\Delta\mathrm{\dot{H}}\right]  \right\rangle }{\left\langle \left(
\Delta\mathrm{H}\right)  ^{2}\right\rangle ^{3}}\nonumber\\
&  =-i\frac{-2\left\langle \mathrm{H}\right\rangle \left\langle \left(
\Delta\mathrm{H}\right)  ^{2}\right\rangle }{\left\langle \left(
\Delta\mathrm{H}\right)  ^{2}\right\rangle ^{3}}\left\langle \left[
\Delta\mathrm{H}\text{, }\Delta\mathrm{\dot{H}}\right]  \right\rangle
\nonumber\\
&  =-i\frac{-2\left(  \mathbf{a\cdot m}\right)  }{\left\langle \left(
\Delta\mathrm{H}\right)  ^{2}\right\rangle ^{2}}2i\left[  \mathbf{a\cdot
}\left(  \mathbf{m\times\dot{m}}\right)  \right] \nonumber\\
&  =-4\frac{\left(  \mathbf{a\cdot m}\right)  \left[  \mathbf{a\cdot}\left(
\mathbf{m\times\dot{m}}\right)  \right]  }{\left[  \mathbf{m}^{2}-\left(
\mathbf{a\cdot m}\right)  ^{2}\right]  ^{2}}\text{,} \label{term4}%
\end{align}
that is,%
\begin{equation}
T_{4}\left(  t\right)  =-4\frac{\left(  \mathbf{a\cdot m}\right)  \left[
\mathbf{a\cdot}\left(  \mathbf{m\times\dot{m}}\right)  \right]  }{\left[
\mathbf{m}^{2}-\left(  \mathbf{a\cdot m}\right)  ^{2}\right]  ^{2}}\text{.}
\label{guarda}%
\end{equation}
Observe that $T_{4}\left(  t\right)  $ in Eq. (\ref{guarda}) equals minus
$T_{3}\left(  t\right)  $ in Eq. (\ref{work5}). Finally, using Eqs.
(\ref{mm}), (\ref{work5}), (\ref{term2}), and (\ref{guarda}), the torsion
coefficient $\tau_{\mathrm{AC}}^{2}\left(  \mathbf{a}\text{, }\mathbf{m}%
\right)  $ equals $T_{1}(t)+T_{2}(t)$ and becomes%
\begin{equation}
\tau_{\mathrm{AC}}^{2}\left(  \mathbf{a}\text{, }\mathbf{m}\right)
=\frac{\left[  \mathbf{m}^{2}\mathbf{\dot{m}}^{2}-\left(  \mathbf{m\cdot
\dot{m}}\right)  ^{2}\right]  -\left[  \left(  \mathbf{a\cdot\dot{m}}\right)
\mathbf{m-}\left(  \mathbf{a\cdot m}\right)  \mathbf{\dot{m}}\right]  ^{2}%
}{\left[  \mathbf{m}^{2}-\left(  \mathbf{a\cdot m}\right)  ^{2}\right]  ^{3}%
}-\frac{\left[  \mathbf{a\cdot}\left(  \mathbf{m\times\dot{m}}\right)
\right]  ^{2}}{\left[  \mathbf{m}^{2}-\left(  \mathbf{a\cdot m}\right)
^{2}\right]  ^{3}}\text{.} \label{torsione}%
\end{equation}
In writing Eq. (\ref{torsione}), we also exploited the fact that
$\mathbf{\dot{a}=}2\mathbf{m\times a}$ implies $\partial_{t}\left(
\mathbf{a\cdot m}\right)  =\mathbf{\dot{a}\cdot m+a\cdot\dot{m}=a\cdot\dot{m}%
}$ since $\mathbf{\dot{a}\cdot m=}0$. At this point, recalling the Lagrange
identity and the vector triple product relations $\left\Vert \mathbf{v}%
_{1}\times\mathbf{v}_{2}\right\Vert ^{2}=\left(  \mathbf{v}_{1}\cdot
\mathbf{v}_{1}\right)  \left(  \mathbf{v}_{2}\cdot\mathbf{v}_{2}\right)
-\left(  \mathbf{v}_{1}\cdot\mathbf{v}_{2}\right)  ^{2}$ and $\mathbf{v}%
_{1}\times\left(  \mathbf{v}_{2}\times\mathbf{v}_{3}\right)  =\left(
\mathbf{v}_{1}\cdot\mathbf{v}_{3}\right)  \mathbf{v}_{2}-\left(
\mathbf{v}_{1}\cdot\mathbf{v}_{2}\right)  \mathbf{v}_{3}$, respectively, we
have%
\begin{equation}
\mathbf{m}^{2}\mathbf{\dot{m}}^{2}-\left(  \mathbf{m\cdot\dot{m}}\right)
^{2}=\left\Vert \mathbf{m}\times\mathbf{\dot{m}}\right\Vert ^{2}\text{,}
\label{poli1}%
\end{equation}
and%
\begin{equation}
\left[  \left(  \mathbf{a\cdot\dot{m}}\right)  \mathbf{m-}\left(
\mathbf{a\cdot m}\right)  \mathbf{\dot{m}}\right]  ^{2}=\left\Vert
\mathbf{a\times}\left(  \mathbf{m}\times\mathbf{\dot{m}}\right)  \right\Vert
^{2}=\left(  \mathbf{a\cdot a}\right)  \left\Vert \mathbf{m}\times
\mathbf{\dot{m}}\right\Vert ^{2}-\left[  \mathbf{a\cdot}\left(
\mathbf{m\times\dot{m}}\right)  \right]  ^{2}=\left\Vert \mathbf{m}%
\times\mathbf{\dot{m}}\right\Vert ^{2}-\left[  \mathbf{a\cdot}\left(
\mathbf{m\times\dot{m}}\right)  \right]  ^{2}\text{.} \label{poli2}%
\end{equation}
Therefore, combing Eqs. (\ref{poli1}) and (\ref{poli2}), we get
\begin{equation}
\left[  \mathbf{m}^{2}\mathbf{\dot{m}}^{2}-\left(  \mathbf{m\cdot\dot{m}%
}\right)  ^{2}\right]  -\left[  \mathbf{a\cdot}\left(  \mathbf{m\times\dot{m}%
}\right)  \right]  ^{2}=\left[  \left(  \mathbf{a\cdot\dot{m}}\right)
\mathbf{m-}\left(  \mathbf{a\cdot m}\right)  \mathbf{\dot{m}}\right]
^{2}\text{,} \label{cul4}%
\end{equation}
and, thus, $\tau_{\mathrm{AC}}^{2}\left(  \mathbf{a}\text{, }\mathbf{m}%
\right)  $ in Eq. (\ref{torsione}) is
%PMA
identically equal to zero%
\begin{equation}
\tau_{\mathrm{AC}}^{2}\left(  \mathbf{a}\text{, }\mathbf{m}\right)  =0\text{.}
\label{YYY}%
\end{equation}
Eq. (\ref{YYY}) ends our derivation by expressing the torsion coefficient
$\kappa_{\mathrm{AC}}^{2}$ as a function of two three-dimensional vectors,
namely the Bloch vector $\mathbf{a}$ and the magnetic field vector
$\mathbf{m}$. Equipped with Eqs. (\ref{XXX}) and (\ref{YYY}), we are now ready
to discuss an illustrative example in the next section.

\section{Illustrative example}

In this section, we consider an exactly soluble time-dependent two-state Rabi
problem specified by a sinusoidal oscillating time-dependent potential.
Specifically, we study the temporal behavior of the time-dependent curvature
coefficient of quantum curves generated by evolving a single-qubit quantum
state under the Hamiltonian $\mathrm{H}\left(  t\right)  $ under distinct
physical regimes. The physical tunable parameters that characterize
$\mathrm{H}\left(  t\right)  $ are the resonance frequency of the atom
$\omega_{0}$, the energy difference $\hslash\omega_{0}$ between the ground and
excited states, the Rabi frequency $\Omega_{0}$ and, finally, the external
driving frequency $\omega$.

\subsection{The Hamiltonian}

We consider a time-dependent two-level quantum system that evolves under the
Hamiltonian \textrm{H}$\left(  t\right)  $ given by,%
\begin{equation}
\mathrm{H}\left(  t\right)  \overset{\text{def}}{=}\frac{\hslash\omega_{0}}%
{2}\sigma_{z}+\hslash\Omega_{0}\left[  \cos\left(  \omega t\right)  \sigma
_{x}+\sin\left(  \omega t\right)  \sigma_{y}\right]  \text{,} \label{time-H}%
\end{equation}
where $\omega_{0}$ is the resonance frequency of the atom, $\Omega_{0}$
denotes the Rabi frequency, and $\omega$ is the external driving frequency.
For convenience, we set $\hslash$ equal to one and recast the Hamiltonian in
Eq. (\ref{time-H}) as%
\begin{equation}
\mathrm{H}\left(  t\right)  =\vec{m}\left(  t\right)  \cdot\vec{\sigma}%
=\Omega_{\mathrm{H}}\hat{m}\left(  t\right)  \cdot\vec{\sigma}\text{,}%
\end{equation}
where $\Omega_{\mathrm{H}}\overset{\text{def}}{=}\sqrt{\Omega_{0}^{2}%
+(\omega_{0}/2)^{2}}$ is the generalized Rabi frequency and $\hat{m}\left(
t\right)  \overset{\text{def}}{=}(1/\Omega_{\mathrm{H}})(\Omega_{0}\cos\left(
\omega t\right)  $, $\Omega_{0}\sin\left(  \omega t\right)  $, $\omega_{0}/2)$
is a time-dependent
%PMA
Bloch unit vector. Interestingly, it is possible to write down the exact
analytical expression of the unitary evolution operator $U(t)$ that
corresponds to the nonstationary Hamiltonian in Eq. (\ref{time-H}) by
employing the rotating frame formalism \cite{rabi37,rabi54}. Specifically,
denote with $U_{\mathrm{RF}}\left(  t\right)  \overset{\text{def}%
}{=}e^{-i\frac{\omega}{2}t\sigma_{z}}$ the unitary frame transformation that
allows to relate states and observables in the laboratory and rotating frames
as $\left\vert \psi\left(  t\right)  \right\rangle =U_{\mathrm{RF}}\left(
t\right)  \left\vert \psi_{\mathrm{RF}}\left(  t\right)  \right\rangle $ and
$O\left(  t\right)  =U_{\mathrm{RF}}\left(  t\right)  O_{\mathrm{RF}}\left(
t\right)  U_{\mathrm{RF}}^{\dagger}\left(  t\right)  $, respectively. The
sub-index \textquotedblleft\textrm{RF}\textquotedblright\ means rotating
frame. Then, the operator $U(t)$ is given by%
\begin{equation}
U(t)=U_{\mathrm{RF}}\left(  t\right)  U_{\mathrm{Rabi}}\left(  t\right)
\text{,} \label{letstalkaboutlove}%
\end{equation}
where $U_{\mathrm{Rabi}}\left(  t\right)  =e^{-i\mathrm{H}_{\mathrm{Rabi}}t}$
and $\mathrm{H}_{\mathrm{Rabi}}\overset{\text{def}}{=}\left[  \left(
\omega_{0}-\omega\right)  /2\right]  \sigma_{z}+\Omega_{0}\sigma_{x}$ is the
time-independent Hamiltonian that describes the evolution of the system in the
rotating frame. Therefore, $U\left(  t\right)  $ can be conveniently rewritten
as%
\begin{equation}
U\left(  t\right)  =e^{-i\frac{\omega}{2}t\sigma_{z}}\left[  \cos\left(
\Omega t\right)  \mathrm{I}-i\sin\left(  \Omega t\right)  \hat{n}\cdot
\vec{\sigma}\right]  \text{,} \label{uni2}%
\end{equation}
where $\Omega\overset{\text{def}}{=}\sqrt{\Omega_{0}^{2}+\left(
\Delta/2\right)  ^{2}}$, $\hat{n}\overset{\text{def}}{=}(1/\Omega)(\Omega_{0}%
$, $0$, $\Delta/2)$, and $\Delta\overset{\text{def}}{=}\omega_{0}-\omega$ is
the detuning. Finally, introducing the adimensional quantities $\tau
\overset{\text{def}}{=}\Omega t$ and $\delta\overset{\text{def}}{=}%
\omega/(2\Omega)$, the unitary evolution operator $U\left(  t\right)  $ in Eq.
(\ref{uni2}) can be conveniently recast as
\begin{equation}
U\left(  \tau\right)  =e^{-i\delta\tau\sigma_{z}}\left[  \cos\left(
\tau\right)  \mathrm{I}-i\sin\left(  \tau\right)  \hat{n}\cdot\vec{\sigma
}\right]  \text{.} \label{uni3}%
\end{equation}
For completeness, we note that $U_{\mathrm{Rabi}}\left(  \tau\right)
=\cos\left(  \tau\right)  \mathrm{I}-i\sin\left(  \tau\right)  \hat{n}%
\cdot\vec{\sigma}$ can be explicitly recast as%
\begin{equation}
U_{\mathrm{Rabi}}\left(  \tau\right)  =\cos\left(  \tau\right)  \mathrm{I}%
-i\sin\left(  \tau\right)  \left[  \frac{\Omega_{0}}{\Omega}\sigma_{x}%
+\frac{\Delta}{2\Omega}\sigma_{z}\right]  \text{.} \label{uni4}%
\end{equation}
We emphasize that, modulo a different choice of notation of the physical
parameters of the problem, $U_{\mathrm{Rabi}}\left(  \tau\right)  $ in Eq.
(\ref{uni4}) coincides with the one used by Wilczek and collaborators in Ref.
\cite{frank20}.

\subsection{The evolution operator}

In what follows, we want to derive the Rabi unitary time evolution operator in
Eq. (\ref{uni4}). We set $\hslash=1$ and begin by observing that the unitary
time evolution in the laboratory frame of reference is specified by the
operator $U\left(  t\right)  $ that corresponds to the Hamiltonian
\textrm{H}$\left(  t\right)  $ in\ Eq. (\ref{time-H}). In the laboratory, the
Schr\"{o}dinger equation is given by $i\partial_{t}\left\vert \psi\left(
t\right)  \right\rangle =$\textrm{H}$\left(  t\right)  \left\vert \psi\left(
t\right)  \right\rangle $. In the rotating reference frame, instead, the
unitary time evolution is specified by the operator $U_{\mathrm{Rabi}}\left(
t\right)  $ that corresponds to the (yet unknown) Hamiltonian \textrm{H}%
$_{\mathrm{Rabi}}\left(  t\right)  $. In the rotating frame of reference, the
Schr\"{o}dinger equation is given by $i\partial_{t}\left\vert \psi
_{\mathrm{RF}}\left(  t\right)  \right\rangle =$\textrm{H}$_{\mathrm{Rabi}%
}\left(  t\right)  \left\vert \psi_{\mathrm{RF}}\left(  t\right)
\right\rangle $. States and observables in these two frames of reference are
related by means of the unitary frame transformation $U_{\mathrm{RF}}\left(
t\right)  \overset{\text{def}}{=}e^{-i\frac{\omega}{2}t\sigma_{z}}$. Indeed,
one has for states and observables the relations $\left\vert \psi\left(
t\right)  \right\rangle =U_{\mathrm{RF}}\left(  t\right)  \left\vert
\psi_{\mathrm{RF}}\left(  t\right)  \right\rangle $ and $O\left(  t\right)
=U_{\mathrm{RF}}\left(  t\right)  O_{\mathrm{RF}}\left(  t\right)
U_{\mathrm{RF}}^{\dagger}\left(  t\right)  $, respectively. To find
\textrm{H}$_{\mathrm{Rabi}}$, we proceed as follows. Note that, $i\partial
_{t}\left\vert \psi_{\mathrm{RF}}\left(  t\right)  \right\rangle =$%
\textrm{H}$_{\mathrm{Rabi}}\left(  t\right)  \left\vert \psi_{\mathrm{RF}%
}\left(  t\right)  \right\rangle $ with%
\begin{align}
i\partial_{t}\left\vert \psi_{\mathrm{RF}}\left(  t\right)  \right\rangle  &
=i\partial_{t}\left[  U_{\mathrm{RF}}^{\dagger}\left(  t\right)  \left\vert
\psi\left(  t\right)  \right\rangle \right] \nonumber\\
&  =iU_{\mathrm{RF}}^{\dagger}\left(  t\right)  \partial_{t}\left\vert
\psi\left(  t\right)  \right\rangle +i\frac{\partial U_{\mathrm{RF}}^{\dagger
}\left(  t\right)  }{\partial t}\left\vert \psi\left(  t\right)  \right\rangle
\nonumber\\
&  =iU_{\mathrm{RF}}^{\dagger}\left(  t\right)  \left[  \frac{1}{i}%
\mathrm{H}\left(  t\right)  \left\vert \psi\left(  t\right)  \right\rangle
\right]  +i\frac{\partial U_{\mathrm{RF}}^{\dagger}\left(  t\right)
}{\partial t}U_{\mathrm{RF}}\left(  t\right)  \left\vert \psi_{\mathrm{RF}%
}\left(  t\right)  \right\rangle \nonumber\\
&  =\left[  U_{\mathrm{RF}}^{\dagger}\left(  t\right)  \mathrm{H}\left(
t\right)  U_{\mathrm{RF}}\left(  t\right)  +i\frac{\partial U_{\mathrm{RF}%
}^{\dagger}\left(  t\right)  }{\partial t}U_{\mathrm{RF}}\left(  t\right)
\right]  \left\vert \psi_{\mathrm{RF}}\left(  t\right)  \right\rangle \text{.}
\label{win1}%
\end{align}
Therefore, from Eq. (\ref{win1}), we get%
\begin{equation}
\mathrm{H}_{\mathrm{Rabi}}\left(  t\right)  =U_{\mathrm{RF}}^{\dagger}\left(
t\right)  \mathrm{H}\left(  t\right)  U_{\mathrm{RF}}\left(  t\right)
+i\frac{\partial U_{\mathrm{RF}}^{\dagger}\left(  t\right)  }{\partial
t}U_{\mathrm{RF}}\left(  t\right)  \text{.} \label{win2}%
\end{equation}
Before evaluating $\mathrm{H}_{\mathrm{Rabi}}$ in\ Eq. (\ref{win2}), we
provide some preliminary useful results that shall help us with this explicit
calculation. Specifically, we remark that $U_{\mathrm{RF}}^{\dagger}\left(
t\right)  \sigma_{\pm}U_{\mathrm{RF}}\left(  t\right)  =e^{\pm i\omega
t}\sigma_{\pm}$ with $\sigma_{\pm}\overset{\text{def}}{=}\sigma_{x}\pm
i\sigma_{y}$. These relations can be explicitly checked by recalling the usual
properties of the Pauli matrices given by $\left[  \sigma_{i}\text{, }%
\sigma_{j}\right]  =2i\epsilon_{ijk}\sigma_{k}$, $\left\{  \sigma_{i}\text{,
}\sigma_{j}\right\}  =2\delta_{ij}\mathrm{I}$, and $\sigma_{i}\sigma
_{j}=i\epsilon_{ijk}\sigma_{k}+\delta_{ij}\mathrm{I}$. Let us verify the
relation $U_{\mathrm{RF}}^{\dagger}\left(  t\right)  \sigma_{+}U_{\mathrm{RF}%
}\left(  t\right)  =e^{i\omega t}\sigma_{+}$. We notice that,%
\begin{align}
U_{\mathrm{RF}}^{\dagger}\left(  t\right)  \sigma_{+}U_{\mathrm{RF}}\left(
t\right)   &  =U_{\mathrm{RF}}^{\dagger}\left(  t\right)  \sigma
_{x}U_{\mathrm{RF}}\left(  t\right)  +iU_{\mathrm{RF}}^{\dagger}\left(
t\right)  \sigma_{y}U_{\mathrm{RF}}\left(  t\right) \nonumber\\
&  =e^{i\omega t}(\sigma_{x}+i\sigma_{y})\nonumber\\
&  =e^{i\omega t}\sigma_{x}+ie^{i\omega t}\sigma_{y}\nonumber\\
&  =\left[  \cos\left(  \omega t\right)  +i\sin(\omega t)\right]  \sigma
_{x}+i\left[  \cos\left(  \omega t\right)  +i\sin\left(  \omega t\right)
\right]  \sigma_{y}\nonumber\\
&  =\left[  \cos\left(  \omega t\right)  \sigma_{x}-\sin\left(  \omega
t\right)  \sigma_{y}\right]  +i\left[  \sin\left(  \omega t\right)  \sigma
_{x}+\cos\left(  \omega t\right)  \sigma_{y}\right]  \text{.} \label{going1}%
\end{align}
Therefore, given Eq. (\ref{going1}), we need to verify that $U_{\mathrm{RF}%
}^{\dagger}\left(  t\right)  \sigma_{x}U_{\mathrm{RF}}\left(  t\right)
=\cos\left(  \omega t\right)  \sigma_{x}-\sin\left(  \omega t\right)
\sigma_{y}$ and $U_{\mathrm{RF}}^{\dagger}\left(  t\right)  \sigma
_{y}U_{\mathrm{RF}}\left(  t\right)  =\sin\left(  \omega t\right)  \sigma
_{x}+\cos\left(  \omega t\right)  \sigma_{y}$. We begin by noting that,%
\begin{align}
U_{\mathrm{RF}}^{\dagger}\left(  t\right)  \sigma_{x}U_{\mathrm{RF}}\left(
t\right)   &  =e^{i\frac{\omega}{2}t\sigma_{z}}\sigma_{x}e^{-i\frac{\omega}%
{2}t\sigma_{z}}\nonumber\\
&  =\left[  \cos\left(  \frac{\omega t}{2}\right)  \mathrm{I}+i\sin\left(
\frac{\omega t}{2}\right)  \sigma_{z}\right]  \sigma_{x}\left[  \cos\left(
\frac{\omega t}{2}\right)  \mathrm{I}-i\sin\left(  \frac{\omega t}{2}\right)
\sigma_{z}\right] \nonumber\\
&  =\left[  \cos\left(  \frac{\omega t}{2}\right)  \sigma_{x}+i\sin\left(
\frac{\omega t}{2}\right)  \sigma_{z}\sigma_{x}\right]  \left[  \cos\left(
\frac{\omega t}{2}\right)  \mathrm{I}-i\sin\left(  \frac{\omega t}{2}\right)
\sigma_{z}\right] \nonumber\\
&  =\cos^{2}\left(  \frac{\omega t}{2}\right)  \sigma_{x}-i\sin\left(
\frac{\omega t}{2}\right)  \cos\left(  \frac{\omega t}{2}\right)  \sigma
_{x}\sigma_{z}+i\sin\left(  \frac{\omega t}{2}\right)  \cos\left(
\frac{\omega t}{2}\right)  \sigma_{z}\sigma_{x}+\nonumber\\
&  +\sin^{2}\left(  \frac{\omega t}{2}\right)  \sigma_{z}\sigma_{x}\sigma
_{z}\text{,}%
\end{align}
that is,%
\begin{equation}
U_{\mathrm{RF}}^{\dagger}\left(  t\right)  \sigma_{x}U_{\mathrm{RF}}\left(
t\right)  =\cos\left(  \omega t\right)  \sigma_{x}-\sin\left(  \omega
t\right)  \sigma_{y}\text{,}%
\end{equation}
since $\sigma_{x}\sigma_{z}=-i\sigma_{y}=-\sigma_{z}\sigma_{x}$ and
$\sigma_{z}\sigma_{x}\sigma_{z}=-\sigma_{x}$. Similarly, we can explicitly
check that the relation $U_{\mathrm{RF}}^{\dagger}\left(  t\right)  \sigma
_{y}U_{\mathrm{RF}}\left(  t\right)  =\sin\left(  \omega t\right)  \sigma
_{x}+\cos\left(  \omega t\right)  \sigma_{y}$ holds true as well. We can now
return to the calculation of $\mathrm{H}_{\mathrm{Rabi}}$ in\ Eq.
(\ref{win2}). Note that,%
\begin{equation}
i\frac{\partial U_{\mathrm{RF}}^{\dagger}\left(  t\right)  }{\partial
t}U_{\mathrm{RF}}\left(  t\right)  =i(i\frac{\omega}{2}\sigma_{z}%
)U_{\mathrm{RF}}^{\dagger}\left(  t\right)  U_{\mathrm{RF}}\left(  t\right)
=-\frac{\omega}{2}\sigma_{z}\text{.} \label{adel1}%
\end{equation}
Moreover, to calculate the term $U_{\mathrm{RF}}^{\dagger}\left(  t\right)
\mathrm{H}\left(  t\right)  U_{\mathrm{RF}}\left(  t\right)  $ by exploiting
the relations $U_{\mathrm{RF}}^{\dagger}\left(  t\right)  \sigma_{\pm
}U_{\mathrm{RF}}\left(  t\right)  =e^{\pm i\omega t}\sigma_{\pm}$, we need to
first recast the Hamiltonian operator $\mathrm{H}\left(  t\right)  $ in terms
of the operators $\sigma_{\pm}\overset{\text{def}}{=}\sigma_{x}\pm i\sigma
_{y}$. We have,%
\begin{equation}
\mathrm{H}\left(  t\right)  =\frac{\omega_{0}}{2}\sigma_{z}+\Omega_{0}\left[
\cos\left(  \omega t\right)  \frac{\sigma_{+}+\sigma_{-}}{2}+\sin\left(
\omega t\right)  \frac{\sigma_{+}-\sigma_{-}}{2i}\right]  \text{.}%
\end{equation}
Therefore, we finally see that%
\begin{align}
U_{\mathrm{RF}}^{\dagger}\left(  t\right)  \mathrm{H}\left(  t\right)
U_{\mathrm{RF}}\left(  t\right)   &  =U_{\mathrm{RF}}^{\dagger}\left(
t\right)  \left\{  \frac{\omega_{0}}{2}\sigma_{z}+\Omega_{0}\left[
\cos\left(  \omega t\right)  \frac{\sigma_{+}+\sigma_{-}}{2}+\sin\left(
\omega t\right)  \frac{\sigma_{+}-\sigma_{-}}{2i}\right]  \right\}
U_{\mathrm{RF}}\left(  t\right) \nonumber\\
&  =U_{\mathrm{RF}}^{\dagger}\left(  t\right)  \left[  \frac{\omega_{0}}%
{2}\sigma_{z}\right]  U_{\mathrm{RF}}\left(  t\right)  +\Omega_{0}\cos\left(
\omega t\right)  U_{\mathrm{RF}}^{\dagger}\left(  t\right)  \left[
\frac{\sigma_{+}+\sigma_{-}}{2}\right]  U_{\mathrm{RF}}\left(  t\right)
+\nonumber\\
&  +\Omega_{0}\sin\left(  \omega t\right)  U_{\mathrm{RF}}^{\dagger}\left(
t\right)  \left[  \frac{\sigma_{+}-\sigma_{-}}{2i}\right]  U_{\mathrm{RF}%
}\left(  t\right) \nonumber\\
&  =\frac{\omega_{0}}{2}\sigma_{z}+\Omega_{0}\cos\left(  \omega t\right)
\left[  \frac{U_{\mathrm{RF}}^{\dagger}\left(  t\right)  \sigma_{+}%
U_{\mathrm{RF}}\left(  t\right)  }{2}+\frac{U_{\mathrm{RF}}^{\dagger}\left(
t\right)  \sigma_{-}U_{\mathrm{RF}}\left(  t\right)  }{2}\right]  +\nonumber\\
&  +\Omega_{0}\sin\left(  \omega t\right)  \left[  \frac{U_{\mathrm{RF}%
}^{\dagger}\left(  t\right)  \sigma_{+}U_{\mathrm{RF}}\left(  t\right)  }%
{2i}-\frac{U_{\mathrm{RF}}^{\dagger}\left(  t\right)  \sigma_{-}%
U_{\mathrm{RF}}\left(  t\right)  }{2i}\right] \nonumber\\
&  =\frac{\omega_{0}}{2}\sigma_{z}+\Omega_{0}\cos\left(  \omega t\right)
\left[  \frac{e^{i\omega t}}{2}\sigma_{+}+\frac{e^{-i\omega t}}{2}\sigma
_{-}\right]  +\nonumber\\
&  +\Omega_{0}\sin\left(  \omega t\right)  \left[  \frac{e^{i\omega t}}%
{2i}\sigma_{+}-\frac{e^{-i\omega t}}{2i}\sigma_{-}\right] \nonumber\\
&  =\frac{\omega_{0}}{2}\sigma_{z}+\Omega_{0}\cos\left(  \omega t\right)
\left[  \cos\left(  \omega t\right)  \sigma_{x}-\sin\left(  \omega t\right)
\sigma_{y}\right]  +\nonumber\\
&  +\Omega_{0}\sin\left(  \omega t\right)  \left[  \sin\left(  \omega
t\right)  \sigma_{x}+\cos\left(  \omega t\right)  \sigma_{y}\right]
\nonumber\\
&  =\frac{\omega_{0}}{2}\sigma_{z}+\Omega_{0}\sigma_{x}\text{,}%
\end{align}
that is,%
\begin{equation}
U_{\mathrm{RF}}^{\dagger}\left(  t\right)  \mathrm{H}\left(  t\right)
U_{\mathrm{RF}}\left(  t\right)  =\frac{\omega_{0}}{2}\sigma_{z}+\Omega
_{0}\sigma_{x}\text{.} \label{adel2}%
\end{equation}
Finally, combining Eqs. (\ref{adel1}) and (\ref{adel2}), we have%
\begin{align}
\mathrm{H}_{\mathrm{Rabi}}  &  =U_{\mathrm{RF}}^{\dagger}\left(  t\right)
\mathrm{H}\left(  t\right)  U_{\mathrm{RF}}\left(  t\right)  +i\frac{\partial
U_{\mathrm{RF}}^{\dagger}\left(  t\right)  }{\partial t}U_{\mathrm{RF}}\left(
t\right) \nonumber\\
&  =\frac{\omega_{0}}{2}\sigma_{z}+\Omega_{0}\sigma_{x}-\frac{\omega}{2}%
\sigma_{z}\nonumber\\
&  =\Omega_{0}\sigma_{x}+\frac{\omega_{0}-\omega}{2}\sigma_{z}\nonumber\\
&  =\Omega_{0}\sigma_{x}+\frac{\Delta}{2}\sigma_{z}\text{,}%
\end{align}
that is,%
\begin{equation}
\mathrm{H}_{\mathrm{Rabi}}=\Omega_{0}\sigma_{x}+\frac{\Delta}{2}\sigma
_{z}\text{,} \label{RabiH}%
\end{equation}
with $\Delta\overset{\text{def}}{=}\omega_{0}-\omega$ being the detuning.
Observe that $\mathrm{H}_{\mathrm{Rabi}}$ in Eq. (\ref{RabiH}) is the
time-independent Hamiltonian and specifies the time evolution of the system in
the rotating frame. We are now in a position of being able to calculate
$U\left(  t\right)  $. We have,%
\begin{align}
U\left(  t\right)   &  =U_{\mathrm{RF}}\left(  t\right)  U_{\mathrm{Rabi}%
}\left(  t\right) \nonumber\\
&  =e^{-i\frac{\omega}{2}t\sigma_{z}}e^{-i\mathrm{H}_{\mathrm{Rabi}}%
t}\nonumber\\
&  =e^{-i\frac{\omega}{2}t\sigma_{z}}e^{-i\left[  \Omega_{0}\sigma_{x}%
+\frac{\Delta}{2}\sigma_{z}\right]  t}\nonumber\\
&  =e^{-i\frac{\omega}{2}t\sigma_{z}}e^{-i\Omega\hat{n}\cdot\vec{\sigma}%
t}\nonumber\\
&  =e^{-i\frac{\omega}{2}t\sigma_{z}}\left[  \cos\left(  \Omega t\right)
\mathrm{I}-i\sin\left(  \Omega t\right)  \hat{n}\cdot\vec{\sigma}\right]
\nonumber\\
&  =e^{-i\delta\tau\sigma_{z}}\left[  \cos\left(  \tau\right)  \mathrm{I}%
-i\sin\left(  \tau\right)  \hat{n}\cdot\vec{\sigma}\right]  \text{,}%
\end{align}
that is, we obtain Eq. (\ref{uni3}) with%
\begin{equation}
U\left(  \tau\right)  =e^{-i\delta\tau\sigma_{z}}\left[  \cos\left(
\tau\right)  \mathrm{I}-i\sin\left(  \tau\right)  \hat{n}\cdot\vec{\sigma
}\right]  \text{,} \label{makeup}%
\end{equation}
where $\tau\overset{\text{def}}{=}\Omega t$, $\delta\overset{\text{def}%
}{=}\omega/(2\Omega)$, $\hat{n}\overset{\text{def}}{=}(1/\Omega)\left(
\Omega_{0}\text{, }0\text{, }\Delta/2\right)  $, and $\Omega
\overset{\text{def}}{=}\sqrt{\Omega_{0}^{2}+(\Delta/2)^{2}}$. For
completeness, we reiterate that the evolution operator $U_{\mathrm{Rabi}%
}\left(  \tau\right)  $ in the rotating reference frame can be explicitly
written as%
\begin{equation}
U_{\mathrm{Rabi}}\left(  \tau\right)  =\cos\left(  \tau\right)  \mathrm{I}%
-i\sin\left(  \tau\right)  \left[  \frac{\Omega_{0}}{\Omega}\sigma_{x}%
+\frac{\Delta}{2\Omega}\sigma_{z}\right]  \text{.} \label{fine}%
\end{equation}
As previously mentioned, despite a different choice of notation of the
physical parameters of the problem, $U_{\mathrm{Rabi}}\left(  \tau\right)  $
in Eq. (\ref{fine}) is the same as the one employed in Ref. \cite{frank20}.

\subsection{The Bloch vector}

Given the Hamiltonian in Eq. (\ref{time-H}) and the unitary time evolution
operators in Eqs. (\ref{makeup}) and (\ref{fine}), we wish to find in this
subsection an analytical expression for the evolution of the Bloch vector
$\vec{a}\left(  t\right)  $ with $\rho\left(  t\right)  \overset{\text{def}%
}{=}\left\vert \psi\left(  t\right)  \right\rangle \left\langle \psi\left(
t\right)  \right\vert =\left[  \mathrm{I}+\vec{a}\left(  t\right)  \cdot
\vec{\sigma}\right]  /2$ that emerges from the time-dependent Hamiltonian in
Eq. (\ref{time-H}), \textrm{H}$\left(  t\right)  \overset{\text{def}}{=}%
\vec{m}\left(  t\right)  \cdot\vec{\sigma}$ with the magnetic field vector
given by (setting $\hslash=1$)%
\begin{equation}
\vec{m}\left(  t\right)  \overset{\text{def}}{=}\Omega_{0}\cos\left(  \omega
t\right)  \hat{x}+\Omega_{0}\sin\left(  \omega t\right)  \hat{y}+\frac
{\omega_{0}}{2}\hat{z}\text{,} \label{field23}%
\end{equation}
or, alternatively, $\vec{m}\left(  t\right)  =\Omega_{\mathrm{H}}\hat
{m}\left(  t\right)  $ with $\Omega_{\mathrm{H}}\overset{\text{def}}{=}%
\sqrt{\Omega_{0}^{2}+(\omega_{0}/2)^{2}}$ being the generalized Rabi frequency
and $\hat{m}\left(  t\right)  \overset{\text{def}}{=}(1/\Omega_{\mathrm{H}%
})(\Omega_{0}\cos\left(  \omega t\right)  $, $\Omega_{0}\sin\left(  \omega
t\right)  $, $\omega_{0}/2)$. From Eq. (\ref{field23}), we note that the
magnitude of $\vec{m}\left(  t\right)  $ is constant in time since $\left\Vert
\vec{m}\left(  t\right)  \right\Vert ^{2}=\hslash^{2}\left[  \Omega_{0}%
^{2}+\left(  \omega_{0}/2\right)  ^{2}\right]  $. However, its corresponding
unit vector $\hat{m}\left(  t\right)  =\vec{m}\left(  t\right)  /\left\Vert
\vec{m}\left(  t\right)  \right\Vert $ is time-dependent. Interestingly, this
means that $\vec{m}\left(  t\right)  $ and $\partial_{t}\left[  \vec{m}\left(
t\right)  \right]  $ are not collinear. Indeed, in our case, we can simply
verify that $\vec{m}\left(  t\right)  $ and $\partial_{t}\left[  \vec
{m}\left(  t\right)  \right]  $ are orthogonal since $\vec{m}\left(  t\right)
\cdot\partial_{t}\left[  \vec{m}\left(  t\right)  \right]  =0$. To find
$\vec{a}\left(  t\right)  $, we proceed as follows. Observe that the pure
state $\rho\left(  t\right)  \overset{\text{def}}{=}\left\vert \psi\left(
t\right)  \right\rangle \left\langle \psi\left(  t\right)  \right\vert
=\left[  \mathrm{I}+\vec{a}\left(  t\right)  \cdot\vec{\sigma}\right]  /2$ can
be rewritten as
\begin{align}
\rho\left(  t\right)   &  =U\left(  t\right)  \rho\left(  0\right)
U^{\dagger}\left(  t\right) \nonumber\\
&  =e^{-i\frac{\alpha}{2}\hat{n}\cdot\vec{\sigma}}\rho\left(  0\right)
e^{i\frac{\alpha}{2}\hat{n}\cdot\vec{\sigma}}\nonumber\\
&  =e^{-i\frac{\alpha}{2}\hat{n}\cdot\vec{\sigma}}\left[  \frac{\mathrm{I}%
+\vec{a}\left(  0\right)  \cdot\vec{\sigma}}{2}\right]  e^{i\frac{\alpha}%
{2}\hat{n}\cdot\vec{\sigma}}\nonumber\\
&  =U_{1}\left(  t\right)  \left[  U_{2}\left(  t\right)  \rho\left(
0\right)  U_{2}^{\dagger}\left(  t\right)  \right]  U_{1}^{\dagger}\left(
t\right) \nonumber\\
&  =e^{-i\frac{\alpha_{1}}{2}\hat{n}_{1}\cdot\vec{\sigma}}\left\{
e^{-i\frac{\alpha_{2}}{2}\hat{n}_{2}\cdot\vec{\sigma}}\left[  \frac
{\mathrm{I}+\vec{a}\left(  0\right)  \cdot\vec{\sigma}}{2}\right]
e^{i\frac{\alpha_{2}}{2}\hat{n}_{2}\cdot\vec{\sigma}}\right\}  e^{i\frac
{\alpha_{1}}{2}\hat{n}_{1}\cdot\vec{\sigma}}\text{.} \label{ricci}%
\end{align}
Notice that $U\left(  t\right)  =U_{1}\left(  t\right)  U_{2}\left(  t\right)
=U_{\mathrm{RF}}\left(  t\right)  U_{\mathrm{Rabi}}\left(  t\right)  $, in our
case. In what follows, we first connect the pair $\left(  \alpha\text{, }%
\hat{n}\right)  $ to the pairs $\left(  \alpha_{1}\text{, }\hat{n}_{1}\right)
$ and $\left(  \alpha_{2}\text{, }\hat{n}_{2}\right)  $ as evident from Eq.
(\ref{ricci}). Then, we calculate $e^{-i\frac{\alpha}{2}\hat{n}\cdot
\vec{\sigma}}\left(  \left[  \mathrm{I}+\vec{a}\left(  0\right)  \cdot
\vec{\sigma}\right]  /2\right)  e^{i\frac{\alpha}{2}\hat{n}\cdot\vec{\sigma}}%
$. We want to express the sequential application of two rotations on a single
qubit in terms of a single rotation. Formally, given that%
\begin{equation}
e^{-i\frac{\alpha_{1}}{2}\hat{n}_{1}\cdot\vec{\sigma}}e^{-i\frac{\alpha_{2}%
}{2}\hat{n}_{2}\cdot\vec{\sigma}}=e^{-i\frac{\alpha}{2}\hat{n}\cdot\vec
{\sigma}}\text{,}%
\end{equation}
we need to find the angle $\alpha$ and the axis of rotation $\hat{n}$. We
proceed as follows. Note that $e^{-i\frac{\alpha}{2}\hat{n}\cdot\vec{\sigma}}$
can be written as,%
\begin{align}
e^{-i\frac{\alpha}{2}\hat{n}\cdot\vec{\sigma}}  &  =e^{-i\frac{\alpha_{1}}%
{2}\hat{n}_{1}\cdot\vec{\sigma}}e^{-i\frac{\alpha_{2}}{2}\hat{n}_{2}\cdot
\vec{\sigma}}\nonumber\\
&  =\left[  \cos\left(  \frac{\alpha_{1}}{2}\right)  \mathrm{I}-i\sin\left(
\frac{\alpha_{1}}{2}\right)  \hat{n}_{1}\cdot\vec{\sigma}\right]  \left[
\cos\left(  \frac{\alpha_{2}}{2}\right)  \mathrm{I}-i\sin\left(  \frac
{\alpha_{2}}{2}\right)  \hat{n}_{2}\cdot\vec{\sigma}\right] \nonumber\\
&  =\cos\left(  \frac{\alpha_{1}}{2}\right)  \cos\left(  \frac{\alpha_{2}}%
{2}\right)  \mathrm{I}-i\cos\left(  \frac{\alpha_{1}}{2}\right)  \sin\left(
\frac{\alpha_{2}}{2}\right)  \hat{n}_{2}\cdot\vec{\sigma}-i\sin\left(
\frac{\alpha_{1}}{2}\right)  \cos\left(  \frac{\alpha_{2}}{2}\right)  \hat
{n}_{1}\cdot\vec{\sigma}+\nonumber\\
&  -\sin\left(  \frac{\alpha_{1}}{2}\right)  \sin\left(  \frac{\alpha_{2}}%
{2}\right)  \left(  \hat{n}_{1}\cdot\vec{\sigma}\right)  \left(  \hat{n}%
_{2}\cdot\vec{\sigma}\right)  \text{.} \label{1}%
\end{align}
From quantum mechanics, recall that $\left(  \hat{n}_{1}\cdot\vec{\sigma
}\right)  \left(  \hat{n}_{2}\cdot\vec{\sigma}\right)  =\left(  \hat{n}%
_{1}\cdot\hat{n}_{2}\right)  \mathrm{I}+i\left(  \hat{n}_{1}\times\hat{n}%
_{2}\right)  \cdot\vec{\sigma}$. Indeed, note that%
\begin{align}
\left(  \hat{n}_{1}\cdot\vec{\sigma}\right)  \left(  \hat{n}_{2}\cdot
\vec{\sigma}\right)   &  =\left(  n_{1i}\sigma_{i}\right)  \left(
n_{2j}\sigma_{j}\right) \nonumber\\
&  =n_{1i}n_{2j}\sigma_{i}\sigma_{j}\nonumber\\
&  =n_{1i}n_{2j}\left(  \frac{1}{2}\left[  \sigma_{i}\text{, }\sigma
_{j}\right]  +\frac{1}{2}\left\{  \sigma_{i}\text{, }\sigma_{j}\right\}
\right) \nonumber\\
&  =n_{1i}n_{2j}\left(  \frac{1}{2}2i\epsilon_{ijk}\sigma_{k}+\frac{1}%
{2}2\delta_{ij}\mathrm{I}\right) \nonumber\\
&  =n_{1i}n_{2j}\delta_{ij}\mathrm{I}+i\epsilon_{ijk}n_{1i}n_{2j}\sigma
_{k}\nonumber\\
&  =\left(  \hat{n}_{1}\cdot\hat{n}_{2}\right)  \mathrm{I}+i\left(  \hat
{n}_{1}\times\hat{n}_{2}\right)  \cdot\vec{\sigma}\text{,} \label{2}%
\end{align}
Substituting Eq. (\ref{2}) into Eq. (\ref{1}), we get%
\begin{align}
e^{-i\frac{\alpha}{2}\hat{n}\cdot\vec{\sigma}}  &  =\cos\left(  \frac
{\alpha_{1}}{2}\right)  \cos\left(  \frac{\alpha_{2}}{2}\right)
\mathrm{I}-i\cos\left(  \frac{\alpha_{1}}{2}\right)  \sin\left(  \frac
{\alpha_{2}}{2}\right)  \hat{n}_{2}\cdot\vec{\sigma}-i\sin\left(  \frac
{\alpha_{1}}{2}\right)  \cos\left(  \frac{\alpha_{2}}{2}\right)  \hat{n}%
_{1}\cdot\vec{\sigma}+\nonumber\\
&  -\sin\left(  \frac{\alpha_{1}}{2}\right)  \sin\left(  \frac{\alpha_{2}}%
{2}\right)  \left(  \hat{n}_{1}\cdot\hat{n}_{2}\right)  \mathrm{I}%
-i\sin\left(  \frac{\alpha_{1}}{2}\right)  \sin\left(  \frac{\alpha_{2}}%
{2}\right)  \left(  \hat{n}_{1}\times\hat{n}_{2}\right)  \cdot\vec{\sigma
}\nonumber\\
& \nonumber\\
&  =\left[  \cos\left(  \frac{\alpha_{1}}{2}\right)  \cos\left(  \frac
{\alpha_{2}}{2}\right)  -\sin\left(  \frac{\alpha_{1}}{2}\right)  \sin\left(
\frac{\alpha_{2}}{2}\right)  \left(  \hat{n}_{1}\cdot\hat{n}_{2}\right)
\right]  \mathrm{I}+\nonumber\\
&  -i\left[  \sin\left(  \frac{\alpha_{1}}{2}\right)  \cos\left(  \frac
{\alpha_{2}}{2}\right)  \hat{n}_{1}+\cos\left(  \frac{\alpha_{1}}{2}\right)
\sin\left(  \frac{\alpha_{2}}{2}\right)  \hat{n}_{2}-\sin\left(  \frac
{\alpha_{1}}{2}\right)  \sin\left(  \frac{\alpha_{2}}{2}\right)  \left(
\hat{n}_{2}\times\hat{n}_{1}\right)  \right]  \cdot\vec{\sigma}\nonumber\\
& \nonumber\\
&  \equiv\left[  \cos\left(  \frac{\alpha}{2}\right)  \mathrm{I}-i\sin\left(
\frac{\alpha}{2}\right)  \hat{n}\cdot\vec{\sigma}\right]  \text{,}%
\end{align}
that is,%
\begin{equation}
\left\{
\begin{array}
[c]{c}%
\cos\left(  \frac{\alpha}{2}\right)  =\cos\left(  \frac{\alpha_{1}}{2}\right)
\cos\left(  \frac{\alpha_{2}}{2}\right)  -\sin\left(  \frac{\alpha_{1}}%
{2}\right)  \sin\left(  \frac{\alpha_{2}}{2}\right)  \left(  \hat{n}_{1}%
\cdot\hat{n}_{2}\right)  \overset{\text{def}}{=}c_{12}\text{,}\\
\\
\sin\left(  \frac{\alpha}{2}\right)  \hat{n}=\sin\left(  \frac{\alpha_{1}}%
{2}\right)  \cos\left(  \frac{\alpha_{2}}{2}\right)  \hat{n}_{1}+\cos\left(
\frac{\alpha_{1}}{2}\right)  \sin\left(  \frac{\alpha_{2}}{2}\right)  \hat
{n}_{2}-\sin\left(  \frac{\alpha_{1}}{2}\right)  \sin\left(  \frac{\alpha_{2}%
}{2}\right)  \left(  \hat{n}_{2}\times\hat{n}_{1}\right)  \overset{\text{def}%
}{=}\vec{v}_{12}%
\end{array}
\right.  \text{.} \label{linki}%
\end{equation}
From Eq. (\ref{linki}), we get%
\begin{equation}
\hat{n}=\frac{\vec{v}_{12}}{\left\Vert \vec{v}_{12}\right\Vert }\text{, and
}\tan^{2}\left(  \frac{\alpha}{2}\right)  =\frac{\vec{v}_{12}\cdot\vec{v}%
_{12}}{c_{12}^{2}}\text{.} \label{linki1}%
\end{equation}
Eq. (\ref{linki1}) expresses the link between the pair the pair $\left(
\alpha\text{, }\hat{n}\right)  $ to the pairs $\left(  \alpha_{1}\text{, }%
\hat{n}_{1}\right)  \overset{\text{def}}{=}\left(  \omega t\text{, }\hat
{z}\right)  $ and $\left(  \alpha_{2}\text{, }\hat{n}_{2}\right)
\overset{\text{def}}{=}\left(  2\Omega t\text{, }\frac{\Omega_{0}\hat{x}%
+\frac{\Delta}{2}\hat{z}}{\Omega}\right)  $. We return now to consider the
calculation of $e^{-i\frac{\alpha}{2}\hat{n}\cdot\vec{\sigma}}\left(  \left[
\mathrm{I}+\vec{a}\left(  0\right)  \cdot\vec{\sigma}\right]  /2\right)
e^{i\frac{\alpha}{2}\hat{n}\cdot\vec{\sigma}}$. For ease of notation, we set
$\vec{a}\left(  0\right)  =\vec{a}_{0}$. We have,%
\begin{align}
\rho\left(  t\right)   &  =U\left(  t\right)  \rho\left(  0\right)
U^{\dagger}\left(  t\right) \nonumber\\
&  =e^{-i\frac{\alpha}{2}\hat{n}\cdot\vec{\sigma}}\rho\left(  0\right)
e^{i\frac{\alpha}{2}\hat{n}\cdot\vec{\sigma}}\nonumber\\
&  =e^{-i\frac{\alpha}{2}\hat{n}\cdot\vec{\sigma}}\left(  \left[
\mathrm{I}+\vec{a}_{0}\cdot\vec{\sigma}\right]  /2\right)  e^{i\frac{\alpha
}{2}\hat{n}\cdot\vec{\sigma}}\nonumber\\
&  =\left[  \cos\left(  \frac{\alpha}{2}\right)  \mathrm{I}-i\sin\left(
\frac{\alpha}{2}\right)  \left(  \hat{n}\cdot\vec{\sigma}\right)  \right]
\left[  \frac{\mathrm{I}+\vec{a}_{0}\cdot\vec{\sigma}}{2}\right]  \left[
\cos\left(  \frac{\alpha}{2}\right)  \mathrm{I}+i\sin\left(  \frac{\alpha}%
{2}\right)  \left(  \hat{n}\cdot\vec{\sigma}\right)  \right] \nonumber\\
&  =\frac{1}{2}\left[  \cos\left(  \frac{\alpha}{2}\right)  \mathrm{I}%
+\cos\left(  \frac{\alpha}{2}\right)  \left(  \vec{a}_{0}\cdot\vec{\sigma
}\right)  -i\sin\left(  \frac{\alpha}{2}\right)  \left(  \hat{n}\cdot
\vec{\sigma}\right)  -i\sin\left(  \frac{\alpha}{2}\right)  \left(  \hat
{n}\cdot\vec{\sigma}\right)  \left(  \vec{a}_{0}\cdot\vec{\sigma}\right)
\right]  \cdot\nonumber\\
&  \cdot\left[  \cos\left(  \frac{\alpha}{2}\right)  \mathrm{I}+i\sin\left(
\frac{\alpha}{2}\right)  \left(  \hat{n}\cdot\vec{\sigma}\right)  \right]
\text{,}%
\end{align}
that is,%
\begin{equation}
\rho\left(  t\right)  =\frac{1}{2}\left[
\begin{array}
[c]{c}%
\cos^{2}\left(  \frac{\alpha}{2}\right)  \mathrm{I}+i\sin\left(  \frac{\alpha
}{2}\right)  \cos\left(  \frac{\alpha}{2}\right)  \left(  \hat{n}\cdot
\vec{\sigma}\right)  +\cos^{2}\left(  \frac{\alpha}{2}\right)  \left(  \vec
{a}_{0}\cdot\vec{\sigma}\right)  +i\sin\left(  \frac{\alpha}{2}\right)
\cos\left(  \frac{\alpha}{2}\right)  \left(  \vec{a}_{0}\cdot\vec{\sigma
}\right)  \left(  \hat{n}\cdot\vec{\sigma}\right)  +\\
\\
-\sin\left(  \frac{\alpha}{2}\right)  \cos\left(  \frac{\alpha}{2}\right)
\left(  \hat{n}\cdot\vec{\sigma}\right)  +\sin^{2}\left(  \frac{\alpha}%
{2}\right)  \left(  \hat{n}\cdot\vec{\sigma}\right)  \left(  \hat{n}\cdot
\vec{\sigma}\right)  -i\sin\left(  \frac{\alpha}{2}\right)  \cos\left(
\frac{\alpha}{2}\right)  \left(  \hat{n}\cdot\vec{\sigma}\right)  \left(
\vec{a}_{0}\cdot\vec{\sigma}\right)  +\\
\\
+\sin^{2}\left(  \frac{\alpha}{2}\right)  \left(  \hat{n}\cdot\vec{\sigma
}\right)  \left(  \vec{a}_{0}\cdot\vec{\sigma}\right)  \left(  \hat{n}%
\cdot\vec{\sigma}\right)  \text{.}%
\end{array}
\right]  \label{pu}%
\end{equation}
To further simplify the expression of $\rho\left(  t\right)  $ in Eq.
(\ref{pu}), we rewrite $\rho\left(  t\right)  $ as%
\begin{equation}
\rho\left(  t\right)  =\frac{1}{2}\left[
\begin{array}
[c]{c}%
\cos^{2}\left(  \frac{\alpha}{2}\right)  \mathrm{I}+i\sin\left(  \frac{\alpha
}{2}\right)  \cos\left(  \frac{\alpha}{2}\right)  \left(  \hat{n}\cdot
\vec{\sigma}\right)  +\cos^{2}\left(  \frac{\alpha}{2}\right)  \left(  \vec
{a}_{0}\cdot\vec{\sigma}\right)  +\\
\\
-\sin\left(  \frac{\alpha}{2}\right)  \cos\left(  \frac{\alpha}{2}\right)
\left(  \hat{n}\cdot\vec{\sigma}\right)  +\sin^{2}\left(  \frac{\alpha}%
{2}\right)  \left(  \hat{n}\cdot\vec{\sigma}\right)  \left(  \hat{n}\cdot
\vec{\sigma}\right)  +Q_{1}+\sin^{2}\left(  \frac{\alpha}{2}\right)
Q_{2}\text{.}%
\end{array}
\right]  \text{,} \label{pupu}%
\end{equation}
with the quantities $Q_{1}$ and $Q_{2}$ defined as,%
\begin{equation}
Q_{1}\overset{\text{def}}{=}i\sin\left(  \frac{\alpha}{2}\right)  \cos\left(
\frac{\alpha}{2}\right)  \left(  \vec{a}_{0}\cdot\vec{\sigma}\right)  \left(
\hat{n}\cdot\vec{\sigma}\right)  -i\sin\left(  \frac{\alpha}{2}\right)
\cos\left(  \frac{\alpha}{2}\right)  \left(  \hat{n}\cdot\vec{\sigma}\right)
\left(  \vec{a}_{0}\cdot\vec{\sigma}\right)  \text{,}%
\end{equation}
and%
\begin{equation}
Q_{2}\overset{\text{def}}{=}\left(  \hat{n}\cdot\vec{\sigma}\right)  \left(
\vec{a}_{0}\cdot\vec{\sigma}\right)  \left(  \hat{n}\cdot\vec{\sigma}\right)
\text{,}%
\end{equation}
respectively. \ Making use of the identities $\left(  \vec{a}_{0}\cdot
\vec{\sigma}\right)  \left(  \hat{n}\cdot\vec{\sigma}\right)  =\left(  \hat
{n}\cdot\vec{a}_{0}\right)  \mathrm{I}-i\left(  \hat{n}\times\vec{a}%
_{0}\right)  \cdot\vec{\sigma}$, $\left(  \hat{n}\cdot\vec{\sigma}\right)
\left(  \hat{n}\cdot\vec{\sigma}\right)  =\mathrm{I}$, and $\left(  \hat
{n}\cdot\vec{\sigma}\right)  \left(  \vec{a}_{0}\cdot\vec{\sigma}\right)
=\left(  \hat{n}\cdot\vec{a}_{0}\right)  \mathrm{I}+i\left(  \hat{n}\times
\vec{a}_{0}\right)  \cdot\vec{\sigma}$, we have
\begin{align}
Q_{1}  &  =i\sin\left(  \frac{\alpha}{2}\right)  \cos\left(  \frac{\alpha}%
{2}\right)  \left(  \vec{a}_{0}\cdot\vec{\sigma}\right)  \left(  \hat{n}%
\cdot\vec{\sigma}\right)  -i\sin\left(  \frac{\alpha}{2}\right)  \cos\left(
\frac{\alpha}{2}\right)  \left(  \hat{n}\cdot\vec{\sigma}\right)  \left(
\vec{a}_{0}\cdot\vec{\sigma}\right) \nonumber\\
&  =i\sin\left(  \frac{\alpha}{2}\right)  \cos\left(  \frac{\alpha}{2}\right)
\left[  \left(  \hat{n}\cdot\vec{a}_{0}\right)  \mathrm{I}-i\left(  \hat
{n}\times\vec{a}_{0}\right)  \cdot\vec{\sigma}\right]  -i\sin\left(
\frac{\alpha}{2}\right)  \cos\left(  \frac{\alpha}{2}\right)  \left[  \left(
\hat{n}\cdot\vec{a}_{0}\right)  \mathrm{I}+i\left(  \hat{n}\times\vec{a}%
_{0}\right)  \cdot\vec{\sigma}\right] \nonumber\\
&  =2\sin\left(  \frac{\alpha}{2}\right)  \cos\left(  \frac{\alpha}{2}\right)
\left(  \hat{n}\times\vec{a}_{0}\right)  \cdot\vec{\sigma}=\sin\left(
\alpha\right)  \left(  \hat{n}\times\vec{a}_{0}\right)  \cdot\vec{\sigma
}\text{,} \label{q1}%
\end{align}
that is,%
\begin{equation}
Q_{1}=2\sin\left(  \frac{\alpha}{2}\right)  \cos\left(  \frac{\alpha}%
{2}\right)  \left(  \hat{n}\times\vec{a}_{0}\right)  \cdot\vec{\sigma}%
=\sin\left(  \alpha\right)  \left(  \hat{n}\times\vec{a}_{0}\right)  \cdot
\vec{\sigma}\text{.} \label{q1B}%
\end{equation}
Furthermore, $Q_{2}$ reduces to%
\begin{align}
Q_{2}  &  =\left(  \hat{n}\cdot\vec{\sigma}\right)  \left(  \vec{a}_{0}%
\cdot\vec{\sigma}\right)  \left(  \hat{n}\cdot\vec{\sigma}\right) \nonumber\\
&  =\left(  \hat{n}\cdot\vec{\sigma}\right)  \left[  \left(  \vec{a}_{0}%
\cdot\hat{n}\right)  \mathrm{I}+i\left(  \vec{a}_{0}\times\hat{n}\right)
\cdot\vec{\sigma}\right] \nonumber\\
&  =\left(  \vec{a}_{0}\cdot\hat{n}\right)  \left(  \hat{n}\cdot\vec{\sigma
}\right)  +i\left\{  \left(  \hat{n}\cdot\vec{\sigma}\right)  \left[  \left(
\vec{a}_{0}\times\hat{n}\right)  \cdot\vec{\sigma}\right]  \right\}
\nonumber\\
&  =\left(  \vec{a}_{0}\cdot\hat{n}\right)  \left(  \hat{n}\cdot\vec{\sigma
}\right)  +i\left\{  \hat{n}\cdot\left(  \vec{a}_{0}\times\hat{n}\right)
+i\left[  \hat{n}\times\left(  \vec{a}_{0}\times\hat{n}\right)  \right]
\cdot\vec{\sigma}\right\} \nonumber\\
&  =\left(  \vec{a}_{0}\cdot\hat{n}\right)  \left(  \hat{n}\cdot\vec{\sigma
}\right)  -\left[  \hat{n}\times\left(  \vec{a}_{0}\times\hat{n}\right)
\right]  \cdot\vec{\sigma}\text{.} \label{q2}%
\end{align}
However, since $\vec{a}\times(\vec{b}\times\vec{c})=\left(  \vec{a}\cdot
\vec{c}\right)  \vec{b}-(\vec{a}\cdot\vec{b})\vec{c}$ for any three vectors in
$%
%TCIMACRO{\U{211d} }%
%BeginExpansion
\mathbb{R}
%EndExpansion
^{3}$, we have
\begin{equation}
\hat{n}\times\left(  \vec{a}_{0}\times\hat{n}\right)  =\left(  \hat{n}%
\cdot\hat{n}\right)  \vec{a}_{0}-\left(  \hat{n}\cdot\vec{a}_{0}\right)
\hat{n}=\vec{a}_{0}-\left(  \vec{a}_{0}\cdot\hat{n}\right)  \hat{n}\text{.}
\label{q3}%
\end{equation}
Therefore, using Eq. (\ref{q3}), $Q_{2}$ in\ Eq. (\ref{q2}) becomes%
\begin{align}
Q_{2}  &  =\left(  \vec{a}_{0}\cdot\hat{n}\right)  \left(  \hat{n}\cdot
\vec{\sigma}\right)  -\left[  \hat{n}\times\left(  \vec{a}_{0}\times\hat
{n}\right)  \right]  \cdot\vec{\sigma}\nonumber\\
&  =\left(  \vec{a}_{0}\cdot\hat{n}\right)  \left(  \hat{n}\cdot\vec{\sigma
}\right)  -\left[  \vec{a}_{0}-\left(  \vec{a}_{0}\cdot\hat{n}\right)  \hat
{n}\right]  \cdot\vec{\sigma}\nonumber\\
&  =\left(  \vec{a}_{0}\cdot\hat{n}\right)  \left(  \hat{n}\cdot\vec{\sigma
}\right)  -\vec{a}_{0}\cdot\vec{\sigma}+\left(  \vec{a}_{0}\cdot\hat
{n}\right)  \left(  \hat{n}\cdot\vec{\sigma}\right) \nonumber\\
&  =2\left(  \vec{a}_{0}\cdot\hat{n}\right)  \left(  \hat{n}\cdot\vec{\sigma
}\right)  -\vec{a}_{0}\cdot\vec{\sigma}\text{,}%
\end{align}
that is%
\begin{equation}
Q_{2}=2\left(  \vec{a}_{0}\cdot\hat{n}\right)  \left(  \hat{n}\cdot\vec
{\sigma}\right)  -\vec{a}_{0}\cdot\vec{\sigma}\text{.} \label{q2B}%
\end{equation}
Substituting Eqs. (\ref{q1B} ) and (\ref{q2B}) into Eq. (\ref{pupu}),
$\rho\left(  t\right)  $ becomes%
\begin{align}
\rho\left(  t\right)   &  =\frac{1}{2}\left[  \mathrm{I}+\vec{a}\left(
t\right)  \cdot\vec{\sigma}\right] \nonumber\\
&  =\frac{1}{2}\left\{  \mathrm{I}+\cos\left(  \alpha\right)  \left(  \vec
{a}_{0}\cdot\vec{\sigma}\right)  +\sin\left(  \alpha\right)  \left(  \hat
{n}\times\vec{a}_{0}\right)  \cdot\vec{\sigma}+\left[  1-\cos\left(
\alpha\right)  \right]  \left(  \vec{a}_{0}\cdot\hat{n}\right)  \left(
\hat{n}\cdot\vec{\sigma}\right)  \right\}  \text{,}%
\end{align}
that is,%
\begin{align}
\vec{a}\left(  t\right)   &  =\cos\left(  \alpha\right)  \vec{a}_{0}%
+\sin\left(  \alpha\right)  \left(  \hat{n}\times\vec{a}_{0}\right)  +\left[
1-\cos\left(  \alpha\right)  \right]  \left(  \vec{a}_{0}\cdot\hat{n}\right)
\hat{n}\nonumber\\
&  =\left(  \vec{a}_{0}\cdot\hat{n}\right)  \hat{n}+\cos\left(  \alpha\right)
\left[  \vec{a}_{0}-\left(  \vec{a}_{0}\cdot\hat{n}\right)  \hat{n}\right]
+\sin\left(  \alpha\right)  \left(  \hat{n}\times\vec{a}_{0}\right)  \text{.}
\label{www}%
\end{align}
Finally, observing that%
\begin{align}
\left(  \hat{n}\times\vec{a}_{0}\right)  \times\hat{n}  &  =\hat{n}%
\times\left(  \vec{a}_{0}\times\hat{n}\right) \nonumber\\
&  =\left(  \hat{n}\cdot\hat{n}\right)  \vec{a}_{0}-\left(  \hat{n}\cdot
\vec{a}_{0}\right)  \hat{n}\nonumber\\
&  =\vec{a}_{0}-\left(  \hat{n}\cdot\vec{a}_{0}\right)  \hat{n}\text{,}%
\end{align}
the Bloch vector $\vec{a}\left(  t\right)  $ becomes%
\begin{equation}
\vec{a}\left(  t\right)  =\left(  \hat{n}\cdot\vec{a}_{0}\right)  \hat{n}%
+\sin\left(  \alpha\right)  \left(  \hat{n}\times\vec{a}_{0}\right)
+\cos\left(  \alpha\right)  \left(  \hat{n}\times\vec{a}_{0}\right)
\times\hat{n}\text{.} \label{coc}%
\end{equation}
%PMA
Note that in Eq.~(\ref{coc}) the triad of unit vectors $(\hat{n}, \frac
{\hat{n}\times\vec{a}_{0}}{||\hat{n}\times\vec{a}_{0}||}, \frac{(\hat{n}%
\times\vec{a}_{0})\times\hat{n}}{||(\hat{n}\times\vec{a}_{0})\times\hat{n}%
||})$ forms an orthonormal moving frame at each instant of time $t$.

At this point, we need to find the explicit expressions for $\alpha
=\alpha\left(  t\right)  $ and $\hat{n}=\hat{n}\left(  t\right)  $ in Eq.
(\ref{coc}). In our specific working example, we have from Eqs. (\ref{linki})
and (\ref{linki1}) that%
\begin{align}
\hat{n}  &  =\frac{\vec{v}_{12}}{\left\Vert \vec{v}_{12}\right\Vert
}\nonumber\\
&  =\frac{\sin\left(  \frac{\alpha_{1}}{2}\right)  \cos\left(  \frac
{\alpha_{2}}{2}\right)  \hat{n}_{1}+\cos\left(  \frac{\alpha_{1}}{2}\right)
\sin\left(  \frac{\alpha_{2}}{2}\right)  \hat{n}_{2}-\sin\left(  \frac
{\alpha_{1}}{2}\right)  \sin\left(  \frac{\alpha_{2}}{2}\right)  \left(
\hat{n}_{2}\times\hat{n}_{1}\right)  }{\left\Vert \sin\left(  \frac{\alpha
_{1}}{2}\right)  \cos\left(  \frac{\alpha_{2}}{2}\right)  \hat{n}_{1}%
+\cos\left(  \frac{\alpha_{1}}{2}\right)  \sin\left(  \frac{\alpha_{2}}%
{2}\right)  \hat{n}_{2}-\sin\left(  \frac{\alpha_{1}}{2}\right)  \sin\left(
\frac{\alpha_{2}}{2}\right)  \left(  \hat{n}_{2}\times\hat{n}_{1}\right)
\right\Vert }\nonumber\\
&  =\frac{\sin\left(  \frac{\omega t}{2}\right)  \cos\left(  \Omega t\right)
\hat{z}+\cos\left(  \frac{\omega t}{2}\right)  \sin\left(  \Omega t\right)
\left(  \frac{\Omega_{0}\hat{x}+\frac{\Delta}{2}\hat{z}}{\Omega}\right)
-\sin\left(  \frac{\omega t}{2}\right)  \sin\left(  \Omega t\right)  \left(
\frac{\Omega_{0}\hat{x}+\frac{\Delta}{2}\hat{z}}{\Omega}\times\hat{z}\right)
}{\left\Vert \sin\left(  \frac{\omega t}{2}\right)  \cos\left(  \Omega
t\right)  \hat{z}+\cos\left(  \frac{\omega t}{2}\right)  \sin\left(  \Omega
t\right)  \left(  \frac{\Omega_{0}\hat{x}+\frac{\Delta}{2}\hat{z}}{\Omega
}\right)  -\sin\left(  \frac{\omega t}{2}\right)  \sin\left(  \Omega t\right)
\left(  \frac{\Omega_{0}\hat{x}+\frac{\Delta}{2}\hat{z}}{\Omega}\times\hat
{z}\right)  \right\Vert }\nonumber\\
&  =\frac{\left\{  \frac{\Omega_{0}}{\Omega}\cos\left(  \frac{\omega t}%
{2}\right)  \sin\left(  \Omega t\right)  \hat{x}+\frac{\Omega_{0}}{\Omega}%
\sin\left(  \frac{\omega t}{2}\right)  \sin\left(  \Omega t\right)  \hat
{y}+\left[  \sin\left(  \frac{\omega t}{2}\right)  \cos\left(  \Omega
t\right)  +\frac{\Delta}{2\Omega}\cos\left(  \frac{\omega t}{2}\right)
\sin\left(  \Omega t\right)  \right]  \hat{z}\right\}  }{\left\Vert \left\{
\frac{\Omega_{0}}{\Omega}\cos\left(  \frac{\omega t}{2}\right)  \sin\left(
\Omega t\right)  \hat{x}+\frac{\Omega_{0}}{\Omega}\sin\left(  \frac{\omega
t}{2}\right)  \sin\left(  \Omega t\right)  \hat{y}+\left[  \sin\left(
\frac{\omega t}{2}\right)  \cos\left(  \Omega t\right)  +\frac{\Delta}%
{2\Omega}\cos\left(  \frac{\omega t}{2}\right)  \sin\left(  \Omega t\right)
\right]  \hat{z}\right\}  \right\Vert }\text{,}%
\end{align}
that is,%
\begin{equation}
\hat{n}\left(  t\right)  =\frac{\left\{  \frac{\Omega_{0}}{\Omega}\cos\left(
\frac{\omega t}{2}\right)  \sin\left(  \Omega t\right)  \hat{x}+\frac
{\Omega_{0}}{\Omega}\sin\left(  \frac{\omega t}{2}\right)  \sin\left(  \Omega
t\right)  \hat{y}+\left[  \sin\left(  \frac{\omega t}{2}\right)  \cos\left(
\Omega t\right)  +\frac{\Delta}{2\Omega}\cos\left(  \frac{\omega t}{2}\right)
\sin\left(  \Omega t\right)  \right]  \hat{z}\right\}  }{\left\Vert \left\{
\frac{\Omega_{0}}{\Omega}\cos\left(  \frac{\omega t}{2}\right)  \sin\left(
\Omega t\right)  \hat{x}+\frac{\Omega_{0}}{\Omega}\sin\left(  \frac{\omega
t}{2}\right)  \sin\left(  \Omega t\right)  \hat{y}+\left[  \sin\left(
\frac{\omega t}{2}\right)  \cos\left(  \Omega t\right)  +\frac{\Delta}%
{2\Omega}\cos\left(  \frac{\omega t}{2}\right)  \sin\left(  \Omega t\right)
\right]  \hat{z}\right\}  \right\Vert }\text{.} \label{UV}%
\end{equation}
From Eq. (\ref{UV}), we observe that $\hat{n}\left(  t\right)  $ correctly
reduces to $\hat{z}$ for any instant $t$ when $\Omega_{0}=0$. Also, we note
that $\hat{n}\left(  0\right)  =\vec{0}$, as it should be. Furthermore, we get
from Eqs. (\ref{linki}) and (\ref{linki1}) that
\begin{align}
\tan^{2}\left(  \frac{\alpha}{2}\right)   &  =\frac{\vec{v}_{12}\cdot\vec
{v}_{12}}{c_{12}^{2}}\nonumber\\
&  =\frac{\left[  \frac{\Omega_{0}}{\Omega}\cos\left(  \frac{\omega t}%
{2}\right)  \sin\left(  \Omega t\right)  \right]  ^{2}+\left[  \frac
{\Omega_{0}}{\Omega}\sin\left(  \frac{\omega t}{2}\right)  \sin\left(  \Omega
t\right)  \right]  ^{2}+\left[  \sin\left(  \frac{\omega t}{2}\right)
\cos\left(  \Omega t\right)  +\frac{\Delta}{2\Omega}\cos\left(  \frac{\omega
t}{2}\right)  \sin\left(  \Omega t\right)  \right]  ^{2}}{\left[  \cos\left(
\frac{\omega t}{2}\right)  \cos\left(  \Omega t\right)  -\frac{\Delta}%
{2\Omega}\sin\left(  \frac{\omega t}{2}\right)  \sin\left(  \Omega t\right)
\right]  ^{2}}\text{,}%
\end{align}
that is,%
\begin{equation}
\tan^{2}\left(  \frac{\alpha\left(  t\right)  }{2}\right)  =\frac{\left[
\frac{\Omega_{0}}{\Omega}\cos\left(  \frac{\omega t}{2}\right)  \sin\left(
\Omega t\right)  \right]  ^{2}+\left[  \frac{\Omega_{0}}{\Omega}\sin\left(
\frac{\omega t}{2}\right)  \sin\left(  \Omega t\right)  \right]  ^{2}+\left[
\sin\left(  \frac{\omega t}{2}\right)  \cos\left(  \Omega t\right)
+\frac{\Delta}{2\Omega}\cos\left(  \frac{\omega t}{2}\right)  \sin\left(
\Omega t\right)  \right]  ^{2}}{\left[  \cos\left(  \frac{\omega t}{2}\right)
\cos\left(  \Omega t\right)  -\frac{\Delta}{2\Omega}\sin\left(  \frac{\omega
t}{2}\right)  \sin\left(  \Omega t\right)  \right]  ^{2}}\text{.} \label{X}%
\end{equation}
From Eq. (\ref{X}), we note that when $\Omega_{0}=0$ we have%
\begin{align}
\tan\left(  \frac{\alpha\left(  t\right)  }{2}\right)   &  =\frac{\sin\left(
\frac{\omega t}{2}\right)  \cos\left(  \Omega t\right)  +\frac{\Delta}%
{2\Omega}\cos\left(  \frac{\omega t}{2}\right)  \sin\left(  \Omega t\right)
}{\cos\left(  \frac{\omega t}{2}\right)  \cos\left(  \Omega t\right)
-\frac{\Delta}{2\Omega}\sin\left(  \frac{\omega t}{2}\right)  \sin\left(
\Omega t\right)  }\nonumber\\
&  =\frac{\sin\left(  \frac{\omega t}{2}\right)  \cos\left(  \frac{\omega
_{0}-\omega}{2}t\right)  +\cos\left(  \frac{\omega t}{2}\right)  \sin\left(
\frac{\omega_{0}-\omega}{2}t\right)  }{\cos\left(  \frac{\omega t}{2}\right)
\cos\left(  \frac{\omega_{0}-\omega}{2}t\right)  -\sin\left(  \frac{\omega
t}{2}\right)  \sin\left(  \frac{\omega_{0}-\omega}{2}t\right)  }\nonumber\\
&  =\tan\left(  \frac{\omega_{0}t}{2}\right)  \text{,}%
\end{align}
that is, $\alpha\left(  t\right)  =\omega_{0}t$, as expected. In summary,
setting $\hslash=1$, the full unitary evolution operator in the laboratory
frame of reference can be recast as,%
\begin{align}
U\left(  t\right)   &  =U_{\mathrm{RF}}\left(  t\right)  U_{\mathrm{Rabi}%
}\left(  t\right) \nonumber\\
&  =e^{-i\frac{\omega}{2}t\sigma_{z}}e^{-i\mathrm{H}_{\mathrm{Rabi}}%
t}\nonumber\\
&  =e^{-i\frac{\omega}{2}t\sigma_{z}}e^{-i\left[  \Omega_{0}\sigma_{x}%
+\frac{\Delta}{2}\sigma_{z}\right]  t}\nonumber\\
&  =e^{-i\frac{\alpha}{2}\hat{n}\cdot\vec{\sigma}}\text{,}%
\end{align}
where $\hat{n}=\hat{n}\left(  t\right)  $ and $\alpha=\alpha\left(  t\right)
$ are explicitly given in Eqs. (\ref{UV}) and (\ref{X}), respectively. In
conclusion, the Bloch vector $\vec{a}\left(  t\right)  $ of the quantum state
at time $t$ in the laboratory frame of reference is specified by $\vec{a}%
_{0}=\vec{a}\left(  0\right)  $, $\hat{n}=\hat{n}\left(  t\right)  $, and
$\alpha=\alpha\left(  t\right)  $ and is given by%
\begin{equation}%
\begin{array}
[c]{c}%
\vec{a}\left(  t\right)  =\left(  \hat{n}\cdot\vec{a}_{0}\right)  \hat{n}%
+\sin\left(  \alpha\right)  \left(  \hat{n}\times\vec{a}_{0}\right)
+\cos\left(  \alpha\right)  \left(  \hat{n}\times\vec{a}_{0}\right)
\times\hat{n}\text{,}\\
\\
\hat{n}\left(  t\right)  =\frac{\left\{  \frac{\Omega_{0}}{\Omega}\cos\left(
\frac{\omega t}{2}\right)  \sin\left(  \Omega t\right)  \hat{x}+\frac
{\Omega_{0}}{\Omega}\sin\left(  \frac{\omega t}{2}\right)  \sin\left(  \Omega
t\right)  \hat{y}+\left[  \sin\left(  \frac{\omega t}{2}\right)  \cos\left(
\Omega t\right)  +\frac{\Delta}{2\Omega}\cos\left(  \frac{\omega t}{2}\right)
\sin\left(  \Omega t\right)  \right]  \hat{z}\right\}  }{\left\Vert \left\{
\frac{\Omega_{0}}{\Omega}\cos\left(  \frac{\omega t}{2}\right)  \sin\left(
\Omega t\right)  \hat{x}+\frac{\Omega_{0}}{\Omega}\sin\left(  \frac{\omega
t}{2}\right)  \sin\left(  \Omega t\right)  \hat{y}+\left[  \sin\left(
\frac{\omega t}{2}\right)  \cos\left(  \Omega t\right)  +\frac{\Delta}%
{2\Omega}\cos\left(  \frac{\omega t}{2}\right)  \sin\left(  \Omega t\right)
\right]  \hat{z}\right\}  \right\Vert }\\
\\
\tan^{2}\left(  \frac{\alpha\left(  t\right)  }{2}\right)  =\frac{\left[
\frac{\Omega_{0}}{\Omega}\cos\left(  \frac{\omega t}{2}\right)  \sin\left(
\Omega t\right)  \right]  ^{2}+\left[  \frac{\Omega_{0}}{\Omega}\sin\left(
\frac{\omega t}{2}\right)  \sin\left(  \Omega t\right)  \right]  ^{2}+\left[
\sin\left(  \frac{\omega t}{2}\right)  \cos\left(  \Omega t\right)
+\frac{\Delta}{2\Omega}\cos\left(  \frac{\omega t}{2}\right)  \sin\left(
\Omega t\right)  \right]  ^{2}}{\left[  \cos\left(  \frac{\omega t}{2}\right)
\cos\left(  \Omega t\right)  -\frac{\Delta}{2\Omega}\sin\left(  \frac{\omega
t}{2}\right)  \sin\left(  \Omega t\right)  \right]  ^{2}}%
\end{array}
\text{.} \label{right}%
\end{equation}
Interestingly, note that there is no need to find an explicit expression for
$\alpha\left(  t\right)  $ since $\sin\left(  \alpha\right)  $ and
$\cos\left(  \alpha\right)  $ in $\vec{a}\left(  t\right)  $ can be completely
expressed in terms of $\tan^{2}\left(  \alpha/2\right)  $. Indeed, from some
basic trigonometry, we find%
\begin{equation}
\sin\left(  \alpha\right)  =\frac{2\tan\left(  \frac{\alpha}{2}\right)
}{1+\tan^{2}\left(  \frac{\alpha}{2}\right)  }\text{, and }\cos\left(
\alpha\right)  =\frac{1-\tan^{2}\left(  \frac{\alpha}{2}\right)  }{1+\tan
^{2}\left(  \frac{\alpha}{2}\right)  }\text{.}%
\end{equation}
The derivation of Eq. (\ref{right}) ends our explicit analytical calculation
for the Bloch vector $\vec{a}\left(  t\right)  $. Note that $\vec{m}\left(
t\right)  $ in Eq. (\ref{field23}) and $\vec{a}\left(  t\right)  $ in Eq.
(\ref{right}) are the only two real three-dimensional vectors needed to
calculate the time-dependent curvature coefficient $\kappa_{\mathrm{AC}}%
^{2}\left(  t\right)  =\kappa_{\mathrm{AC}}^{2}\left(  \mathbf{a}\left(
t\right)  \text{, }\mathbf{m}\left(  t\right)  \right)  $,
\begin{equation}
\kappa_{\mathrm{AC}}^{2}\left(  \mathbf{a}\left(  t\right)  \text{,
}\mathbf{m}\left(  t\right)  \right)  \overset{\text{def}}{=}4\frac{\left(
\mathbf{a\cdot m}\right)  ^{2}}{\mathbf{m}^{2}-\left(  \mathbf{a\cdot
m}\right)  ^{2}}+\frac{\left[  \mathbf{m}^{2}\mathbf{\dot{m}}^{2}-\left(
\mathbf{m\cdot\dot{m}}\right)  ^{2}\right]  -\left[  \left(  \mathbf{a\cdot
\dot{m}}\right)  \mathbf{m-}\left(  \mathbf{a\cdot m}\right)  \mathbf{\dot{m}%
}\right]  ^{2}}{\left[  \mathbf{m}^{2}-\left(  \mathbf{a\cdot m}\right)
^{2}\right]  ^{3}}+4\frac{\left(  \mathbf{a\cdot m}\right)  \left[
\mathbf{a\cdot}\left(  \mathbf{m\times\dot{m}}\right)  \right]  }{\left[
\mathbf{m}^{2}-\left(  \mathbf{a\cdot m}\right)  ^{2}\right]  ^{2}}\text{,}
\label{B52}%
\end{equation}
with $\mathbf{a}\overset{\text{def}}{=}\vec{a}$ and $\mathbf{m}%
\overset{\text{def}}{=}\vec{m}$ in Eq. (\ref{B52}). For completeness, we note
that we can also present an alternative parametrization of the curvature
coefficient in terms of $\kappa_{\mathrm{AC}}^{2}\left(  \tau\right)
=\kappa_{\mathrm{AC}}^{2}\left(  \mathbf{a}\left(  \tau\right)  \text{,
}\mathbf{m}\left(  \tau\right)  \right)  $ with $\tau\overset{\text{def}%
}{=}\Omega t$, $\Omega\overset{\text{def}}{=}\sqrt{\Omega_{0}^{2}+\left(
\Delta/2\right)  ^{2}}$, and $\Delta\overset{\text{def}}{=}\omega_{0}-\omega
$.
%===============================================

\begin{figure}[th]%
\begin{tabular}
[c]{cc}%
\includegraphics[width=6.0in,height=1.75in]{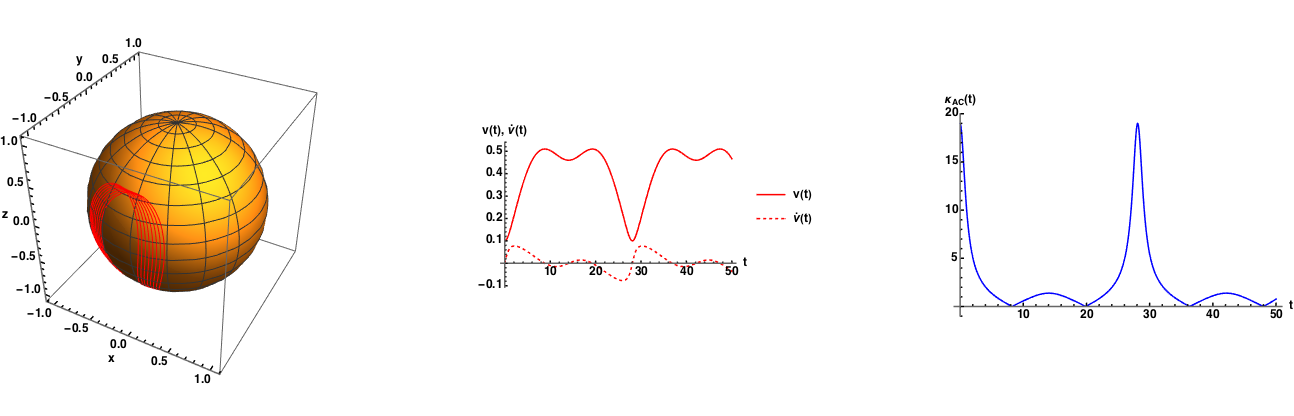} & \\
\includegraphics[width=6.0in,height=1.75in]{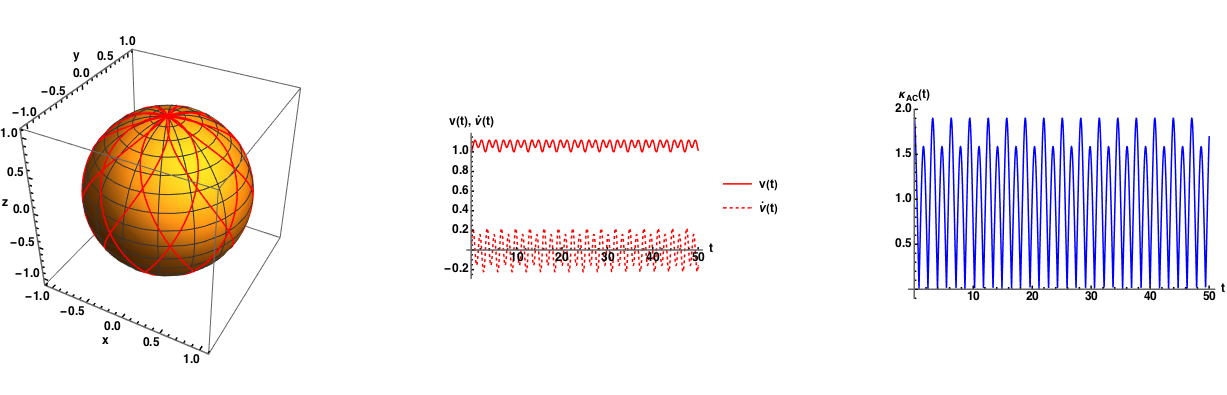} &
\end{tabular}
\caption{(Color online)(left) Bloch vector orbit $\mathbf{a}(t)$, (center)
velocity $v(t)$ and acceleration $\dot{v}(t)$ and (right) $\kappa_{AC}(t)$ for
near resonance coupling $(\omega_{0}=1.0, \omega= 0.9)$ for an initial
spin-down state $\mathbf{a}(0)$, for (top) weak driving: $\Omega_{0} = 0.1$,
(bottom) strong driving: $\Omega_{0} = 1.0$. }%
\label{fig:orbits:near:resonance:spindown}%
\end{figure}

\begin{figure}[th]%
\begin{tabular}
[c]{cc}%
\includegraphics[width=6.0in,height=1.75in]{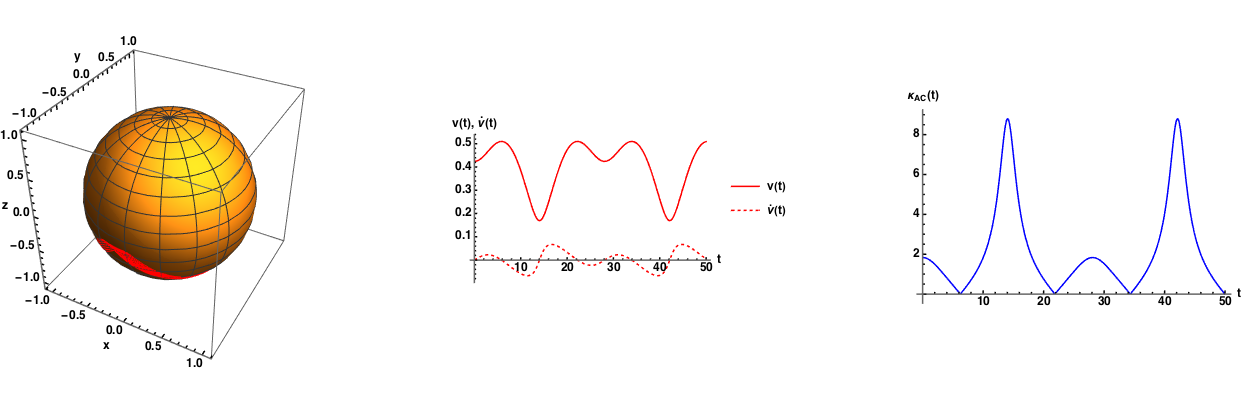} & \\
\includegraphics[width=6.0in,height=1.75in]{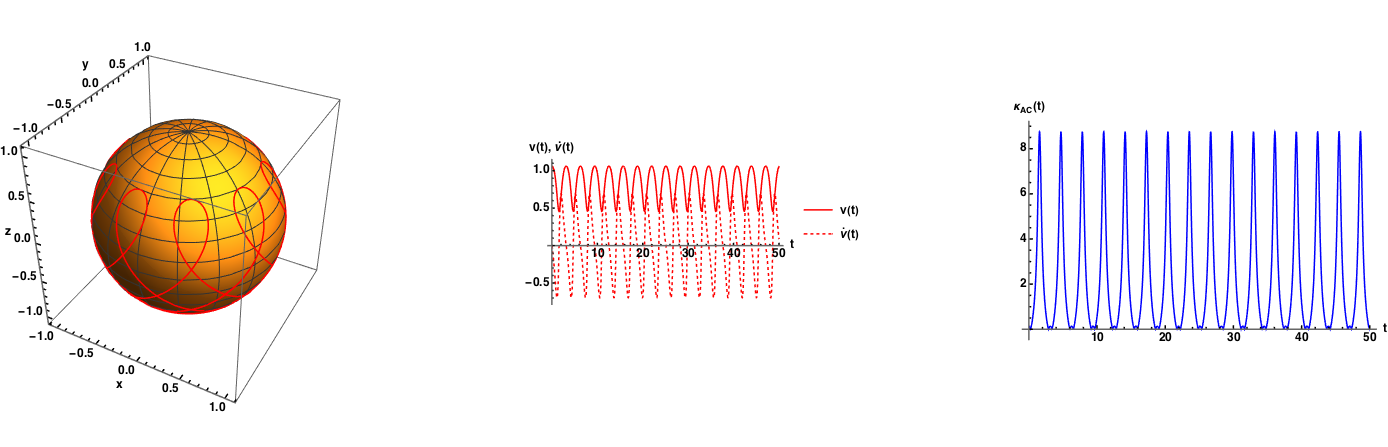} &
\end{tabular}
\caption{(Color online)(left) Bloch vector orbit $\mathbf{a}(t)$, (center)
velocity $v(t)$ and acceleration $\dot{v}(t)$ and (right) $\kappa_{AC}(t)$ for
near resonance coupling $(\omega_{0}=1.0, \omega= 0.9)$ for an initial state
$\mathbf{a}(0)$ with $(\theta_{0},\phi_{0}) = (3\pi/4,0)$, for (top) weak
driving: $\Omega_{0} = 0.1$, (bottom) strong driving: $\Omega_{0} = 1.0$. }%
\label{fig:orbits:near:resonance:theta0:3pidiv4}%
\end{figure}

%===============================================
%PMA added text and Bloch orbit figures: 27Nov2023
%===============================================
In Fig.~\ref{fig:orbits:near:resonance:spindown} we show the (left) Bloch
vector orbit $\mathbf{a}(t)$ Eq.~(\ref{right}), (center) velocity $v(t)$ and
acceleration $\dot{v}(t)$ Eq.~(\ref{vvdot}), and (right) $\kappa_{AC}(t)$
(square root of Eq.~(\ref{B52})) for near resonance coupling with atomic
frequency (normalized to) $\omega_{0}=1.0$, and driving frequency $\omega=
0.9$ for an initial spin-down state $\mathbf{a}(0)$ with $(\theta_{0},\phi
_{0}) = (\pi,0)$, for (top) weak driving: $\Omega_{0} = 0.1$, (bottom) strong
driving: $\Omega_{0} = 1.0$. In
Fig.~\ref{fig:orbits:near:resonance:theta0:3pidiv4} we show the same plots as
in Fig.~\ref{fig:orbits:near:resonance:spindown}, but now with an initial
state $\mathbf{a}(0)$ with $(\theta_{0},\phi_{0}) = (3\pi/4,0)$.

It is interesting to note that for strong driving (bottom row of figures)
$\Omega_{0}=\omega_{0}=1.0$ the velocity $v(t)$ of the orbits fluctuates about
a nearly constant value, so that the acceleration $\dot{v}(t)$ fluctuates
about zero, with an amplitude that appears dependent on the initial state. The
curvature $\kappa_{AC}$ exhibits a periodic or doubly periodic structure that
also appears to be initial state dependent. On the other hand, for weak
driving (top row of figures) $\Omega_{0}=0.1, \omega_{0}=1.0$ velocity $v(t)$
of the orbits show periodic regions of ``near'' constancy, so again the
acceleration $\dot{v}(t)$ fluctuates about zero periodically, though with
larger amplitude. The curvature $\kappa_{AC}$ periodically spikes about the
turning points of the orbits (i.e. around the reversal of south-to-north pole
motion to north-to-south pole motion).

%===============================================
%===============================================
%===============================================

%===============================================
\clearpage
\newpage

\section{Final Remarks}

%===============================================
In this paper, we proposed a geometric perspective on how to quantify the
bending and the twisting of quantum curves traced by state vectors evolving
under nonstationary Hamiltonians. Specifically, relying on the existing
geometric viewpoint for stationary Hamiltonians as in Ref. \cite{alsing1}, we
discussed the generalization of our theoretical construct to nonstationary
quantum-mechanical scenarios where both time-varying curvature and torsion
coefficients play a key role. Specifically, we presented a quantum version of
the Frenet-Serret apparatus for a quantum trajectory in projective Hilbert
space traced out by a parallel-transported pure quantum state evolving
unitarily under a time-dependent Hamiltonian specifying the Schr\"{o}dinger
evolution equation. The time-varying curvature coefficient is defined by the
magnitude squared of the covariant derivative of the tangent vector
$\left\vert T\right\rangle $ to the state vector $\left\vert \Psi\right\rangle
$ and measures the bending of the quantum curve (Eq. (\ref{peggio})). The
time-varying torsion coefficient, instead, is specified by the magnitude
squared of the projection of the covariant derivative of the tangent vector
$\left\vert T\right\rangle $ to the state vector $\left\vert \Psi\right\rangle
$, orthogonal to $\left\vert T\right\rangle $ and $\left\vert \Psi
\right\rangle $ and, in addition, measures the twisting of the quantum curve
Eq. ((\ref{torcina})).

A summary of our main findings can be described as follows:

\begin{enumerate}
\item[{[i]}] We extended our geometric approach, originally developed for
stationary Hamiltonians in Ref. \cite{alsing1}, to nonstationary Hamiltonians
of arbitrary finite-dimensional quantum systems. Specifically, the
time-dependent expressions for the curvature and torsion coefficients of a
quantum evolution appear in Eqs. (\ref{curvatime}) and (\ref{torsiontime}), respectively.

\item[{[ii]}] We discussed the richer statistical structure underlying the
concepts of curvature and torsion of a quantum evolution governed by a
nonstationary Hamiltonian, going beyond the concepts of skewness and kurtosis
reported in Ref. \cite{alsing1}. The statistical expressions for the curvature
and torsion coefficients of a quantum evolution appear in Eqs. (\ref{form1})
and (\ref{form2}), respectively. \ In addition, we discussed the vanishing of
the torsion coefficient expressed in terms of generalized covariance
$\left\vert \Sigma\left(  \mathrm{H}\text{, }\mathrm{\dot{H}}\right)
\right\vert $ (Eq. (\ref{myheart})) for two-level systems in Appendix A.

\item[{[iii]}] We provided in Eqs. (\ref{XXX}) and (\ref{YYY}) a closed-form
expression for the curvature and torsion coefficients, respectively, of a
two-level system that evolves under a time-dependent Hamiltonian in terms of
only two real three-dimensional vectors, i.e., the Bloch vector $\mathbf{a}%
\left(  t\right)  $ and the magnetic field vector $\mathbf{m}\left(  t\right)
$. This is very helpful because of the computational difficulties that arise
in the nonstationary scenario when one attempts to express the physical time
parameter in terms of the arc length parameter. Interestingly, the fact that
the Bloch vector representation of qubits provides a very intuitive picture of
the dynamics of the quantum system was already emphasized by Feynman and
collaborators in Ref. \cite{dick57} while \ dealing with maser problems by
studying the behavior of an ensemble of noninteracting two-level quantum
systems in the presence of an external perturbation.

\item[{[iv]}] We provided three alternative ways to verify that the torsion
coefficient of a quantum evolution specified by an arbitrary time-dependent
Hamiltonian for a $2$ -level system is zero. The first verification is based
upon the projection operators formalism, a natural extension of the one used
in the time-independent setting \cite{alsing1}. In this case, the operator
\textrm{P}$^{\left(  T\right)  }\mathrm{P}^{\left(  \Psi\right)  }$ becomes
the null operator $\mathcal{O}$ and, as a consequence, $\tau_{\mathrm{AC}}%
^{2}\left(  s\right)  =\left\Vert \mathrm{P}^{\left(  T\right)  }%
\mathrm{P}^{\left(  \Psi\right)  }\left\vert T^{\prime}\left(  s\right)
\right\rangle \right\Vert ^{2}=0$. This approach leads to the conclusion that,
even though it is possible to bend the evolution of a $2$-level quantum
system, there is not enough \textquotedblleft room\textquotedblright\ for
twisting its evolution. The second verification relies on a vector algebra
approach where the two key vectors are the unit Bloch vector $\mathbf{a\ }%
$\ and the magnetic field vector $\mathbf{m}$ (Eq. (\ref{YYY}) in Subsection B
in Section IV). The insight that emerges from this second approach is that as
long as the pure state of the two-level system evolves on the unit-sphere with
$\left\Vert \mathbf{a}\right\Vert =1$, the torsion coefficient $\tau
_{\mathrm{AC}}^{2}\left(  s\right)  $ remains identically equal to zero during
the quantum evolution. Loosely speaking, dimensionality arguments forbid the
$2$-level system to leave the quantum version of the osculating plane spanned
by the tangent and normal vectors. The third verification relies on
statistical arguments underlying quantum theory (Appendix A). After showing
that for $2$-level systems the torsion coefficient can be expressed in terms
of the generalized variance (i.e., the determinant of the covariance matrix
between \textrm{H} and \textrm{\.{H}}),we concluded that the vanishing of the
torsion coefficient is a consequence of the perfect correlation between the
operators \textrm{H} and \textrm{\.{H}}. This perfect correlation, in turn,
yields the absence of an overall joint dispersion between \textrm{H} and
\textrm{\.{H}}. Finally, we arrived at the conclusion that the lack of an
overall joint dispersion in the pair $\left(  \mathrm{H}\text{, }%
\mathrm{\dot{H}}\right)  $ implies no twisting for $2$-level systems that
evolve under nonstationary Hamiltonians of the form \textrm{H}$\left(
t\right)  =\mathbf{m}\left(  t\right)  \cdot\vec{\sigma}$.

\item[{[V]}] We discussed in Section V the physical insights into the dynamics
of two-level quantum systems that can be obtained by applying our geometric
approach to an exactly soluble time-dependent two-state Rabi problem specified
by a sinusoidal oscillating time-dependent potential (Eq. (\ref{time-H})). The
physical tunable parameters that characterize the Hamiltonian $\mathrm{H}%
\left(  t\right)  $ in Eq. (\ref{time-H}) are the resonance frequency of the
atom $\omega_{0}$, the energy difference $\hslash\omega_{0}$ between the
ground and excited states, the Rabi frequency $\Omega_{0}$ and, finally, the
external driving frequency $\omega$. This example is particularly significant
since a time-dependent problem has a much richer structure than a
time-independent one Ref. \cite{alsing1}. In many ways, it is more
representative since, for instance, the curvature coefficient in Eq.
(\ref{B52}) expressed in terms of the magnetic field vector $\mathbf{m}\left(
t\right)  $ in Eq. (\ref{field23}) and the Bloch vector $\mathbf{a}\left(
t\right)  $ in Eq. (\ref{right}) can generally exhibit rather complicated
temporal behaviors. Indeed, we specialized in this example on the
characterization of the temporal behavior of the curvature coefficient in
different dynamical scenarios, including off-resonance (i.e., $\left\vert
\omega_{0}-\omega\right\vert \gg\Omega_{0}$) and on-resonance regimes (i.e.,
$\left\vert \omega_{0}-\omega\right\vert \ll\Omega_{0}$) and, in addition,
strong (i.e., $\Omega_{0}\gg\omega_{0}$) and weak (i.e., $\Omega_{0}\ll
\omega_{0}$) driving configurations. We illustrate the temporal behavior of
the Bloch vector orbit $a\left(  t\right)  $, the speed of quantum evolution
$v\left(  t\right)  $, its rate of change $\dot{v}\left(  t\right)  $ and,
finally, the curvature $\kappa_{\mathrm{AC}}\left(  t\right)  $ in specific
physical conditions (i.e., near resonance, weak driving, strong driving, and
given initial states) in Fig. $1$ and Fig. $2$.
\end{enumerate}

\smallskip

In our paper, the expressions for $\kappa_{\mathrm{AC}}^{2}\left(  s\right)  $
and $\tau_{\mathrm{AC}}^{2}\left(  s\right)  $ are formally applicable to any
$d$-level quantum system that evolves under a nonstationary Hamiltonian.
However, the discussion of our illustrative example was limited here to
$2$-level systems only. In this simpler case, unlike what happens in
higher-dimensional scenarios, the concepts of Bloch vectors and Bloch spheres
have a clear interpretation and geometric visualization. Remarkably, the
notion of Bloch vector can be properly defined for $4$-level two-qubit systems
\cite{jakob01} and for arbitrary $d$-dimensional quantum-mechanical systems
with $d>2$ (i.e., qudits) \cite{kimura03,krammer08,kurzy11}. For single-qubit
systems, the utility of the concept of Bloch vector is twofold. First, an
orbit on the Bloch sphere captures essential aspects of the unitary temporal
evolution of a single-qubit quantum state. Moreover, an orbit offers a neat
visualization of the quantum-mechanical unitary temporal evolution. Second,
the Bloch vector is defined in terms of real components that are expressed as
expectation values of experimentally measurable observables characterized by
Hermitian operators (i.e., the Pauli operators for $2$-level quantum systems).
When transitioning from $2$-level quantum systems to higher dimensional
systems, the physical significance of the Bloch vector conserves its utility.
Regrettably, its geometric visualization is not as transparent as in the case
of single-qubit quantum systems. Indeed, peculiar features of quantum theory
appear in higher-dimensional systems, including the simplest but non-trivial
case specified by $3$-level quantum systems (i.e., qutrits) \cite{kurzy11}.
These odd quantum properties make it complicated to grasp the geometry of
multidimensional quantum systems \cite{xie20,siewert21}. For example, a major
difference between $2$-level systems and higher-dimensional quantum systems
can be describes as follows. For single qubits, any point on the Bloch sphere
or inside the Bloch ball represents a physical state (i.e., a pure and a mixed
quantum state, respectively). Alternatively, not every point on the
\textquotedblleft Bloch sphere\textquotedblright\ in dimensions $d^{2}-1$
represents a physical state for $d$-dimensional qudit systems. In spite of the
fact that it is possible to establish a unique Bloch vector for any physical
state, not every Bloch vector defines a quantum state. Specifically, there are
Bloch vectors in higher-dimensions corresponding to unphysical states
characterized by density matrices with negative eigenvalues (even if of trace
one, \cite{kurzy11}). We suggest Ref. \cite{gamel16} for a detailed
presentation on Bloch vector representations of single-qubit systems,
single-qutrit systems, and two-qubit systems in terms of Pauli, Gell-Mann, and
Dirac matrices, respectively.

\smallskip

In conclusion, despite its current limitations \cite{alsing1}, we sincerely
hope our work will stimulate other scientists and pave the way to further
investigations on the interplay between geometry and quantum mechanics. For
the time being, we leave a more in-depth quantitative discussion on geometric
extensions \cite{jakob01,kimura03,krammer08,kurzy11,xie20,siewert21,gamel16}
of our analytical findings, including generalizations to mixed state geometry
and time-dependent quantum evolutions of higher-dimensional physical systems,
to future scientific efforts.

\begin{acknowledgments}
P.M.A. acknowledges support from the Air Force Office of Scientific Research
(AFOSR). C.C. is grateful to the United States Air Force Research Laboratory
(AFRL) Summer Faculty Fellowship Program for providing support for this work.
Any opinions, findings and conclusions or recommendations expressed in this
material are those of the author(s) and do not necessarily reflect the views
of the Air Force Research Laboratory (AFRL).
\end{acknowledgments}

\pagebreak

\appendix

\section{Statistical interpretation of a vanishing torsion}

We divide this Appendix in two parts. In the first part, we show that the
time-dependent torsion coefficient $\tau_{\mathrm{AC}}^{2}\left(  t\right)  $
that specifies the quantum evolution of two-level systems can be expressed in
terms of the generalized variance $\left\vert \Sigma\left(  \mathrm{H}\text{,
}\mathrm{\dot{H}}\right)  \right\vert $. The latter, in turn, is nothing but
the determinant of the covariance matrix $\Sigma\left(  \mathrm{H}\text{,
}\mathrm{\dot{H}}\right)  $ between the Hamiltonian operator \textrm{H }and
its first derivative \textrm{\.{H}}. In the second part, instead, we show that
$\tau_{\mathrm{AC}}^{2}\left(  t\right)  $ is identically zero. The vanishing
of the torsion coefficient was also clear from the other two proofs, namely
the one based on the projector formalism and the one based on the two real
three-dimensional vectors $\mathbf{a}\left(  t\right)  $ and $\mathbf{m}%
\left(  t\right)  $. However, the statistical proof that we present here
offers an alternative perspective than complements the other two perspectives.

\subsection{Torsion as generalized variance}

In the case of two-level quantum systems, the general expression of the
time-dependent torsion coefficient $\tau_{\mathrm{AC}}^{2}\left(  s\right)
$,
\begin{align}
\tau_{\mathrm{AC}}^{2}\left(  s\right)   &  =\left\langle \left(  \Delta
h\right)  ^{4}\right\rangle -\left\langle \left(  \Delta h\right)
^{2}\right\rangle ^{2}-\left\langle \left(  \Delta h\right)  ^{3}\right\rangle
^{2}+\left[  \left\langle \left(  \Delta h^{\prime}\right)  ^{2}\right\rangle
-\left\langle \Delta h^{\prime}\right\rangle ^{2}-\left\langle \left(  \Delta
h^{\prime}\right)  \left(  \Delta h\right)  \right\rangle \left\langle \left(
\Delta h\right)  \left(  \Delta h^{\prime}\right)  \right\rangle \right]
+\nonumber\\
&  +i\left[  \left\langle \left[  \left(  \Delta h\right)  ^{2}\text{, }\Delta
h^{\prime}\right]  \right\rangle -\left\langle \left(  \Delta h\right)
^{3}\right\rangle \left\langle \left[  \Delta h\text{, }\Delta h^{\prime
}\right]  \right\rangle \right]  \text{,} \label{c1}%
\end{align}
can be significantly simplified. First, we reiterate that the term
$\left\langle \left(  \Delta h\right)  ^{4}\right\rangle -\left\langle \left(
\Delta h\right)  ^{2}\right\rangle ^{2}-\left\langle \left(  \Delta h\right)
^{3}\right\rangle ^{2}$ in Eq. (\ref{c1}) equals zero and this is a
consequence of the fact that $\left\langle \left(  \Delta\mathrm{H}\right)
^{3}\right\rangle =-2\left\langle \mathrm{H}\right\rangle \left\langle \left(
\Delta\mathrm{H}\right)  ^{2}\right\rangle $ when the Hamiltonian is given by
\textrm{H}$\left(  t\right)  =\mathbf{m}\left(  t\right)  \cdot\vec{\sigma}$.
Indeed, we have%
\begin{align}
\left\langle \left(  \Delta\mathrm{H}\right)  ^{3}\right\rangle  &
=\left\langle \mathrm{H}^{3}\right\rangle -3\left\langle \mathrm{H}%
^{2}\right\rangle \left\langle \mathrm{H}\right\rangle +2\left\langle
\mathrm{H}\right\rangle ^{3}\nonumber\\
&  =\mathbf{m}^{2}\left\langle \mathrm{H}\right\rangle -3\mathbf{m}%
^{2}\left\langle \mathrm{H}\right\rangle +2\left\langle \mathrm{H}%
\right\rangle ^{3}\nonumber\\
&  =-2\mathbf{m}^{2}\left\langle \mathrm{H}\right\rangle +2\left\langle
\mathrm{H}\right\rangle ^{3}\nonumber\\
&  =-2\left\langle \mathrm{H}\right\rangle \left[  \mathbf{m}^{2}-\left\langle
\mathrm{H}\right\rangle ^{2}\right] \nonumber\\
&  =-2\left\langle \mathrm{H}\right\rangle \left[  \left\langle \mathrm{H}%
^{2}\right\rangle -\left\langle \mathrm{H}\right\rangle ^{2}\right]
\nonumber\\
&  =-2\left\langle \mathrm{H}\right\rangle \left\langle \left(  \Delta
\mathrm{H}\right)  ^{2}\right\rangle \text{,}%
\end{align}
that is,%
\begin{equation}
\left\langle \left(  \Delta\mathrm{H}\right)  ^{3}\right\rangle
=-2\left\langle \mathrm{H}\right\rangle \left\langle \left(  \Delta
\mathrm{H}\right)  ^{2}\right\rangle \text{.} \label{c3}%
\end{equation}
Using Eq. (\ref{c3}),
%PMA
the first three terms of Eq.(\ref{c1}) $\left\langle \left(  \Delta h\right)
^{4}\right\rangle -\left\langle \left(  \Delta h\right)  ^{2}\right\rangle
^{2}-\left\langle \left(  \Delta h\right)  ^{3}\right\rangle ^{2}$ becomes
zero since%
\begin{align}
\left\langle \left(  \Delta h\right)  ^{4}\right\rangle -\left\langle \left(
\Delta h\right)  ^{2}\right\rangle ^{2}-\left\langle \left(  \Delta h\right)
^{3}\right\rangle ^{2}  &  =\frac{\left\langle \left(  \Delta\mathrm{H}%
\right)  ^{4}\right\rangle -\left\langle \left(  \Delta\mathrm{H}\right)
^{2}\right\rangle ^{2}}{\left\langle \left(  \Delta\mathrm{H}\right)
^{2}\right\rangle ^{2}}-\frac{\left\langle \left(  \Delta\mathrm{H}\right)
^{3}\right\rangle ^{2}}{\left\langle \left(  \Delta\mathrm{H}\right)
^{2}\right\rangle ^{3}}\nonumber\\
&  =\frac{\left[  \left\langle \mathrm{H}^{4}\right\rangle -4\left\langle
\mathrm{H}^{3}\right\rangle \left\langle \mathrm{H}\right\rangle
+6\left\langle \mathrm{H}^{2}\right\rangle \left\langle \mathrm{H}%
\right\rangle ^{2}-3\left\langle \mathrm{H}\right\rangle ^{4}\right]  -\left[
\left\langle \mathrm{H}^{2}\right\rangle -\left\langle \mathrm{H}\right\rangle
^{2}\right]  ^{2}}{\left\langle \left(  \Delta\mathrm{H}\right)
^{2}\right\rangle ^{2}}-\frac{\left\langle \left(  \Delta\mathrm{H}\right)
^{3}\right\rangle ^{2}}{\left\langle \left(  \Delta\mathrm{H}\right)
^{2}\right\rangle ^{3}}\nonumber\\
&  =\frac{\mathbf{m}^{4}-4\mathbf{m}^{2}\left\langle \mathrm{H}\right\rangle
^{2}+6\mathbf{m}^{2}\left\langle \mathrm{H}\right\rangle ^{2}-3\left\langle
\mathrm{H}\right\rangle ^{4}-\mathbf{m}^{4}-\left\langle \mathrm{H}%
\right\rangle ^{4}+2\mathbf{m}^{2}\left\langle \mathrm{H}\right\rangle ^{2}%
}{\left\langle \left(  \Delta\mathrm{H}\right)  ^{2}\right\rangle ^{2}}%
-\frac{\left\langle \left(  \Delta\mathrm{H}\right)  ^{3}\right\rangle ^{2}%
}{\left\langle \left(  \Delta\mathrm{H}\right)  ^{2}\right\rangle ^{3}%
}\nonumber\\
&  =\frac{4\mathbf{m}^{2}\left\langle \mathrm{H}\right\rangle ^{2}%
-4\left\langle \mathrm{H}\right\rangle ^{4}}{\left\langle \left(
\Delta\mathrm{H}\right)  ^{2}\right\rangle ^{2}}-\frac{\left\langle \left(
\Delta\mathrm{H}\right)  ^{3}\right\rangle ^{2}}{\left\langle \left(
\Delta\mathrm{H}\right)  ^{2}\right\rangle ^{3}}\nonumber\\
&  =\frac{4\left\langle \mathrm{H}\right\rangle ^{2}\left[  \mathbf{m}%
^{2}-\left\langle \mathrm{H}\right\rangle ^{2}\right]  }{\left\langle \left(
\Delta\mathrm{H}\right)  ^{2}\right\rangle ^{2}}-\frac{\left\langle \left(
\Delta\mathrm{H}\right)  ^{3}\right\rangle ^{2}}{\left\langle \left(
\Delta\mathrm{H}\right)  ^{2}\right\rangle ^{3}}\nonumber\\
&  =\frac{4\left\langle \mathrm{H}\right\rangle ^{2}\left\langle \left(
\Delta\mathrm{H}\right)  ^{2}\right\rangle }{\left\langle \left(
\Delta\mathrm{H}\right)  ^{2}\right\rangle ^{2}}-\frac{\left\langle \left(
\Delta\mathrm{H}\right)  ^{3}\right\rangle ^{2}}{\left\langle \left(
\Delta\mathrm{H}\right)  ^{2}\right\rangle ^{3}}\nonumber\\
&  =4\frac{\left\langle \mathrm{H}\right\rangle ^{2}}{\left\langle \left(
\Delta\mathrm{H}\right)  ^{2}\right\rangle }-\frac{\left(  -2\left\langle
\mathrm{H}\right\rangle \left\langle \left(  \Delta\mathrm{H}\right)
^{2}\right\rangle \right)  ^{2}}{\left\langle \left(  \Delta\mathrm{H}\right)
^{2}\right\rangle ^{3}}\nonumber\\
&  =4\frac{\left\langle \mathrm{H}\right\rangle ^{2}}{\left\langle \left(
\Delta\mathrm{H}\right)  ^{2}\right\rangle }-4\frac{\left\langle
\mathrm{H}\right\rangle ^{2}}{\left\langle \left(  \Delta\mathrm{H}\right)
^{2}\right\rangle }\nonumber\\
&  =0\text{.} \label{kissme1}%
\end{align}
Let us focus on the term $i\left[  \left\langle \left[  \left(  \Delta
h\right)  ^{2}\text{, }\Delta h^{\prime}\right]  \right\rangle -\left\langle
\left(  \Delta h\right)  ^{3}\right\rangle \left\langle \left[  \Delta
h\text{, }\Delta h^{\prime}\right]  \right\rangle \right]  $ in
%PMA
the second line of Eq.(\ref{c1}). We have,%
\begin{align}
i\left[  \left\langle \left[  \left(  \Delta h\right)  ^{2}\text{, }\Delta
h^{\prime}\right]  \right\rangle -\left\langle \left(  \Delta h\right)
^{3}\right\rangle \left\langle \left[  \Delta h\text{, }\Delta h^{\prime
}\right]  \right\rangle \right]   &  =i\left[  \frac{\left\langle \left[
\left(  \Delta\mathrm{H}\right)  ^{2}\text{, }\Delta\mathrm{\dot{H}}\right]
\right\rangle }{\left\langle \left(  \Delta\mathrm{H}\right)  ^{2}%
\right\rangle ^{2}}-\left\langle \left(  \Delta\mathrm{H}\right)
^{3}\right\rangle \frac{\left\langle \left[  \Delta\mathrm{H}\text{, }%
\Delta\mathrm{\dot{H}}\right]  \right\rangle }{\left\langle \left(
\Delta\mathrm{H}\right)  ^{2}\right\rangle ^{3}}\right] \nonumber\\
&  =i\frac{\left\langle \left[  \left(  \Delta\mathrm{H}\right)  ^{2}\text{,
}\Delta\mathrm{\dot{H}}\right]  \right\rangle }{\left\langle \left(
\Delta\mathrm{H}\right)  ^{2}\right\rangle ^{2}}-i\left\langle \left(
\Delta\mathrm{H}\right)  ^{3}\right\rangle \frac{\left\langle \left[
\Delta\mathrm{H}\text{, }\Delta\mathrm{\dot{H}}\right]  \right\rangle
}{\left\langle \left(  \Delta\mathrm{H}\right)  ^{2}\right\rangle ^{3}%
}\nonumber\\
&  =i\frac{\left\langle \left[  \mathrm{H}^{2}\text{, }\mathrm{\dot{H}%
}\right]  \right\rangle -2\left\langle \mathrm{H}\right\rangle \left\langle
\left[  \mathrm{H}\text{, }\mathrm{\dot{H}}\right]  \right\rangle
}{\left\langle \left(  \Delta\mathrm{H}\right)  ^{2}\right\rangle ^{2}%
}-i\left(  -2\left\langle \mathrm{H}\right\rangle \left\langle \left(
\Delta\mathrm{H}\right)  ^{2}\right\rangle \right)  \frac{\left\langle \left[
\mathrm{H}\text{, }\mathrm{\dot{H}}\right]  \right\rangle }{\left\langle
\left(  \Delta\mathrm{H}\right)  ^{2}\right\rangle ^{3}}\nonumber\\
&  =-2i\frac{\left\langle \mathrm{H}\right\rangle \left\langle \left[
\mathrm{H}\text{, }\mathrm{\dot{H}}\right]  \right\rangle }{\left\langle
\left(  \Delta\mathrm{H}\right)  ^{2}\right\rangle ^{2}}+2i\frac{\left\langle
\mathrm{H}\right\rangle \left\langle \left[  \mathrm{H}\text{, }%
\mathrm{\dot{H}}\right]  \right\rangle }{\left\langle \left(  \Delta
\mathrm{H}\right)  ^{2}\right\rangle ^{2}}\nonumber\\
&  =0\text{ ,} \label{kissme}%
\end{align}
where we used in the last step in Eq. (\ref{kissme}) the fact that
$\left\langle \left[  \mathrm{H}^{2}\text{, }\mathrm{\dot{H}}\right]
\right\rangle =0,$
%PMA
since $H^{2}= \vec{m}^{2}\, I$. Therefore, using Eqs. (\ref{kissme1}) and
(\ref{kissme}) and recalling that $\left\langle \Delta h^{\prime}\right\rangle
^{2}=0$, the general expression for $\tau_{\mathrm{AC}}^{2}\left(  s\right)  $
in Eq. (\ref{c1}) reduces to%
\begin{equation}
\tau_{\mathrm{AC}}^{2}\left(  s\right)  =\left\langle \left(  \Delta
h^{\prime}\right)  ^{2}\right\rangle -\left\langle \left(  \Delta h^{\prime
}\right)  \left(  \Delta h\right)  \right\rangle \left\langle \left(  \Delta
h\right)  \left(  \Delta h^{\prime}\right)  \right\rangle \text{.}
\label{metoo1}%
\end{equation}
To further simplify $\tau_{\mathrm{AC}}^{2}\left(  s\right)  $ in Eq.
(\ref{metoo1}), we note that%
\begin{align}
\tau_{\mathrm{AC}}^{2}\left(  s\right)   &  =\left\langle \left(  \Delta
h^{\prime}\right)  ^{2}\right\rangle -\left\langle \left(  \Delta h^{\prime
}\right)  \left(  \Delta h\right)  \right\rangle \left\langle \left(  \Delta
h\right)  \left(  \Delta h^{\prime}\right)  \right\rangle \nonumber\\
&  =\left\langle \left(  \frac{\Delta\mathrm{\dot{H}}}{v^{2}}-\frac
{\Delta\mathrm{H}}{v^{3}}\dot{v}\right)  ^{2}\right\rangle -\left\langle
\left(  \frac{\Delta\mathrm{\dot{H}}}{v^{2}}-\frac{\Delta\mathrm{H}}{v^{3}%
}\dot{v}\right)  \left(  \frac{\Delta\mathrm{H}}{v}\right)  \right\rangle
\left\langle \left(  \frac{\Delta\mathrm{H}}{v}\right)  \left(  \frac
{\Delta\mathrm{\dot{H}}}{v^{2}}-\frac{\Delta\mathrm{H}}{v^{3}}\dot{v}\right)
\right\rangle \nonumber\\
&  =\frac{\left\langle \left(  \Delta\mathrm{\dot{H}}\right)  ^{2}%
\right\rangle -\dot{v}^{2}}{v^{4}}-\left\langle \left(  \frac{\Delta
\mathrm{\dot{H}}}{v^{2}}-\frac{\Delta\mathrm{H}}{v^{3}}\dot{v}\right)  \left(
\frac{\Delta\mathrm{H}}{v}\right)  \right\rangle \left\langle \left(
\frac{\Delta\mathrm{H}}{v}\right)  \left(  \frac{\Delta\mathrm{\dot{H}}}%
{v^{2}}-\frac{\Delta\mathrm{H}}{v^{3}}\dot{v}\right)  \right\rangle \text{,}
\label{c5}%
\end{align}
where $\tilde{Q}\left(  t\right)  \overset{\text{def}}{=}\left\langle \left(
\frac{\Delta\mathrm{\dot{H}}}{v^{2}}-\frac{\Delta\mathrm{H}}{v^{3}}\dot
{v}\right)  \left(  \frac{\Delta\mathrm{H}}{v}\right)  \right\rangle
\left\langle \left(  \frac{\Delta\mathrm{H}}{v}\right)  \left(  \frac
{\Delta\mathrm{\dot{H}}}{v^{2}}-\frac{\Delta\mathrm{H}}{v^{3}}\dot{v}\right)
\right\rangle $ can be written as%
\begin{align}
\tilde{Q}\left(  t\right)   &  =\left\langle \frac{\left(  \Delta
\mathrm{\dot{H}}\right)  \left(  \Delta\mathrm{H}\right)  }{v^{3}}%
-\frac{\left(  \Delta\mathrm{H}\right)  ^{2}}{v^{4}}\dot{v}\right\rangle
\left\langle \frac{\left(  \Delta\mathrm{H}\right)  \left(  \Delta
\mathrm{\dot{H}}\right)  }{v^{3}}-\frac{\left(  \Delta\mathrm{H}\right)  ^{2}%
}{v^{4}}\dot{v}\right\rangle \nonumber\\
&  =\left[  \frac{\left\langle \left(  \Delta\mathrm{\dot{H}}\right)  \left(
\Delta\mathrm{H}\right)  \right\rangle }{v^{3}}-\frac{\left\langle \left(
\Delta\mathrm{H}\right)  ^{2}\right\rangle }{v^{4}}\dot{v}\right]  \left[
\frac{\left\langle \left(  \Delta\mathrm{H}\right)  \left(  \Delta
\mathrm{\dot{H}}\right)  \right\rangle }{v^{3}}-\frac{\left\langle \left(
\Delta\mathrm{H}\right)  ^{2}\right\rangle }{v^{4}}\dot{v}\right] \nonumber\\
&  =\left[  \frac{\left\langle \left(  \Delta\mathrm{\dot{H}}\right)  \left(
\Delta\mathrm{H}\right)  \right\rangle }{v^{3}}-\frac{\dot{v}}{v^{2}}\right]
\left[  \frac{\left\langle \left(  \Delta\mathrm{H}\right)  \left(
\Delta\mathrm{\dot{H}}\right)  \right\rangle }{v^{3}}-\frac{\dot{v}}{v^{2}%
}\right] \nonumber\\
&  =\frac{\left\langle \left(  \Delta\mathrm{\dot{H}}\right)  \left(
\Delta\mathrm{H}\right)  \right\rangle \left\langle \left(  \Delta
\mathrm{H}\right)  \left(  \Delta\mathrm{\dot{H}}\right)  \right\rangle
}{v^{6}}-\frac{\dot{v}}{v^{5}}\left\langle \left(  \Delta\mathrm{\dot{H}%
}\right)  \left(  \Delta\mathrm{H}\right)  \right\rangle -\frac{\dot{v}}%
{v^{5}}\left\langle \left(  \Delta\mathrm{H}\right)  \left(  \Delta
\mathrm{\dot{H}}\right)  \right\rangle +\frac{\dot{v}^{2}}{v^{4}}\nonumber\\
&  =\frac{\left\langle \left(  \Delta\mathrm{\dot{H}}\right)  \left(
\Delta\mathrm{H}\right)  \right\rangle \left\langle \left(  \Delta
\mathrm{H}\right)  \left(  \Delta\mathrm{\dot{H}}\right)  \right\rangle
}{v^{6}}-\frac{\dot{v}}{v^{5}}\left\langle \left(  \Delta\mathrm{\dot{H}%
}\right)  \left(  \Delta\mathrm{H}\right)  +\left(  \Delta\mathrm{H}\right)
\left(  \Delta\mathrm{\dot{H}}\right)  \right\rangle +\frac{\dot{v}^{2}}%
{v^{4}}\nonumber\\
&  =\frac{\left\langle \left(  \Delta\mathrm{\dot{H}}\right)  \left(
\Delta\mathrm{H}\right)  \right\rangle \left\langle \left(  \Delta
\mathrm{H}\right)  \left(  \Delta\mathrm{\dot{H}}\right)  \right\rangle
}{v^{6}}-\frac{\dot{v}}{v^{5}}\left(  2v\dot{v}\right)  +\frac{\dot{v}^{2}%
}{v^{4}}\nonumber\\
&  =\frac{\left\langle \left(  \Delta\mathrm{\dot{H}}\right)  \left(
\Delta\mathrm{H}\right)  \right\rangle \left\langle \left(  \Delta
\mathrm{H}\right)  \left(  \Delta\mathrm{\dot{H}}\right)  \right\rangle
}{v^{6}}-\frac{\dot{v}^{2}}{v^{4}}\text{,}%
\end{align}
that is,%
\begin{equation}
\tilde{Q}\left(  t\right)  \overset{\text{def}}{=}\left\langle \left(
\frac{\Delta\mathrm{\dot{H}}}{v^{2}}-\frac{\Delta\mathrm{H}}{v^{3}}\dot
{v}\right)  \left(  \frac{\Delta\mathrm{H}}{v}\right)  \right\rangle
\left\langle \left(  \frac{\Delta\mathrm{H}}{v}\right)  \left(  \frac
{\Delta\mathrm{\dot{H}}}{v^{2}}-\frac{\Delta\mathrm{H}}{v^{3}}\dot{v}\right)
\right\rangle =\frac{\left\langle \left(  \Delta\mathrm{\dot{H}}\right)
\left(  \Delta\mathrm{H}\right)  \right\rangle \left\langle \left(
\Delta\mathrm{H}\right)  \left(  \Delta\mathrm{\dot{H}}\right)  \right\rangle
}{v^{6}}-\frac{\dot{v}^{2}}{v^{4}}\text{.} \label{q3b}%
\end{equation}
Therefore, combing Eqs. (\ref{q3b}) and (\ref{c5}), $\tau_{\mathrm{AC}}%
^{2}\left(  s\right)  =\left\langle \left(  \Delta h^{\prime}\right)
^{2}\right\rangle -\left\langle \left(  \Delta h^{\prime}\right)  \left(
\Delta h\right)  \right\rangle \left\langle \left(  \Delta h\right)  \left(
\Delta h^{\prime}\right)  \right\rangle $ becomes%
\begin{align}
\tau_{\mathrm{AC}}^{2}\left(  t\right)   &  =\left(  \frac{\left\langle
\left(  \Delta\mathrm{\dot{H}}\right)  ^{2}\right\rangle -\dot{v}^{2}}{v^{4}%
}\right)  -\left(  \frac{\left\langle \left(  \Delta\mathrm{\dot{H}}\right)
\left(  \Delta\mathrm{H}\right)  \right\rangle \left\langle \left(
\Delta\mathrm{H}\right)  \left(  \Delta\mathrm{\dot{H}}\right)  \right\rangle
}{v^{6}}-\frac{\dot{v}^{2}}{v^{4}}\right) \nonumber\\
&  =\frac{\left\langle \left(  \Delta\mathrm{\dot{H}}\right)  ^{2}%
\right\rangle }{v^{4}}-\frac{\dot{v}^{2}}{v^{4}}-\frac{\left\langle \left(
\Delta\mathrm{\dot{H}}\right)  \left(  \Delta\mathrm{H}\right)  \right\rangle
\left\langle \left(  \Delta\mathrm{H}\right)  \left(  \Delta\mathrm{\dot{H}%
}\right)  \right\rangle }{v^{6}}+\frac{\dot{v}^{2}}{v^{4}}\nonumber\\
&  =\frac{v^{2}\left\langle \left(  \Delta\mathrm{\dot{H}}\right)
^{2}\right\rangle -\left\langle \left(  \Delta\mathrm{\dot{H}}\right)  \left(
\Delta\mathrm{H}\right)  \right\rangle \left\langle \left(  \Delta
\mathrm{H}\right)  \left(  \Delta\mathrm{\dot{H}}\right)  \right\rangle
}{v^{6}}\nonumber\\
&  =\frac{\left\langle \left(  \Delta\mathrm{H}\right)  ^{2}\right\rangle
\left\langle \left(  \Delta\mathrm{\dot{H}}\right)  ^{2}\right\rangle
-\left\langle \left(  \Delta\mathrm{\dot{H}}\right)  \left(  \Delta
\mathrm{H}\right)  \right\rangle \left\langle \left(  \Delta\mathrm{H}\right)
\left(  \Delta\mathrm{\dot{H}}\right)  \right\rangle }{v^{6}}\nonumber\\
&  =\frac{\sigma_{\mathrm{HH}}^{2}\sigma_{\mathrm{\dot{H}\dot{H}}}^{2}%
-\sigma_{\mathrm{\dot{H}H}}^{2}\sigma_{\mathrm{H\dot{H}}}^{2}}{v^{6}}\text{,}%
\end{align}
that is,%
\begin{equation}
\tau_{\mathrm{AC}}^{2}\left(  t\right)  =\frac{\sigma_{\mathrm{HH}}^{2}%
\sigma_{\mathrm{\dot{H}\dot{H}}}^{2}-\sigma_{\mathrm{H\dot{H}}}^{2}%
\sigma_{\mathrm{\dot{H}H}}^{2}}{v^{6}}=\frac{1}{v^{6}}\det\left(
\begin{array}
[c]{cc}%
\sigma_{\mathrm{HH}}^{2} & \sigma_{\mathrm{H\dot{H}}}^{2}\\
\sigma_{\mathrm{\dot{H}H}}^{2} & \sigma_{\mathrm{\dot{H}\dot{H}}}^{2}%
\end{array}
\right)  =\frac{\left\vert \Sigma\left(  \mathrm{H}\text{, }\mathrm{\dot{H}%
}\right)  \right\vert }{v^{6}} \label{myheart}%
\end{equation}
Eq. (\ref{myheart}) expresses the fact that $\tau_{\mathrm{AC}}^{2}\left(
t\right)  \geq0$ is, modulo the $1/v^{6}$ term, equal to the generalized
variance $\left\vert \Sigma\left(  \mathrm{H}\text{, }\mathrm{\dot{H}}\right)
\right\vert $, the determinant of the covariance matrix $\Sigma\left(
\mathrm{H}\text{, }\mathrm{\dot{H}}\right)  $ between the Hamiltonian operator
\textrm{H }and its first derivative \textrm{\.{H}}. The derivation of Eq.
(\ref{myheart}) ends our first step. In the next step, we show that
$\tau_{\mathrm{AC}}^{2}\left(  t\right)  $ in\ Eq. (\ref{myheart}) vanishes.

\subsection{Vanishing torsion}

In the quantum setting, the generalized variance (GV) \cite{wilks32} is a
scalar quantity defined as the determinant of the covariance matrix for a
vector of quantum observables. The GV\ is an overall measure of dispersion
that takes into account the correlations among all pairs of observables that
specify the vector of quantum observables. The GV is large (small,
respectively) when there is weak (strong, respectively) correlation among the
observables. We shall see that $\left\vert \Sigma\left(  \mathrm{H}\text{,
}\mathrm{\dot{H}}\right)  \right\vert $ vanishes when \textrm{H}$\left(
t\right)  =\mathbf{m}\left(  t\right)  \cdot\vec{\sigma}$, regardless of of
the specific form of the vector $\mathbf{m}\left(  t\right)  $. As a
preliminary remark, we point out that $\mathrm{H}$ and $\mathrm{\dot{H}}$ are
incompatible observables, in general, since $\left[  \mathrm{H}\text{,
\textrm{\.{H}}}\right]  =2i\left(  \mathbf{m\times\dot{m}}\right)  \cdot
\vec{\sigma}$ is a nonvanishing operator unless $\mathbf{m}$ and
$\mathbf{\dot{m}}$ are collinear (i.e., unless the external magnetic field
changes only in magnitude but not in direction). Observe that the covariance
matrix $\Sigma\left(  \mathrm{H}\text{, }\mathrm{\dot{H}}\right)  $ can be
recast as,%
\begin{align}
\Sigma\left(  \mathrm{H}\text{, }\mathrm{\dot{H}}\right)   &  =\left(
\begin{array}
[c]{cc}%
\sigma_{\mathrm{HH}}^{2} & \sigma_{\mathrm{H\dot{H}}}^{2}\\
\sigma_{\mathrm{\dot{H}H}}^{2} & \sigma_{\mathrm{\dot{H}\dot{H}}}^{2}%
\end{array}
\right) \nonumber\\
&  =\left(
\begin{array}
[c]{cc}%
\left\langle (\Delta\mathrm{H})^{2}\right\rangle  & \left\langle
(\Delta\mathrm{H})\left(  \Delta\mathrm{\dot{H}}\right)  \right\rangle \\
\left\langle \left(  \Delta\mathrm{\dot{H}}\right)  (\Delta\mathrm{H}%
)\right\rangle  & \left\langle (\Delta\mathrm{\dot{H}})^{2}\right\rangle
\end{array}
\right) \nonumber\\
&  =\left(
\begin{array}
[c]{cc}%
\left\langle \mathrm{H}^{2}\right\rangle -\left\langle \mathrm{H}\right\rangle
^{2} & \left\langle \mathrm{H\dot{H}}\right\rangle -\left\langle
\mathrm{H}\right\rangle \left\langle \mathrm{\dot{H}}\right\rangle \\
\left\langle \mathrm{\dot{H}H}\right\rangle -\left\langle \mathrm{\dot{H}%
}\right\rangle \left\langle \mathrm{H}\right\rangle  & \left\langle
\mathrm{\dot{H}}^{2}\right\rangle -\left\langle \mathrm{\dot{H}}\right\rangle
^{2}%
\end{array}
\right) \nonumber\\
&  =\left(
\begin{array}
[c]{cc}%
\left\langle \mathrm{H}^{2}\right\rangle -\left\langle \mathrm{H}\right\rangle
^{2} & \frac{\left\langle \left[  \mathrm{H}\text{, }\mathrm{\dot{H}}\right]
+\left\{  \mathrm{H}\text{, }\mathrm{\dot{H}}\right\}  \right\rangle }%
{2}-\left\langle \mathrm{H}\right\rangle \left\langle \mathrm{\dot{H}%
}\right\rangle \\
\frac{\left\langle \left[  \mathrm{\dot{H}}\text{, }\mathrm{H}\right]
+\left\{  \mathrm{\dot{H}}\text{, }\mathrm{H}\right\}  \right\rangle }%
{2}-\left\langle \mathrm{H}\right\rangle \left\langle \mathrm{\dot{H}%
}\right\rangle  & \left\langle \mathrm{\dot{H}}^{2}\right\rangle -\left\langle
\mathrm{\dot{H}}\right\rangle ^{2}%
\end{array}
\right) \nonumber\\
&  =\left(
\begin{array}
[c]{cc}%
\left\langle \mathrm{H}^{2}\right\rangle -\left\langle \mathrm{H}\right\rangle
^{2} & \frac{\left\langle \left\{  \mathrm{H}\text{, }\mathrm{\dot{H}%
}\right\}  \right\rangle +\left\langle \left[  \mathrm{H}\text{, }%
\mathrm{\dot{H}}\right]  \right\rangle }{2}-\left\langle \mathrm{H}%
\right\rangle \left\langle \mathrm{\dot{H}}\right\rangle \\
\frac{\left\langle \left\{  \mathrm{H}\text{, }\mathrm{\dot{H}}\right\}
\right\rangle -\left\langle \left[  \mathrm{H}\text{, }\mathrm{\dot{H}%
}\right]  \right\rangle }{2}-\left\langle \mathrm{H}\right\rangle \left\langle
\mathrm{\dot{H}}\right\rangle  & \left\langle \mathrm{\dot{H}}^{2}%
\right\rangle -\left\langle \mathrm{\dot{H}}\right\rangle ^{2}%
\end{array}
\right) \nonumber\\
&  =\left(
\begin{array}
[c]{cc}%
\left\langle \mathrm{H}^{2}\right\rangle -\left\langle \mathrm{H}\right\rangle
^{2} & \frac{\left\langle \left\{  \mathrm{H}\text{, }\mathrm{\dot{H}%
}\right\}  \right\rangle -2\left\langle \mathrm{H}\right\rangle \left\langle
\mathrm{\dot{H}}\right\rangle }{2}+\frac{\left\langle \left[  \mathrm{H}%
\text{, }\mathrm{\dot{H}}\right]  \right\rangle }{2}\\
\frac{\left\langle \left\{  \mathrm{H}\text{, }\mathrm{\dot{H}}\right\}
\right\rangle -2\left\langle \mathrm{H}\right\rangle \left\langle
\mathrm{\dot{H}}\right\rangle }{2}-\frac{\left\langle \left[  \mathrm{H}%
\text{, }\mathrm{\dot{H}}\right]  \right\rangle }{2} & \left\langle
\mathrm{\dot{H}}^{2}\right\rangle -\left\langle \mathrm{\dot{H}}\right\rangle
^{2}%
\end{array}
\right)  \text{,}%
\end{align}
that is,%
\begin{equation}
\Sigma\left(  \mathrm{H}\text{, }\mathrm{\dot{H}}\right)  =\left(
\begin{array}
[c]{cc}%
\left\langle \mathrm{H}^{2}\right\rangle -\left\langle \mathrm{H}\right\rangle
^{2} & \frac{\left\langle \left\{  \mathrm{H}\text{, }\mathrm{\dot{H}%
}\right\}  \right\rangle -2\left\langle \mathrm{H}\right\rangle \left\langle
\mathrm{\dot{H}}\right\rangle }{2}+\frac{\left\langle \left[  \mathrm{H}%
\text{, }\mathrm{\dot{H}}\right]  \right\rangle }{2}\\
\frac{\left\langle \left\{  \mathrm{H}\text{, }\mathrm{\dot{H}}\right\}
\right\rangle -2\left\langle \mathrm{H}\right\rangle \left\langle
\mathrm{\dot{H}}\right\rangle }{2}-\frac{\left\langle \left[  \mathrm{H}%
\text{, }\mathrm{\dot{H}}\right]  \right\rangle }{2} & \left\langle
\mathrm{\dot{H}}^{2}\right\rangle -\left\langle \mathrm{\dot{H}}\right\rangle
^{2}%
\end{array}
\right)  \text{,} \label{cm}%
\end{equation}
where $\left\langle \left\{  \mathrm{H}\text{, }\mathrm{\dot{H}}\right\}
\right\rangle $ and $\left\langle \left[  \mathrm{H}\text{, }\mathrm{\dot{H}%
}\right]  \right\rangle $ are real and purely imaginary quantities,
respectively. Note that the off-diagonal terms of the covariance matrix
$\Sigma\left(  \mathrm{H}\text{, }\mathrm{\dot{H}}\right)  $ are the complex
conjugate of each other. From Eq. (\ref{cm}), we have that the GV is given by%
\begin{align}
\left\vert \Sigma\left(  \mathrm{H}\text{, }\mathrm{\dot{H}}\right)
\right\vert  &  =\left[  \left\langle \mathrm{H}^{2}\right\rangle
-\left\langle \mathrm{H}\right\rangle ^{2}\right]  \left[  \left\langle
\mathrm{\dot{H}}^{2}\right\rangle -\left\langle \mathrm{\dot{H}}\right\rangle
^{2}\right]  -\left(  \frac{\left\langle \left\{  \mathrm{H}\text{,
}\mathrm{\dot{H}}\right\}  \right\rangle -2\left\langle \mathrm{H}%
\right\rangle \left\langle \mathrm{\dot{H}}\right\rangle }{2}+\frac
{\left\langle \left[  \mathrm{H}\text{, }\mathrm{\dot{H}}\right]
\right\rangle }{2}\right)  \cdot\nonumber\\
&  \cdot\left(  \frac{\left\langle \left\{  \mathrm{H}\text{, }\mathrm{\dot
{H}}\right\}  \right\rangle -2\left\langle \mathrm{H}\right\rangle
\left\langle \mathrm{\dot{H}}\right\rangle }{2}-\frac{\left\langle \left[
\mathrm{H}\text{, }\mathrm{\dot{H}}\right]  \right\rangle }{2}\right)
\nonumber\\
& \nonumber\\
&  =\left\langle \mathrm{H}^{2}\right\rangle \left\langle \mathrm{\dot{H}}%
^{2}\right\rangle -\left\langle \mathrm{H}^{2}\right\rangle \left\langle
\mathrm{\dot{H}}\right\rangle ^{2}-\left\langle \mathrm{H}\right\rangle
^{2}\left\langle \mathrm{\dot{H}}^{2}\right\rangle +\left\langle
\mathrm{H}\right\rangle ^{2}\left\langle \mathrm{\dot{H}}\right\rangle
^{2}-\frac{\left\vert \left\langle \left\{  \mathrm{H}\text{, }\mathrm{\dot
{H}}\right\}  \right\rangle \right\vert ^{2}}{4}-\left\langle \mathrm{H}%
\right\rangle ^{2}\left\langle \mathrm{\dot{H}}\right\rangle ^{2}+\nonumber\\
&  +\left\langle \left\{  \mathrm{H}\text{, }\mathrm{\dot{H}}\right\}
\right\rangle \left\langle \mathrm{H}\right\rangle \left\langle \mathrm{\dot
{H}}\right\rangle -\frac{\left\vert \left\langle \left[  \mathrm{H}\text{,
}\mathrm{\dot{H}}\right]  \right\rangle \right\vert ^{2}}{4}\text{.}
\label{cul1}%
\end{align}
Interestingly, note that%
\begin{equation}
\left\langle \mathrm{H}^{2}\right\rangle \left\langle \mathrm{\dot{H}%
}\right\rangle ^{2}+\left\langle \mathrm{H}\right\rangle ^{2}\left\langle
\mathrm{\dot{H}}^{2}\right\rangle -\left\langle \left\{  \mathrm{H}\text{,
}\mathrm{\dot{H}}\right\}  \right\rangle \left\langle \mathrm{H}\right\rangle
\left\langle \mathrm{\dot{H}}\right\rangle =\left\langle \left(  \left\langle
\mathrm{\dot{H}}\right\rangle \mathrm{H}\mathbb{-}\left\langle \mathrm{H}%
\right\rangle \mathrm{\dot{H}}\right)  ^{2}\right\rangle \text{.} \label{cul2}%
\end{equation}
Therefore, using Eqs. (\ref{cul1}) and (\ref{cul2}), we conclude that
$\left\vert \Sigma\left(  \mathrm{H}\text{, }\mathrm{\dot{H}}\right)
\right\vert =0$ if and only if%
\begin{equation}
\left[  \left\langle \mathrm{H}^{2}\right\rangle \left\langle \mathrm{\dot{H}%
}^{2}\right\rangle -\frac{\left\vert \left\langle \left\{  \mathrm{H}\text{,
}\mathrm{\dot{H}}\right\}  \right\rangle \right\vert ^{2}}{4}\right]
-\frac{\left\vert \left\langle \left[  \mathrm{H}\text{, }\mathrm{\dot{H}%
}\right]  \right\rangle \right\vert ^{2}}{4}=\left\langle \left(  \left\langle
\mathrm{\dot{H}}\right\rangle \mathrm{H}\mathbb{-}\left\langle \mathrm{H}%
\right\rangle \mathrm{\dot{H}}\right)  ^{2}\right\rangle \text{,} \label{cul3}%
\end{equation}
that is,%
\begin{equation}
\mu_{\mathrm{H}^{2}}\mu_{\mathrm{\dot{H}}^{2}}=\left\langle \left(
\mu_{\mathrm{\dot{H}}}\mathrm{H}-\mu_{\mathrm{H}}\mathrm{\dot{H}}\right)
^{2}\right\rangle +\frac{\left\vert \mu_{\left\{  \mathrm{H}\text{,
}\mathrm{\dot{H}}\right\}  }\right\vert ^{2}+\left\vert \mu_{\left\langle
\left[  \mathrm{H}\text{, }\mathrm{\dot{H}}\right]  \right\rangle }\right\vert
^{2}}{4}\text{,} \label{cul33}%
\end{equation}
where $\mu_{\mathrm{Q}}\overset{\text{def}}{=}\left\langle \mathrm{Q}%
\right\rangle =\mathrm{tr}\left(  \rho\mathrm{Q}\right)  $ denotes the
expectation value of the operator $\mathrm{Q}$. The statistical relation in
Eq. (\ref{cul33}) happens to be satisfied for any Hamiltonian of the form
\textrm{H}$\left(  t\right)  =\mathbf{m}\left(  t\right)  \cdot\vec{\sigma}$.
Indeed, assuming to evaluate the expectation values with respect to the
density operator $\rho\left(  t\right)  \overset{\text{def}}{=}\left\vert
\psi\left(  t\right)  \right\rangle \left\langle \psi\left(  t\right)
\right\vert =\left[  \mathrm{I}+\mathbf{a}\left(  t\right)  \cdot\vec{\sigma
}\right]  /2$ with $\mathbf{a\cdot a}=1$, Eq. (\ref{cul3}) reduces to%
\begin{equation}
\left[  \mathbf{m}^{2}\mathbf{\dot{m}}^{2}-\left(  \mathbf{m\cdot\dot{m}%
}\right)  ^{2}\right]  -\left[  \mathbf{a\cdot}\left(  \mathbf{m\times\dot{m}%
}\right)  \right]  ^{2}=\left[  \left(  \mathbf{a\cdot\dot{m}}\right)
\mathbf{m-}\left(  \mathbf{a\cdot m}\right)  \mathbf{\dot{m}}\right]
^{2}\text{.} \label{cul4a}%
\end{equation}
Finally, to verify the correctness of the vector identity in Eq.
(\ref{cul4a}), we exploit the Lagrange identity and the vector triple product
relations (i. e., $\left\Vert \mathbf{v}_{1}\times\mathbf{v}_{2}\right\Vert
^{2}=\left(  \mathbf{v}_{1}\cdot\mathbf{v}_{1}\right)  \left(  \mathbf{v}%
_{2}\cdot\mathbf{v}_{2}\right)  -\left(  \mathbf{v}_{1}\cdot\mathbf{v}%
_{2}\right)  ^{2}$ and $\mathbf{v}_{1}\times\left(  \mathbf{v}_{2}%
\times\mathbf{v}_{3}\right)  =\left(  \mathbf{v}_{1}\cdot\mathbf{v}%
_{3}\right)  \mathbf{v}_{2}-\left(  \mathbf{v}_{1}\cdot\mathbf{v}_{2}\right)
\mathbf{v}_{3}$, respectively). We then have that $\mathbf{m}^{2}%
\mathbf{\dot{m}}^{2}-\left(  \mathbf{m\cdot\dot{m}}\right)  ^{2}=\left\Vert
\mathbf{m}\times\mathbf{\dot{m}}\right\Vert ^{2}$, and
%PMA
from the term on the right-hand side of Eq.~(\ref{cul4a}) we have $\left[
\left(  \mathbf{a\cdot\dot{m}}\right)  \mathbf{m-}\left(  \mathbf{a\cdot
m}\right)  \mathbf{\dot{m}}\right]  ^{2}=\left\Vert \mathbf{a\times}\left(
\mathbf{m}\times\mathbf{\dot{m}}\right)  \right\Vert ^{2}=\left(
\mathbf{a\cdot a}\right)  \left\Vert \mathbf{m}\times\mathbf{\dot{m}%
}\right\Vert ^{2}-\left[  \mathbf{a\cdot}\left(  \mathbf{m\times\dot{m}%
}\right)  \right]  ^{2}=\left\Vert \mathbf{m}\times\mathbf{\dot{m}}\right\Vert
^{2}-\left[  \mathbf{a\cdot}\left(  \mathbf{m\times\dot{m}}\right)  \right]
^{2}$. Therefore, we conclude that $\left\vert \Sigma\left(  \mathrm{H}\text{,
}\mathrm{\dot{H}}\right)  \right\vert =0$ and, thus, the torsion coefficient
of any time-dependent quantum evolution for qubit systems is zero. The link in
Eq. (\ref{myheart}) between torsion coefficient $\tau_{\mathrm{AC}}^{2}\left(
t\right)  $ and generalized variance $\left\vert \Sigma\left(  \mathrm{H}%
\text{, }\mathrm{\dot{H}}\right)  \right\vert $ that we found leads to a clear
conclusion. The absence of an overall joint dispersion between the pair
$\left(  \mathrm{H}\text{, \textrm{\.{H}}}\right)  $ as evident from the
perfect correlation between the time-dependent qubit Hamiltonian
\textrm{H}$\left(  t\right)  =\mathbf{m}\left(  t\right)  \cdot\vec{\sigma}$
and its first derivative \textrm{\.{H}}$=\mathbf{\dot{m}}\left(  t\right)
\cdot\vec{\sigma}$ assures that, at all times, there is no deviation of the
evolution state vector from the plane of evolution. In short, the absence of
an overall joint dispersion between the operator responsible for the
generation of quantum motion (i.e., \textrm{H}$\left(  t\right)  $) and its
rate of change (i.e., \textrm{\.{H}}$\left(  t\right)  $) leads to the absence
of any deviation of the state vector from the plane of evolution.

\section{Derivation of $\mathbf{\dot{a}=}2\mathbf{m\times a}$}

In this Appendix, we provide a derivation of the equation $\mathbf{\dot{a}%
=}2\mathbf{m\times a}$. As mentioned in Section IV, assume that the
nonstationary Hamiltonian is given by \textrm{H}$\left(  t\right)
=\mathbf{m}\left(  t\right)  \cdot\vec{\sigma}$, while the pure state is
$\rho\left(  t\right)  =\left\vert \psi\left(  t\right)  \right\rangle
\left\langle \psi\left(  t\right)  \right\vert =\left[  \mathrm{I}%
+\mathbf{a}\left(  t\right)  \cdot\mathbf{\sigma}\right]  /2$. Recall that an
arbitrary qubit observable $Q=q_{0}\mathrm{I}+\mathbf{q\cdot}\vec{\sigma}$
with $q_{0}\in%
%TCIMACRO{\U{211d} }%
%BeginExpansion
\mathbb{R}
%EndExpansion
$ and $\mathbf{q\in%
%TCIMACRO{\U{211d} }%
%BeginExpansion
\mathbb{R}
%EndExpansion
}^{3}$ has a corresponding expectation value given by $\left\langle
Q\right\rangle _{\rho}=q_{0}+\mathbf{a\cdot q}$. Then, the expectation value
$\left\langle \mathrm{\dot{H}}\right\rangle $ becomes%
\begin{equation}
\left\langle \mathrm{\dot{H}}\right\rangle =\mathrm{tr}\left[  \rho
\mathrm{\dot{H}}\right]  =\mathbf{a\cdot\dot{m}}\text{.} \label{yo1}%
\end{equation}
Moreover, employing standard quantum mechanics rules, we obtain
\begin{equation}
\left\langle \mathrm{\dot{H}}\right\rangle =\partial_{t}\left\langle
\mathrm{H}\right\rangle =\partial_{t}\left(  \mathbf{a\cdot m}\right)
=\mathbf{\dot{a}\cdot m+a\cdot\dot{m}}\text{.} \label{yo2}%
\end{equation}
From Eqs. (\ref{yo1}) and (\ref{yo2}), we seem to observe an apparent
incompatibility. Fortunately, this is a false impression since $\mathbf{\dot
{a}\cdot m}=0$ because one can show that $\mathbf{\dot{a}=}2\mathbf{m\times
a}$ and, thus, $\mathbf{\dot{a}}$ is orthogonal
%PMA
to $\mathbf{m}$ (and $\mathbf{a}$) so that $\mathbf{\dot{a}\cdot m=}0$.
Therefore, it is necessary for us to check that $\mathbf{\dot{a}%
=}2\mathbf{m\times a}$. We begin by observing that $\left\langle \vec{\sigma
}\right\rangle $ equals,
\begin{equation}
\left\langle \vec{\sigma}\right\rangle =\mathrm{tr}\left[  \rho\vec{\sigma
}\right]  =\frac{1}{2}\mathrm{tr}\left[  \left(  \mathbf{a\cdot}\vec{\sigma
}\right)  \vec{\sigma}\right]  =\mathbf{a}\text{.} \label{sware}%
\end{equation}
Therefore, exploiting the Schr\"{o}dinger evolution equation $i\hslash
\partial_{t}\left\vert \psi\left(  t\right)  \right\rangle =\mathrm{H}\left(
t\right)  \left\vert \psi\left(  t\right)  \right\rangle $ and putting
$\hslash=1$, the time derivative $\partial_{t}\left\langle \vec{\sigma
}\right\rangle =\mathbf{a}$ of $\left\langle \vec{\sigma}\right\rangle $
in\ Eq. (\ref{sware}) reduces to%
\begin{align}
\mathbf{\dot{a}}  &  =\partial_{t}\left\langle \vec{\sigma}\right\rangle
\nonumber\\
&  =\partial_{t}\left(  \left\langle \psi\left\vert \vec{\sigma}\right\vert
\psi\right\rangle \right) \nonumber\\
&  =\left\langle \dot{\psi}\left\vert \vec{\sigma}\right\vert \psi
\right\rangle +\left\langle \psi\left\vert \vec{\sigma}\right\vert \dot{\psi
}\right\rangle \nonumber\\
&  =i\left\langle \psi\left\vert \mathrm{H}\vec{\sigma}\right\vert
\psi\right\rangle -i\left\langle \psi\left\vert \vec{\sigma}\mathrm{H}%
\right\vert \dot{\psi}\right\rangle \nonumber\\
&  =i\left\langle \psi\left\vert \left[  \mathrm{H}\text{, }\vec{\sigma
}\right]  \right\vert \psi\right\rangle \text{,}%
\end{align}
that is, the time derivative $\mathbf{\dot{a}}$ of the time-varying Bloch
vector $\mathbf{a}$ becomes%
\begin{equation}
\mathbf{\dot{a}}=i\left\langle \psi\left\vert \left[  \mathrm{H}\text{, }%
\vec{\sigma}\right]  \right\vert \psi\right\rangle \text{.} \label{a-dot}%
\end{equation}
Eq. (\ref{a-dot}) can be further simplified by noting that the commutator
$\left[  \mathrm{H}\text{, }\sigma_{j}\right]  $ can be rewritten as
\begin{align}
\left[  \mathrm{H}\text{, }\sigma_{j}\right]   &  =\left[  \mathbf{m}\cdot
\vec{\sigma}\text{, }\sigma_{j}\right] \nonumber\\
&  =\left[  m_{i}\sigma_{i}\text{, }\sigma_{j}\right] \nonumber\\
&  =m_{i}\left[  \sigma_{i}\text{, }\sigma_{j}\right] \nonumber\\
&  =m_{i}\left(  2i\epsilon_{ijk}\sigma_{k}\right) \nonumber\\
&  =-2i\epsilon_{ikj}m_{i}\sigma_{k}\text{,}%
\end{align}
that is to say, for any $1\leq j\leq3$,
\begin{equation}
\left[  \mathrm{H}\text{, }\sigma_{j}\right]  =-2i\epsilon_{ikj}m_{i}%
\sigma_{k}\text{.} \label{b-dot}%
\end{equation}
Therefore, using Eqs. (\ref{a-dot}) and (\ref{b-dot}), we get%
\begin{align}
\left(  \mathbf{\dot{a}}\right)  _{j}  &  \mathbf{=}i\left(  -2i\epsilon
_{ikj}m_{i}\left\langle \sigma_{k}\right\rangle \right) \nonumber\\
&  =2\epsilon_{ikj}m_{i}a_{k}\nonumber\\
&  =2\left(  \mathbf{m\times a}\right)  _{j}\text{,}%
\end{align}
that is, we obtain the equality $\mathbf{\dot{a}=}2\mathbf{m\times a}$
%PMA
which states that $\mathbf{a}(t)$ rotates about the instantaneous ``magnetic
field'' $\mathbf{m}(t)$ defined by the Hamiltonian $H(t) =\mathbf{m}%
(t)\cdot\mathbf{\boldsymbol{\sigma}}$ at each instant of time $t$. Finally, we
can conclude that $\partial_{t}\left(  \mathbf{a\cdot m}\right)
=\mathbf{\dot{a}\cdot m+a\cdot\dot{m}=a\cdot\dot{m}}$ since $\mathbf{\dot
{a}\cdot m=}0$ because $\mathbf{\dot{a}=}2\mathbf{m\times a}$ is perpendicular
to the vector $\mathbf{m}$.

\begin{thebibliography}{99}
\bibitem {footnote1}This result follows from the fact that $\partial_{t}
\langle H(t) \rangle\equiv\partial_{t} \langle\psi(t)| H(t) |\psi(t)\rangle=
\langle\partial_{t}\psi(t)| H(t) |\psi(t)\rangle+ \langle\psi(t)| H(t)
|\partial_{t}\psi(t)\rangle+ \langle\psi(t)| \partial_{t}H(t) |\psi(t)\rangle=
\langle\dot{H}(t)\rangle$ since the first two terms cancel via Schr\"odinger's
equation $i \partial_{t}|\psi(t)\rangle| \equiv H(t) |\psi(t)\rangle$. Note,
for an arbitrary time-dependent operator $\mathcal{O}(t)$ one has in general
$\partial_{t} \langle\mathcal{O}(t) \rangle= i\, \langle[H(t),\mathcal{O}(t)]
\rangle+ \langle\partial_{t} \mathcal{O}(t)\rangle$
\bibitem {footnote2}The expressions on the right-hand side of Eq.(\ref{form1})
and Eq.(\ref{form2}) could be further simplified if one instead used the
symmetric covariance between two operators $A$ and $B$ defined as
$\text{cov}^{(sym)}(A,B)=\frac{1}{2}\langle AB+BA\rangle-\langle
A\rangle\langle B\rangle$.
\end{thebibliography}
\end{document}